\documentclass[a4paper,11pt]{article}
\usepackage{jheppub} 
\usepackage[T1]{fontenc} 
\usepackage{graphicx}
\usepackage{bm}
\usepackage{amsmath}
\usepackage{amssymb}
\usepackage{url}
\usepackage{epsfig}
\usepackage[T1]{fontenc}
\usepackage[latin9]{inputenc}
\usepackage{multirow}
\usepackage{xspace}
\usepackage{xcolor}
\usepackage{mathtools}
\usepackage[justification=raggedright]{caption}
\usepackage{hyperref}
\usepackage{float}
\usepackage{microtype}
\usepackage{cprotect}
\usepackage{makecell}
\usepackage{float}

\usepackage{listings}
\newcommand{\di}{i}          

\newcommand{\val}{\mathrm{val}}
\newcommand{\sea}{\mathrm{sea}}

\newcommand{\herwig}{H\protect\scalebox{0.8}{ERWIG}7\xspace}
\newcommand{\pythia}{P\protect\scalebox{0.8}{YTHIA}\xspace}
\newcommand{\sherpa}{S\protect\scalebox{0.8}{HERPA}\xspace}

\newcommand*\phantomrel[1]{\mathrel{\phantom{#1}}}

\def\beq{\begin{equation}}
\def\eeq{\end{equation}}
\def\bear{\begin{eqnarray}}
\def\eear{\end{eqnarray}}

\title{\Large \textbf{On  sum rules for double and triple parton distribution functions and \pythia's model of multiple parton interactions}}

\author[a]{Oleh Fedkevych,} 
\author[b]{Jonathan R. Gaunt.}

\affiliation[a]{Dipartimento di Fisica, Universit\`a di Genova and INFN, Sezione di Genova,\\ Via Dodecaneso 33, 16146, Genoa, Italy}
\affiliation[b]{Department of Physics and Astronomy, University of Manchester, Manchester, M13 9PL,United Kingdom}

\emailAdd{oleh.fedkevych@ge.infn.it}
\emailAdd{jonathan.gaunt@manchester.ac.uk}

\abstract{Multi-parton distributions in a proton, the nonperturbative quantities needed to make predictions for multiple scattering rates, are poorly constrained from theory and data and must be modelled. 
All Monte Carlo event generators that simulate multiple parton interactions (\textit{e.g.} \pythia) contain such a model of multi-parton PDFs. 
One important theoretical constraint for the case of double parton distributions is provided by the so-called number and momentum sum rules. In this paper we investigate to what extent the double parton distribution functions used in the \pythia event generator obey these sum rules. 
We also derive the number and momentum sum rules for the triple parton distribution functions and discuss how one can use the \pythia code to construct triple parton distribution functions which approximately satisfy these sum rules.}

\begin{document}

\begin{flushright}
  LU-TP-20-17\\
  MCNET-22-15
\end{flushright}
\vspace{1cm}

\maketitle

\setcounter{footnote}{0}

\section{Introduction}
\label{s:intro}
\textit{Double parton scattering} (DPS) is a  process where two hard
interactions occur in one hadron-hadron collision. 
Such processes have been measured in various experiments 
for \mbox{\textit{proton-proton} (pp)}, \mbox{\mbox{\textit{proton-antiproton}}} (p$\bar{\rm p}$)  and \mbox{\mbox{\textit{proton-nucleus}}} (pA) collisions~\cite{AxialFieldSpectrometer:1986dfj, UA2:1991apc, CDF:1993sbj,
CDF:1997lmq,  CDF:1997yfa, D0:2009apj, ATLAS:2016ydt, LHCb:2012aiv,
ATLAS:2013aph, CMS:2013huw, D0:2014vql, D0:2014owy, ATLAS:2014ofp, CMS:2013slh, LHCb:2015wvu, D0:2015dyx, D0:2015rpo, ATLAS:2016rnd, CMS:2017han, CMS:2021ijt, LHCb:2020jse, CMS:2021lxi, CMS:2022pio}. 
DPS processes are sensitive to partonic correlations of various types \textit{e.g.}~\cite{Korotkikh:2004bz,Cattaruzza:2005nu,Gaunt:2009re,Calucci:1997ii,Frankfurt:2003td,Calucci:1999yz,Calucci:2009sv,Calucci:2010wg,Rogers:2009ke, Domdey:2009bg,Flensburg:2011kj,Blok:2012mw,Seymour:2013sya,Blok:2013bpa, Rinaldi:2018slz} and their study can provide a deeper insight into hadron structure.

\textit{Double parton distribution functions} (dPDFs) are crucial ingredients of theoretical predictions for DPS, encoding the correlations between the two partons arising from a given hadron\footnote{
Here, we shall use the term ``dPDF'' to refer to the double parton density integrated over the partonic transverse separation $\bm{y}$. 
We will discuss the definition in detail in Section~\ref{s:gs09_pythia}.
}.
Unfortunately, due to the absence of experimental fits, the dPDFs remain unknown. 
Therefore, in phenomenological studies of the DPS  phenomenon one still has to rely on certain model dependent assumptions about the shape of dPDFs  and their relation to standard \textit{single parton distribution functions} (sPDFs). 
There are different approaches to modelling dPDFs. 
To name some we mention the Light-Front approach~\cite{Rinaldi:2014ddl,Rinaldi:2016mlk,Ceccopieri:2017oqe,Rinaldi:2018bsf,Courtoy:2019cxq, Broniowski:2019rmu}, the ``bag model'' of a proton~\cite{Chang:2012nw}, the AdS/QCD method~\cite{Traini:2016jru, Rinaldi:2020ybv}, chiral quark models~\cite{Broniowski:2019rmu} and the models based on the solution of so-called ``double'' DGLAP evolution equations~\cite{Gaunt:2009re, Nagar:2019gij, Snigirev:2003cq, Korotkikh:2004bz, Diehl:2017kgu}. First-principles calculations of the lowest Mellin moments of double parton distributions have also now been performed in lattice QCD~\cite{Bali:2018nde, Bali:2020mij, Bali:2021gel}.

Apart from  the aforementioned approaches to the dPDFs there exist a broad  class of \textit{Monte Carlo} (MC) models of \textit{multiple partonic interactions} (MPI) as implemented in general purpose event  generators, \textit{e.g.} \pythia~\cite{Sjostrand:2006za,Sjostrand:2007gs, Bierlich:2022pfr},  \herwig~\cite{Bellm:2015jjp} and \sherpa~\cite{Gleisberg:2008ta}.
The  detailed description of MPI models being used in \pythia and \herwig can be found in corresponding publications~\cite{Sjostrand:1985vv, Sjostrand:1987su, Sjostrand:2004pf,Corke:2009tk,Corke:2011yy,Bahr:2008dy,Gieseke:2016fpz}. 
The MPI model of  \sherpa  is based upon the old MPI model of \pythia  \cite{Sjostrand:1987su} with some modifications that are described in  Ref.~\cite{Gleisberg:2008ta}\footnote{
There is also a new MPI model under development, see Ref.~\citep{Martin:2012nm}.
}. 
These MPI models are also widely used to simulate DPS 
processes. 

Despite the variety of the models of dPDFs available on the market, there are common criteria dictated by first principles QCD which, to some extent, have to be preserved by each model of dPDFs.  
One such criteria is that the dPDFs in which the two partons are probed at the same scale (`equal-scale dPDFs') should satisfy the sum rules originally proposed by Gaunt and Stirling (\textit{GS sum rules})~\cite{Gaunt:2009re, Blok:2013bpa, Ceccopieri:2014ufa, Diehl:2018kgr}. 
In this paper we  demonstrate how one can use the MPI model of \pythia to construct sets of equal-scale dPDFs which approximately obey the GS sum rules, and quantify the extent to which these sum rules are satisfied. The \pythia model always considers MPI as an ordered sequence of interactions, and thus naturally yields dPDFs that are not symmetric in the first and second parton arguments, even in the equal scale case. We find that these `asymmetric' dPDFs satisfy the GS sum rules when integrating over the $x$ fraction of the `second struck' parton at the level of a few percent  when the other $x$ argument is in between $10^{-6}$ and $0.8$. Since the equal-scale dPDFs should in fact be symmetric in the first and second parton arguments, we then generate symmetrised versions of these dPDFs via a simple procedure.
We find that these dPDFs obey the GS sum rules (when integrating over either $x$ argument) at about $20\%$ accuracy level except for  $x$ fractions bigger than 0.4. 
As a benchmark we use the  GS09 \citep{Gaunt:2009re} set of dPDFs which represents solutions of the inhomogeneous ``double'' DGLAP evolution equations for an input which was constructed to approximately respect the GS sum rules.

Whereas the dPDFs and the corresponding sum rules have been extensively studied in the literature,  the  case of \textit{triple parton distribution functions} (tPDFs) has  received rather less  attention. 
The evolution of the triple parton densities was studied in Ref.~\cite{Snigirev:2016uaq} and theoretical aspects of correlations in MPI have been studied in Refs.~\cite{Calucci:2009sv, Calucci:2009ea, Treleani:2012zi}, however, the  available phenomenological studies of \textit{triple parton scattering} (TPS)~\cite{Maina:2009sj, Maciula:2017wpe, dEnterria:2017yhd, Maciula:2017meb, dEnterria:2016yhy,  dEnterria:2016ids} rely on factorization of tPDFs into products of three sPDFs which essentially neglects correlations between partons\footnote{The recent CMS measurements~\cite{CMS:2021qsn} suggest a non-negligible  TPS contribution  to  triple $J/\psi$ production.
}.  
Moreover, to the best of our knowledge, no set of tPDFs  satisfying the tPDF evolution equations has so far been constructed.
Therefore, in the absence of such ``first principles'' sets of tPDFs, it is useful to have a method to build tPDFs that account for correlations in $x$-space related to momentum and valence number constraints  in an approximate way. 
In this work we demonstrate how one can use the \pythia  code to accomplish this task.

This paper is organized as follows: in \mbox{Section \ref{s:gs09_pythia}} we provide a short description of the GS09 model and the model being used in the \pythia event generator and in Sections~\ref{s:results_dPDFs}--\ref{s:results_tPDFs} we provide our findings. 
Namely, in Section~\ref{s:results_dPDFs}  we demonstrate how one can use \pythia  code to construct asymmetric and symmetric dPDFs and study how well they satisfy the GS sum rules as well as perform some toy phenomenological studies for the double Drell-Yan process. Then, in Section~\ref{s:results_tPDFs} we derive number and momentum sum rules for the case of TPS, and demonstrate how one  can use the \pythia code to create tPDFs which approximately obey such sum rules. 
Finally, in Section \ref{s:summary} we give our conclusions and some closing remarks.

\section{GS09 and \pythia models of dPDFs}
\label{s:gs09_pythia}
\subsection{Double parton distribution functions and the GS09 dPDFs}
\label{ss:gs09_intro}

In this subsection we introduce the theoretical definition of dPDFs, and discuss their key properties including the GS sum rules. 
We will also briefly describe the GS09 dPDF set~\cite{Gaunt:2009re}, and discuss simplifying assumptions that have been used in the past in order to utilise dPDFs directly in DPS cross section predictions.

First of all, consider the total DPS cross section  for the production of final states $A$ and $B$ in the collision of hadrons $h_A$ and $h_B$:\footnote{
The parton model derivation of the expression for the total DPS cross section was given in Refs.~\citep{Paver:1982yp, Mekhfi:1983az, Diehl:2011tt, Diehl:2011yj} 
and the full QCD derivation was provided in Refs.~\citep{Diehl:2015bca, Diehl:2017kgu, Diehl:2018wfy, Vladimirov:2017ksc}.
}
\begin{align}
	\sigma^{\rm DPS}_{AB}  &=
	\frac{1}{1 + \delta_{AB}}
	\sum\limits_{j_1, \, j_2, \, j_3, j_4}
	\int_{|\bm{y}| > \max(Q^{-1}_1,Q^{-1}_2)} \, \prod\limits^{4}_{i = 1} dx_i \, d^2y \,
	\Gamma_{j_1 \, j_2 / h_A}(x_1, x_2, \bm{y}, Q_1, Q_2) \nonumber\\
	&\phantomrel{=}\times\, 	
	\Gamma_{j_3 \, j_4 / h_B}(x_3, x_4, \bm{y}, Q_1, Q_2) \, 
	\hat{\sigma}_{j_1 \, j_3 \rightarrow A} \,
	\hat{\sigma}_{j_2 \, j_4 \rightarrow B}.
	\label{eq:dps_cross_secton_non_factorized}
\end{align}
Following \cite{Diehl:2011tt, Diehl:2011yj,Diehl:2017kgu}, we shall refer to the objects {\small $\Gamma_{j_1 \, j_2 /h}(x_1, x_2, \bm{y}, Q_1, Q_2)$} as the double parton distributions (DPDs) (they are also referred to as ``two parton generalized parton distributions'' or 2pGPDs in the literature \cite{Blok:2011bu, Blok:2012mw, Gaunt:2012dd}). Loosely speaking, these can be interpreted as a probability to find two partons $j_1$ and $j_2$ with longitudinal momentum fractions $x_1$ and $x_2$
separated by transverse distance $\bm{y}$ in a hadron $h$.  
A lower cut-off on the transverse separation $|\bm{y}|$ is included since the region of very small $|\bm{y}|$ values is more appropriately described by single parton scattering (for more details see Ref.~\cite{Diehl:2017kgu}).

Let us define the dPDFs as the integral of the DPD functions over $\bm{y}$, as follows:
\begin{equation} \label{eq:dPDFdef}
D_{j_1 \, j_2 / h}(x_1, x_2, Q_1, Q_2) \equiv \int_{|\bm{y}| > \max(Q^{-1}_1,Q^{-1}_2)}  d^2 y \, \Gamma_{j_1 \, j_2 / h}(x_1, x_2, \bm{y}, Q_1, Q_2). 
\end{equation}

The dPDFs in Eq.~\eqref{eq:dPDFdef} taken at equal factorization scales ($Q_1 = Q_2 = Q \gg \Lambda_{QCD}$) obey  so called \textit{inhomogeneous double DGLAP} (dDGLAP) evolution equations~\cite{Kirschner:1979im, Shelest:1982dg, Zinovev:1982be} which, at the leading logarithmic level, have the form
\begin{align}
	\frac{d D_{j_1 j_2} (x_1, x_2, t)}{dt} &= 
	\frac{\alpha_s(t)}{2\pi} \sum\limits_{j^\prime_1} \, \int\limits^{1 - x_2}_{x_1} \, \frac{dx^\prime_1}{x^\prime_1} \, 
	P_{j^\prime_1 \rightarrow j_1}\left(\frac{x_1}{x^\prime_1}\right)  \, D_{j^\prime_1 j_2}(x^\prime_1, x_2, t) \nonumber\\
	&\phantomrel{=}+\,\frac{\alpha_s(t)}{2\pi} \sum\limits_{j^\prime_2} \, \int\limits^{1 - x_1}_{x_2} \, \frac{dx^\prime_2}{x^\prime_2} \, 
	P_{j^\prime_2 \rightarrow j_2}\left(\frac{x_2}{x^\prime_2}\right) \, D_{j_2 j^\prime_2}(x_1, x^\prime_2, t) \nonumber\\
	&\phantomrel{=}+\,\frac{\alpha_s(t)}{2\pi} \,\sum\limits_{j^\prime} \, 
	P_{j^\prime \rightarrow j_1 j_2}\left(\frac{x_1}{x_1 + x_2}\right) \,f_{j^\prime}(x_1 + x_2, t) \, \frac{1}{x_1 + x_2},
	\label{eq:double_dglap}
\end{align}
where $f_{j^\prime}(x, t)$ are ``standard'' collinear sPDFs and we have dropped the subscript $h_A$ in $D_{j_1 j_2 / h_A}$ in order not to overload the notation\footnote{
Similar evolution equations for quark fragmentation functions were derived in Refs.~\cite{Puhala:1980ms, Sukhatme:1980vs}.
}. 
Here we use  a dimensionless evolution parameter $t = \log(Q^2 / Q^2_0)$  where $Q_0$ is an arbitrary reference scale.
The  first two terms on the right hand side of Eq.~\eqref{eq:double_dglap}  involve standard DGLAP leading order splitting functions $P_{j^\prime_1 \rightarrow j_1}(x)$ and, despite the presence of two Bjorken-$x$'es, essentially have the same structure as the standard DGLAP splitting kernels. 
The last term, however, is new and is linked to the reduction of the cut-off in Eq.~\eqref{eq:dPDFdef} when $Q$ is increased. 
It is an inhomogeneous term that involves the standard  sPDFs and depends on ``$1 \rightarrow 2$'' splitting functions $P_{j^\prime \rightarrow j_1 j_2}(x)$ which are defined at the leading order in $\alpha_s$ via
\begin{eqnarray}
	P_{i \rightarrow j k}(x) = P^R_{i \rightarrow j}(x) \, \delta_{k \kappa(i,j)}
	\label{eq:new_and_old_splitting}
\end{eqnarray}
where $\kappa(i, j)$ is the only parton that can be emitted at leading order when parton $i$ splits to $j$ (\textit{e.g.} if $i = g$ and $j = q$ then $\kappa(i, j) = \bar{q}$) and $P^R_{i \rightarrow j}(x)$ are given by the real-emission parts of the DGLAP splitting functions, see Ref.~\citep{Gaunt:2009re}.
Therefore, in the following, we refer to the third term on the right hand side of Eq.~\eqref{eq:double_dglap} as the ``splitting'' or $1\rightarrow2$ term. 
As we have mentioned before, the dPDFs in Eq.~\eqref{eq:double_dglap} are taken at the same values of the factorization scales. 
The unequal scale dPDFs  can be obtained from the equal scale dPDFs via a single DGLAP evolution step as described in Ref.~\cite{Gaunt:2009re}. 

The dPDFs obey the following sum rules:
\begin{eqnarray}
	&&\sum\limits_{j_2}\int\limits^{1 - x_1}_0 dx_2 \, x_2 \, D_{j_1 j_2} (x_1, x_2, Q) = (1 - x_1)\,f_{j_1} (x_1, Q),
	\label{eq:gs_momentum_rule_inv}\\
	&&\int\limits^{1 - x_1}_0 dx_2 \,  D_{j_1 j_{2v}}(x_1, x_2, Q) = \left( N_{j_{2v}} - \delta_{j_1 j_2} + \delta_{j_1 \bar{j}_2} \right) f_{j_1}(x_1, Q).
	\label{eq:gs_number_rule_inv}
\end{eqnarray}

These dPDF sum rules were first introduced by Gaunt and Stirling in Ref.~\cite{Gaunt:2009re}, who also established that Eqs.~\eqref{eq:gs_momentum_rule_inv}, \eqref{eq:gs_number_rule_inv} are preserved by dDGLAP evolution provided that the dPDFs obey the sum rules at the initial evolution scale $Q_0$ (see also Ref.~\cite{Blok:2013bpa}). 
Thus, we refer to Eq.~\eqref{eq:gs_momentum_rule_inv} and Eq.~\eqref{eq:gs_number_rule_inv} as the {\em GS sum rules}. 
The fundamental character of the GS sum rules was recently demonstrated by Diehl, Pl\"{o}{\ss}l and Sch\"{a}fer in Ref.~\cite{Diehl:2018kgr} where these relations were proved in all orders of perturbation theory for bare dPDFs and for the dPDFs renormalized according to the $\overline{\rm MS}$ scheme (see also Ref.~\cite{Gaunt:2012tfk}). 
The $\overline{\rm MS}$-renormalized dPDFs coincide with the dPDFs in Eq.~\eqref{eq:dPDFdef} up to corrections suppressed by $\mathcal{O}(\alpha_s)$ or $\mathcal{O}(\Lambda_{\rm QCD}^2/Q^2)$ \cite{Diehl:2018kgr}, such that the dPDFs in Eq.~\eqref{eq:dPDFdef} also satisfy the GS sum rules up to these corrections.
%
In Ref.~\cite{Gaunt:2009re}, a set of dPDFs was constructed which approximately satisfies the GS sum rules at the initial scale $Q_0 = 1$ GeV. 
These starting conditions were then evolved according to Eq.~\eqref{eq:double_dglap} to higher scales, up to $Q^2 = 10^9$ GeV$^2$ and for $x_{1,2} \geq 10^{-6}$. 
The dPDFs generated by this procedure are referred to as the GS09 set\footnote{For some phenomenological applications of the GS09 dPDFs see, \textit{e.g.} Refs.~\cite{Gaunt:2010pi, Maina:2010vh, Fedkevych:2020cmd}.}.
 
If one assumes that {\small $\Gamma_{j_1 \, j_2 / h}(x_1, x_2, \bm{y}, Q_1, Q_2)$} can be written as a product of a longitudinal and transverse-dependent pieces, with the transverse piece being some smoothly varying function with a width on the order of the proton radius, then one can write the DPD in terms of the dPDF as follows:
\begin{eqnarray}
	\Gamma_{j_1 \, j_2 / h}(x_1, x_2, \bm{y}, Q_1, Q_2) \approx  D_{j_1 \, j_2 / h}(x_1, x_2, Q_1, Q_2) \, F(\bm{y}),
	\label{eq:factorization_of_dPDF_in_x_and_b}
\end{eqnarray}
provided that $F(\bm{y})$ is normalised as $\int d^2y F(\bm{y}) = 1$. 
Under Eq.~\eqref{eq:factorization_of_dPDF_in_x_and_b}, the DPS cross section in Eq.~\eqref{eq:dps_cross_secton_non_factorized} may be simplified as follows:
\begin{align}
	\sigma^{\rm DPS}_{AB} &=
	\frac{1}{\sigma_{\rm eff}}
	\frac{1}{1 + \delta_{AB}}
	\sum\limits_{j_1, \, j_2, \, j_3, j_4}
	\int \, \prod\limits^{4}_{i = 1} dx_i \,
	D_{j_1 \, j_2 / h_A}(x_1, x_2, Q_1, Q_2) \,
	D_{j_3 \, j_4 / h_B}(x_3, x_4, Q_1, Q_2) \nonumber\\
	&\phantomrel{=}\times\,\hat{\sigma}_{j_1 \, j_3 \rightarrow A} \,
	\hat{\sigma}_{j_2 \, j_4 \rightarrow B},
	\label{eq:dps_cross_secton_dPDFs}
\end{align}
where {\small$\sigma_{\rm eff} = \left[\int d^2y \, F^2(\bm{y})\right]^{-1}$}and has a dimension of area.

Nowadays, it is known that the factorization of the DPDs into the dPDFs and a transverse-dependent piece cannot hold~\cite{Blok:2011bu, Gaunt:2011xd, Blok:2013bpa, Gaunt:2012dd, Ryskin:2011kk,  Ryskin:2012qx,  Manohar:2012pe, Diehl:2017kgu}. 
Actually,  the discussion above already indicates this -- if Eq.~\eqref{eq:factorization_of_dPDF_in_x_and_b} held, then the cut-off in Eq.~\eqref{eq:dPDFdef} would only have a power-suppressed effect, and there would be no inhomogeneous term in Eq.~\eqref{eq:double_dglap}\footnote{A further indication of the inadequacy of Eq.~\eqref{eq:factorization_of_dPDF_in_x_and_b} comes from experimental measurements of DPS. If we take $F(y)$ in Eq.~\eqref{eq:factorization_of_dPDF_in_x_and_b} to have a width of order of the proton radius, and further assume that the dPDFs in Eq.~\eqref{eq:dps_cross_secton_dPDFs} factorise into a product of single PDFs, then we obtain DPS cross sections which are roughly half as large as those obtained experimentally (see e.g. \cite{Blok:2010ge}). Inclusion of perturbative transverse correlation effects yields DPS cross sections in better agreement with data \cite{Blok:2013bpa}.}. 
Nevertheless, when using the GS09 dPDF set in cross section calculations below we will use Eq.~\eqref{eq:dps_cross_secton_dPDFs} -- this is because we would like to make the most direct comparison against the predictions of \pythia, which uses a framework for DPS/MPI that is most closely aligned with Eq.~\eqref{eq:dps_cross_secton_dPDFs}.

It is important to remark that although dPDFs satisfying the GS sum rules do not directly appear in the DPS cross section Eq.~\eqref{eq:dps_cross_secton_non_factorized}, they are linked to the DPDs that do, via Eq.~\eqref{eq:dPDFdef}. 
Any procedure that yields dPDFs that (approximately) satisfy the sum rules is thus of interest and utility even under the theoretically rigorous approach associated with Eq.~\eqref{eq:dps_cross_secton_non_factorized}.
It is also worth remarking that it is possible to adapt techniques used to construct dPDFs satisfying the sum rules to construct DPDs which have both the appropriate transverse dependence and which integrate to dPDFs satisfying the sum rules -- in Ref.~\cite{Diehl:2020xyg} the techniques used to construct the GS09 dPDFs were adapted, extended and improved to yield such a set of DPDs.

\subsection{The PYTHIA model of multiple parton distribution functions}
\label{ss:pythia_model_of_MPI}
Now let us  briefly sketch the way  the \pythia event generator models \textit{$n$-parton distribution functions} (nPDFs) and, more specifically, dPDFs. 
We remind the reader that these dPDFs (and $n$PDFs) are applied to compute cross sections according to a framework that resembles Eq.~\eqref{eq:dps_cross_secton_dPDFs}; the transverse dependence is assumed to essentially factor out. 
The generation of the first hard interaction is performed in the usual way, for example, for the case of massless $2\rightarrow2$ scattering we have
\begin{eqnarray}
	\frac{d\sigma}{d p^2_\perp} = \sum_{j_1 \, j_2 \, j_3 \, j_4} \int d x_1 \, d x_2 \, d \hat{t} \, 
	f_{j_1} (x_1, Q) \, f_{j_2} (x_2, Q) \, 
	\frac{ d\sigma_{j_1 \, j_2 \rightarrow j_3 \, j_4} }{ d\hat{t} } \,
	\delta\left(p^2_\perp - \frac{\hat{t}\hat{u}}{\hat{s}}\right).
	\label{eq:pythia_2_to_2}
\end{eqnarray}
After the successful generation of the first hard interaction, one has to generate other subsequent interactions by taking into account MPIs already generated in a given \mbox{pp} collision. In the general MPI model of \pythia these interactions are generated at successively lower scales. There is also an option to generate DPS events specifically (controlled by the option ``{\tt SecondHard}'') -- here both orderings of the two processes are simulated  (see the beginning of Section \ref{ss:double_DY_pheno} for more details).
To generate the additional interaction(s) \pythia dynamically modifies sPDFs in Eq.~\eqref{eq:pythia_2_to_2}  according to a history of  previous interactions. 
In the following, while  describing \pythia's framework, we use subscripts  
``$r$'' and ``$m$'' to distinguish between unmodified  (``raw'') and modified sPDFs.

In the first model of MPI~\citep{Sjostrand:1985vv, Sjostrand:1987su} \pythia could only preserve overall energy and momentum by 
`squeezing' the PDFs for the $n$th interaction such that instead of lying between $0$ and $1$ in the momentum fraction $x_n$, they lie in between $0$ and $X_n \equiv 1 - \sum^{n -1}_{i = 1} x_i$, where $\sum^{n -1}_{i = 1} x_i$ is the total longitudinal momentum taken by previous interactions. This squeezing is achieved by rescaling the $x$-argument in the PDFs to be
\begin{eqnarray}
	 x^\prime_n = \frac{x_n}{X_n}  \,
	\label{eq:pythia_squezing}
\end{eqnarray}
Since the $x$-argument in a PDF cannot exceed $1$, this implies $x_n < X_n < 1$. Aside from this argument rescaling, one also multiplies the PDF by $1/X_n$.

In the case $n=2$ we thus have a dPDF defined by:
\begin{eqnarray}
	D_{j_1 \, j_2}(x_1, x_2, Q) =  f_{j_1}^{r}(x_1, Q) \, f_{j_2}^{m\leftarrow j_1, x_1}(x_2, Q),
	\label{eq:pythia_dPDFs}
\end{eqnarray}
where $f_{j_1}^{r}(x_1, Q)$ is the unmodified sPDF and 
\begin{eqnarray}
	f_{j_2}^{m\leftarrow j_1, x_1}(x_2, Q) = 
	\frac{1}{1 - x_1}f_{j_2}^{r}\left(\frac{x_2}{1 - x_1}, Q\right)
	\label{eq:pythia_squezing_1}
\end{eqnarray}
is the sPDF modified according to the first interaction involving parton $j_1$ with momentum fraction $x_1$. Since the $x$-argument of $f_{j}^{r}(x, Q)$ cannot exceed $1$, this construction implies that the dPDF is only nonzero when $x_1+x_2 \le 1$.

\begin{figure}[h!]
\begin{minipage}[h]{0.48\linewidth}
\center{\includegraphics[width=1.0\linewidth]{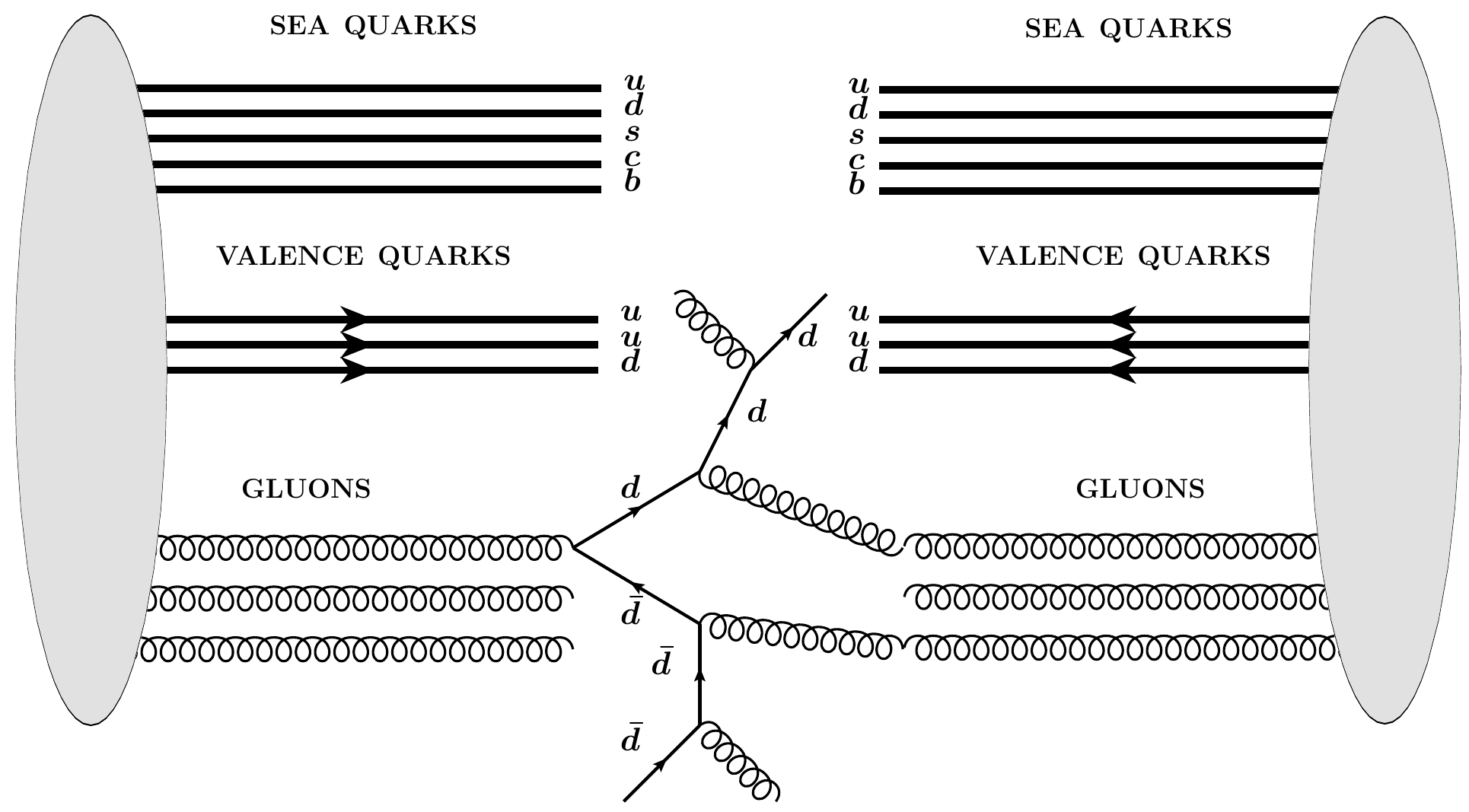}} a)
\end{minipage}
\hfill
\begin{minipage}[h]{0.48\linewidth}
\center{\includegraphics[width=1.0\linewidth]{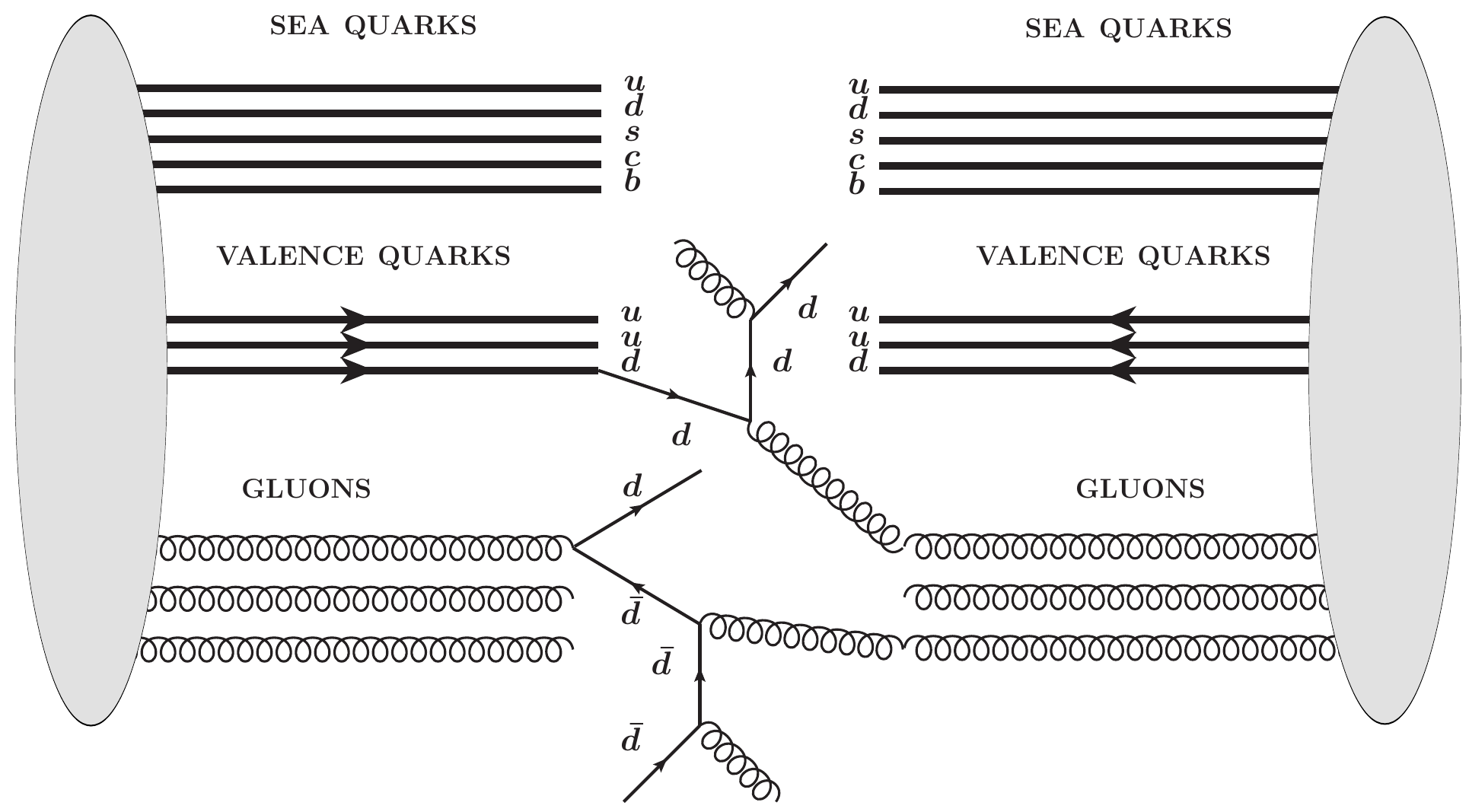}} b)
\end{minipage}
\caption{Two different ways to generate $\left(d \, g \rightarrow d \, g\right) \otimes \left(\bar{d} \, g \rightarrow \bar{d} \, g\right)$ DPS process in the \pythia  event generator: a) the initial state $d$-quark comes from a perturbative $g \rightarrow d \, \bar{d}$ splitting process, b) the initial state  $d$-quark comes from a raw $d$-quark sPDF. }
\label{fig:Pythia_1v2_contribution}
\end{figure}

Note that for the second interaction we have:
\begin{eqnarray}
    &&\sum\limits_{j_2} \,
    \int\limits^{1}_0 dx_2 \, \theta(1 - x_1 - x_2) \,
    x_2 \,
    f_{j_2}^{m\leftarrow j_1, x_1}(x_2, Q) 
    = 1 - x_1.
\end{eqnarray}
That is, the momentum of all partons after the first interaction is $1-x_1$, as desired. In general, for the $n$th interaction we will have:
 \begin{eqnarray}
 	\int\limits^{X_n}_0 dx \, x\left(\sum\limits_f \, q_{fn}(x, Q) + g_n(x, Q)\right) = X_n.
 	\label{eq:pythia_momentum_rule}
 \end{eqnarray}
This approach, however, does not take into account changes in the quark content of a proton.
For example, if one probes a  ``valence'' \mbox{$d$-quark} in the first hard interaction then for the second interaction the number of ``valence'' $d$-quarks in a beam remnant should be reduced to zero. 
Furthermore, if one probes  a \mbox{$d$-quark} sPDF at low-$x$  it means that with  a high  probability the $d$-quark originates from a perturbative $1\rightarrow2$ splitting of a gluon  into a quark-antiquark pair,  which, in turn, implies that there is a left-over ``companion''  antiquark which should be included in a beam remnant, see Fig.~\ref{fig:Pythia_1v2_contribution}.
These effects, obviously, apply to all quark flavours.  

In 2004 the MPI model of \pythia was significantly improved: $n$-parton PDFs of arbitrarily high order $n$ were constructed such that changes in quark content due to higher-scale interactions were taken into account.
For the first interaction  a standard sPDF is assumed to be
valid, as fitted to data. 
The sPDF for the second interaction
thereafter depends on the properties of the first interaction, and so on~\cite{Sjostrand:2004pf}.
The starting point of this approach is to split sPDFs into valence and sea parts as 
\begin{eqnarray}
	f(x, Q) = f_{\val}(x, Q) + f_{\sea}(x, Q),
\end{eqnarray}
where the flavour-antiflavour symmetry in the sea is assumed. 
If we denote the raw valence sPDF for flavour $f$ by $q_{fv0}(x, Q)$, then the valence sPDF after $n$ interactions is given by
\begin{eqnarray}
	q_{fvn}(x, Q) = \frac{N_{fvn}}{N_{fv0}} \, \frac{1}{X_n} \, q_{fv0}(x / X_n, Q),
	\label{eq:pythia_val_quark_reweight}
\end{eqnarray}
where  $N_{fv0}$ is an original number of valence quarks of a flavour $f$,  $q_{fvn}$ and $N_{fvn}$  are distribution and number of  valence quarks of  the flavour $f$ after $n$ interactions correspondingly. 
One can see that modification in Eq.~\eqref{eq:pythia_val_quark_reweight} changes the standard number rule
\begin{eqnarray}
	\int\limits^1_0 dx \, q_{fv0}(x, Q) = N_{fv0}
\end{eqnarray}
to 
\begin{eqnarray}
	\int\limits^1_0 dx \, q_{fvn}(x, Q) \, \theta(X_n - x) = \frac{N_{fvn}}{N_{fv0}} \, \int\limits^{X_n}_0 \, \frac{dx}{X_n} \, q_{fv0}(x / X_n, Q) = N_{fvn}.
\end{eqnarray}

Equation \eqref{eq:pythia_val_quark_reweight} allows to account for changes in the number of valence quarks. 
However, it cannot be used on its own since a subtraction of a valence
quark also removes an average momentum fraction carried by $q_{fvn}(x, Q)$  
\begin{eqnarray}
	\langle x_{fvn}(Q)\rangle \equiv \frac{\int\limits^{X_n}_0 dx \, x \, q_{fvn}(x, Q)}{\int\limits^{X_n}_0 dx \, q_{fvn}(x, Q)} = X_n \langle x_{fv0}(Q) \rangle,
	\label{eq:av_mom_xdvn}
\end{eqnarray}
and this will cause the algorithm to violate the momentum rule Eq.~\eqref{eq:pythia_momentum_rule} (which would be preserved if we only rescaled the momentum fraction in Eq.~\eqref{eq:pythia_val_quark_reweight}).
We will describe how this problem is resolved shortly; however, let us first describe how the $1 \to 2$ splitting effects described earlier are accounted for in the \pythia code.
Let us say that in a particular interaction we find a sea quark $q_s$ in the proton.
This must have been produced in conjunction with a companion (anti-)quark, $q_c$:
\begin{eqnarray}
	g \rightarrow q_s + q_c.
\end{eqnarray}
Let us denote the momentum fraction of the parent gluon by $x_g$, that of the sea quark by $x_s$ and that of the companion by $x = x_g - x_s$. Then, according to Ref.~\cite{Sjostrand:2004pf}, we introduce an sPDF for the companion quark, which is given by\footnote{
Note the resemblance between the structure of the right hand side of Eq.~\eqref{eq:companion_quark} and the splitting term on the right hand side of dDGLAP evolution equations, Eq.~\eqref{eq:double_dglap} -- this is of course no coincidence, as both refer to a perturbative $1 \to 2$ splitting.}
\begin{eqnarray}
	q_c(x, x_s) = C \, P_{g\rightarrow q\bar{q}}\left(\frac{x_s}{x_s + x}\right) \frac{g(x_s + x)}{x_s + x},
	\label{eq:companion_quark}
\end{eqnarray}
where $P_{g\rightarrow q\bar{q}}\left( z \right) = \frac{1}{2}\left[z^2 + (1 - z)^2\right]$ and $C$ is a normalization constant 
which is fixed by
\begin{eqnarray}
	\int\limits^{1 - x_s}_0 dx \, q_c(x, x_s)  = 1.
\end{eqnarray}
Note that Eq.~\eqref{eq:companion_quark} is defined assuming no momentum loss due to the previous hard interactions and hence has to be rescaled accordingly.

Using \mbox{Eq.~\eqref{eq:companion_quark}} one can compute an average momentum fraction introduced by a companion quark to a beam remnant $X_n\langle x_{fc0}(x_s) \rangle$ defined in analogy with Eq.~\eqref{eq:av_mom_xdvn} as 
\begin{eqnarray}
	X_n \langle x_{fc0}(x_s) \rangle \equiv \frac{\int\limits^{X_n}_0 dx \, x \, q_{fcn}(x, x_s)}{\int\limits^{X_n}_0 dx \, q_{fcn}(x, x_s)} = 
	{X_n} \int\limits^1_0 dx \, x q_{fc0}(x, x_s), 
	\label{eq:av_mom_xdcn}
\end{eqnarray}
where index $f$ is used to distinguish between different flavours and $q_{fcn}$ is linked to  $q_{fc0}$ via the momentum rescaling: $q_{fcn}(x, x_s) \equiv q_{fc0}(x/X_n, x_s)/X_n$.
The average momentum introduced by a companion quark together with  the average momenta subtracted due to the removal of valence quarks $X_n \langle x_{fv0}(Q)$ leads to the violation of Eq.~\eqref{eq:pythia_momentum_rule}. 
The way to circumvent this issue is to let the sea and gluon distributions fluctuate in such a way that both contributions are compensated and the momentum rule Eq.~\eqref{eq:pythia_momentum_rule} is preserved together with the number rules
\begin{eqnarray}
	&&\int\limits^X_0 dx \, q_{fvn}(x, Q) = N_{fvn}, 
	\label{eq:pythia_n_constr_1}	\\
	&&\int\limits^X_0 dx \, q_{fc_jn}(x, x_j)
    = 1, \forall j,
	\label{eq:pythia_n_constr_2}
\end{eqnarray}
where index $j$ is used to distinguish between different companion quarks of the flavour $f$. The corresponding modifications of the sea and gluon PDFs are given by
\begin{eqnarray}
	q_{fs}(x, Q) \rightarrow aq_{fs}(x, Q),
	\label{eq:pythia_sea_gluon_rescale_1}\\
	g(x, Q) \rightarrow ag(x, Q),
	\label{eq:pythia_sea_gluon_rescale_2}
\end{eqnarray}
which allows one to rewrite the momentum rule as 
\begin{eqnarray}
	1 &=& 
	\frac{1}{X_n} \int\limits^{X_n}_0 dx \, x\left(\sum\limits_f \left[q_{fvn}(x, Q) + \sum\limits_j q_{fc_jn}(x, x_j) + aq_{fs}(x, Q) \right] + ag_{n}(x, Q)\right) =
	\nonumber\\
	&=& a\left(1 - \sum\limits_f N_{fv0}\langle x_{fv0} \rangle\right) + 
	    \sum\limits_{f}N_{fvn}\langle x_{fv0} \rangle + 
	    \sum\limits_{f,j}\langle x_{fc_j0} \rangle, 
	\label{eq:normalisation_a}
\end{eqnarray}
where $\langle x_{fv0} \rangle$ and $\langle x_{fc_j0} \rangle$ are defined according to Eqs.~\eqref{eq:av_mom_xdvn} and \eqref{eq:av_mom_xdcn} respectively.
Equation~\eqref{eq:normalisation_a} fixes the value of the constant $a$ in Eqs.~\eqref{eq:pythia_sea_gluon_rescale_1} and \eqref{eq:pythia_sea_gluon_rescale_2}  which allows to build a closed system of equations valid for all sPDFs used to simulate an  $n$'th interaction. 

\section{Comparison between the GS09 and \pythia models of dPDFs}
\label{s:results_dPDFs}
\subsection{Asymmetric double parton distribution functions}
\label{ss:results_dPDFs_asym}
Now let us study in detail  how  well dPDFs  constructed out of the sPDFs being used in the MPI model of \pythia satisfy  the GS sum rules. 
We shall note that one can use raw  and modified  sPDFs  to construct
asymmetric \pythia dPDFs as in  Eq.~\eqref{eq:pythia_dPDFs}, or one can construct symmetric dPDFs according to
\begin{eqnarray}
	D^{\rm sym}_{j_1 j_2}(x_1, x_2, Q) \equiv \frac{D_{j_1 j_2}(x_1, x_2, Q) + D_{j_2 j_1}(x_2, x_1, Q)}{2},
	\label{eq:pythia_dPDFs_sym}
\end{eqnarray}
where $D_{j_1 j_2}(x_1, x_2, Q)$ and $D_{j_2 j_1}(x_2, x_1, Q)$ are given by Eq.~\eqref{eq:pythia_dPDFs}.
The asymmetric dPDFs  have a strict ordering of the arguments since the second sPDF $f^{m\leftarrow j_1, x_1}_{j_2}(x_2, Q)$ is always modified according to the first hard interaction involving a parton $j_1$ carrying a  momentum fraction $x_1$. 
Such ordering appears because in the  MPI model of \pythia the MPIs are considered as semi-hard corrections to the first hard process. 
However, one should keep in mind that the dPDFs should be symmetric under simultaneous interchange of flavours, Bjorken-$x$'es and factorization scales, and hence the GS sum rules should be satisfied when either $x$ fraction is integrated over for the equal scale dPDFs.
Nonetheless, it is interesting to see to what extent the asymmetric dPDFs satisfy the sum rules when integrating over the momentum fraction of the second ``modified'' parton, as this is the scenario where, by design, the \pythia dPDFs should best satisfy the sum rules.
Therefore, we start the discussion with the  asymmetric dPDFs  as in  Eq.~\eqref{eq:pythia_dPDFs}  and then consider effects of symmetrization. 
Since the sum rules apply to the case where the scales in the dPDF are equal, we set $Q_1=Q_2$ and set this common scale to  $91$ GeV.

First of all, let us check how the ansatz given by Eq.~\eqref{eq:pythia_dPDFs} satisfies the momentum sum rule given by Eq.~\eqref{eq:gs_momentum_rule_inv} (when integrating over the second parton only). 
By substituting Eq.~\eqref{eq:pythia_dPDFs} into Eq.~\eqref{eq:gs_momentum_rule_inv} we get 
\begin{eqnarray}
	\sum\limits_{j_2}\int\limits^{1 - x_1}_0 dx_2 \, x_2 \, f_{j_1}^r(x_1, Q) f_{j_2}^{m\leftarrow j_1, x_1}(x_2, Q) \overset{?}{=} (1 - x_1)\,f_{j_1}^r(x_1, Q),
\end{eqnarray}
or
\begin{eqnarray}
	\sum\limits_{j_2}\int\limits^{1 - x_1}_0 dx_2 \, x_2 \, f_{j_2}^{m\leftarrow j_1, x_1}(x_2, Q) \overset{?}{=} 1 - x_1.
	\label{eq:pythia_momentum_rule_intermediate}
\end{eqnarray}
where we use the symbol $\overset{?}{=}$ to denote that this relation should be true for the sum rule to be satisfied. 
Since $j_2$ in  $\sum_{j_2}$ runs over all parton species one can write Eq.~\eqref{eq:pythia_momentum_rule_intermediate} as
\begin{eqnarray}
	\int\limits^{1 - x_1}_0 dx_2 \, x_2 \, \left( \sum\limits_{j = q, \bar{q}} f_j^{m\leftarrow j_1, x_1}(x_2, Q) +  f_g^{m\leftarrow j_1, x_1}(x_2, Q) \right) \overset{?}{=} 1 - x_1.
	\label{eq:pythia_momentum_rule_final}
\end{eqnarray}
Looking at \mbox{Eq.~\eqref{eq:pythia_momentum_rule}}  with $n = 2$, $X = 1 - x_1$, we see that this relation does indeed hold. 
That is, the asymmetric \pythia dPDFs satisfy the momentum sum rule, when integrating over the second parton, by design. 
Later in this section we will investigate how well Eq.~(\ref{eq:gs_momentum_rule_inv}) is satisfied by the asymmetric \pythia dPDFs in practice.
However, before doing that, let us first consider the number sum rule given by  Eq.~(\ref{eq:gs_number_rule_inv}).
It is convenient to express  
Eq.~\eqref{eq:gs_number_rule_inv} in terms of a \textit{response function} $R_{j_1 j_2}$ defined as 
\begin{eqnarray}
	R_{j_1 j_2}(x_1, x_2, Q) \equiv x_2 \, \frac{ D_{j_{1} j_2}(x_1, x_2, Q) - D_{j_1 \bar{j_2}}(x_1, x_2, Q) }{f_{j_1}^{r}(x_1, Q)}.
	\label{eq:responce_function_def}
\end{eqnarray}
The meaning of $R_{j_1 j_2}(x_1, x_2, Q)$ becomes clear if one neglects momentum and number conservation and substitutes  $D_{j_1 j_2}(x_1, x_2, Q) = f_{j_1}^{r}(x_1, Q) f_{j_2}^{r}(x_2, Q)$  in Eq.~\eqref{eq:responce_function_def}. 
The function  $R_{j_1 j_2}(x_1, x_2, Q)$ then turns simply into a raw valence quark sPDF multiplied by a corresponding Bjorken-$x$:
\begin{eqnarray}
	R_{j_1 j_2}(x_1, x_2, Q) = x_2 \left[f_{j_2}^{r}(x_2, Q) - f_{\bar{j_2}}^{r}(x_2, Q)\right] = x_2 f^{r}_{j_{2v} } (x_2, Q).
\end{eqnarray}
Using $R_{j_1 j_2}(x_1, x_2, Q)$ one can write Eq.~\eqref{eq:gs_number_rule_inv} as 
\begin{eqnarray}
	\int\limits^{1 - x_1}_0 \frac{dx_2}{x_2} \, R_{j_1 j_2}(x_1, x_2, Q)
	\overset{?}{=}  N_{j_{2v}} - \delta_{j_1 j_2} + \delta_{j_1 \bar{j}_2}.
	\label{eq:gs_number_rule_R}
\end{eqnarray}
Since in the \pythia model $D_{j_1 j_2} (x_1, x_2, Q) = f_{j_1}^{r}(x_1, Q) f_{j_2}^{m\leftarrow j_1, x_1}(x_2, Q)$ one can write $R_{j_1 j_2}(x_1, x_2, Q)$ as 
\begin{eqnarray}
	R_{j_1 j_2}(x_1, x_2, Q) = 
	x_2 \, 
	\left[f_{j_2}^{m\leftarrow j_1, x_1}(x_2, Q) - f_{\bar{j_2}}^{m\leftarrow j_1, x_1}(x_2, Q)\right],
\end{eqnarray}
which turns  Eq.~\eqref{eq:gs_number_rule_R}  into 
\begin{eqnarray}
		\int\limits^{1 - x_1}_0 dx_2 \, \left[f_{j_2}^{m\leftarrow j_1, x_1}(x_2, Q) - f_{\bar{j_2}}^{m\leftarrow j_1, x_1}(x_2, Q)\right]
				\overset{?}{=}   N_{j_{2v}} - \delta_{j_1 j_2} + \delta_{j_1 \bar{j}_2}, 
		\label{eq:gs_number_rule_R_Pythia}
\end{eqnarray}
where one can consider the difference $f_{j_2}^{m\leftarrow j_1, x_1}(x_2, Q) - f^{m\leftarrow j_1, x_1}_{\bar{j_2}}(x_2, Q)$ 
as the response of the sPDF of a valence quark of a flavour $j_2$ to the first hard interaction  involving a parton of the flavour $f_1$ probed at the Bjorken-$x$ equal to $x_1$.

In order to be able to perform numerical integration in  
Eq.~\eqref{eq:gs_number_rule_R_Pythia} one has to be able to access sPDFs modified by 
\pythia to simulate a second hard interaction.
Normally,  it is not possible since the corresponding part of the  MPI model is hidden from a standard \pythia user. 
However, it can be achieved by  instantiation of objects which are members of the  \verb|BeamParticle| class as  explained in Appendix \ref{appendix:how_to_call_modified_PDFs}.
Each call of the   routine  \verb|second_xPDF| from  
\mbox{Appendix~\ref{appendix:how_to_call_modified_PDFs}} returns the value of the modified sPDF being used for the second interaction which allows us to perform numerical integration in Eq.~\eqref{eq:gs_number_rule_R_Pythia}.
One important comment has to be made: the aforementioned method to get the value of modified sPDF  is essentially probabilistic which implies that the modified sPDF being plotted as functions of $x_2$ will have strong oscillations 
which make numerical integration in Eq.~\eqref{eq:gs_number_rule_R_Pythia}  unstable.
Such oscillations appear due to the different treatment of valence and sea sPDFs in the  DPS (MPI) machinery of \pythia.
Therefore, depending on the value of $x_1$ \pythia  decides in a probabilistic manner if $f^{m\leftarrow j_1, x_1}_{j_2}(x_2, Q)$ 
should be evaluated assuming $f^{r}_{j_1} (x_1, Q)$ to be either valence or sea sPDF. 
The value of $f^{m\leftarrow j_1, x_1}_{j_2}(x_2, Q)$ will differ for each case which  leads to the oscillations of $f^{m\leftarrow j_1, x_1}_{j_2}(x_2, Q)$  as soon as $x_1$ approaches the region where the valence and sea sPDFs overlap.
However, the  problem  can be solved  by considering the values of $f^{m\leftarrow j_1, x_1}_{j_2}(x_2, Q)$ averaged over a sufficiently high number of calls of \verb|second_xPDF|  routine at given values of $x_1$ and $x_2$.  
Therefore, in the following, we always average  $f^{m\leftarrow j_1, x_1}_{j_2}(x_2, Q)$ before performing numerical integration (we use $10^6$ and $10^4$ averaging calls for the checks of the number and momentum rules correspondingly). 

Apart from the dPDFs constructed out of the \pythia sPDFs we also consider the GS09 dPDFs. 
In order to check how well GS09 dPDFs satisfy the number sum rule we use the publicly available GS09 grids with 600 grid points \cite{GS09web}. 
Since the  GS09 dPDFs  are based upon the MSTW2008 LO PDF set~\cite{Martin:2009iq} we use it in all our computations involving sPDFs. 

Additionally, as a baseline, we use so-called ``naive'' model of dPDFs where one replaces a dPDF by a product of two sPDFs and a $\theta$-function to enforce the kinematic constraint \mbox{$x_1 + x_2 \leq 1$}:
\begin{eqnarray}
	D_{j_1 j_2} (x_1, x_2, Q) = f^{r}_{j_1} (x_1, Q) \, f^{r}_{j_2} (x_2, Q) \, \theta(1 - x_1 - x_2).
	\label{eq:naive_dPDFs}
\end{eqnarray}
This approach neglects correlations in $x$-space and violates the GS sum rules.
Moreover, as was shown in Refs.~\citep{Kirschner:1979im, Shelest:1982dg, Zinovev:1982be}, this ansatz  does not satisfy the dDGLAP evolution equations. 

\begin{figure}
\begin{minipage}[h]{0.49\linewidth}
\center{\includegraphics[width=1.0\linewidth]{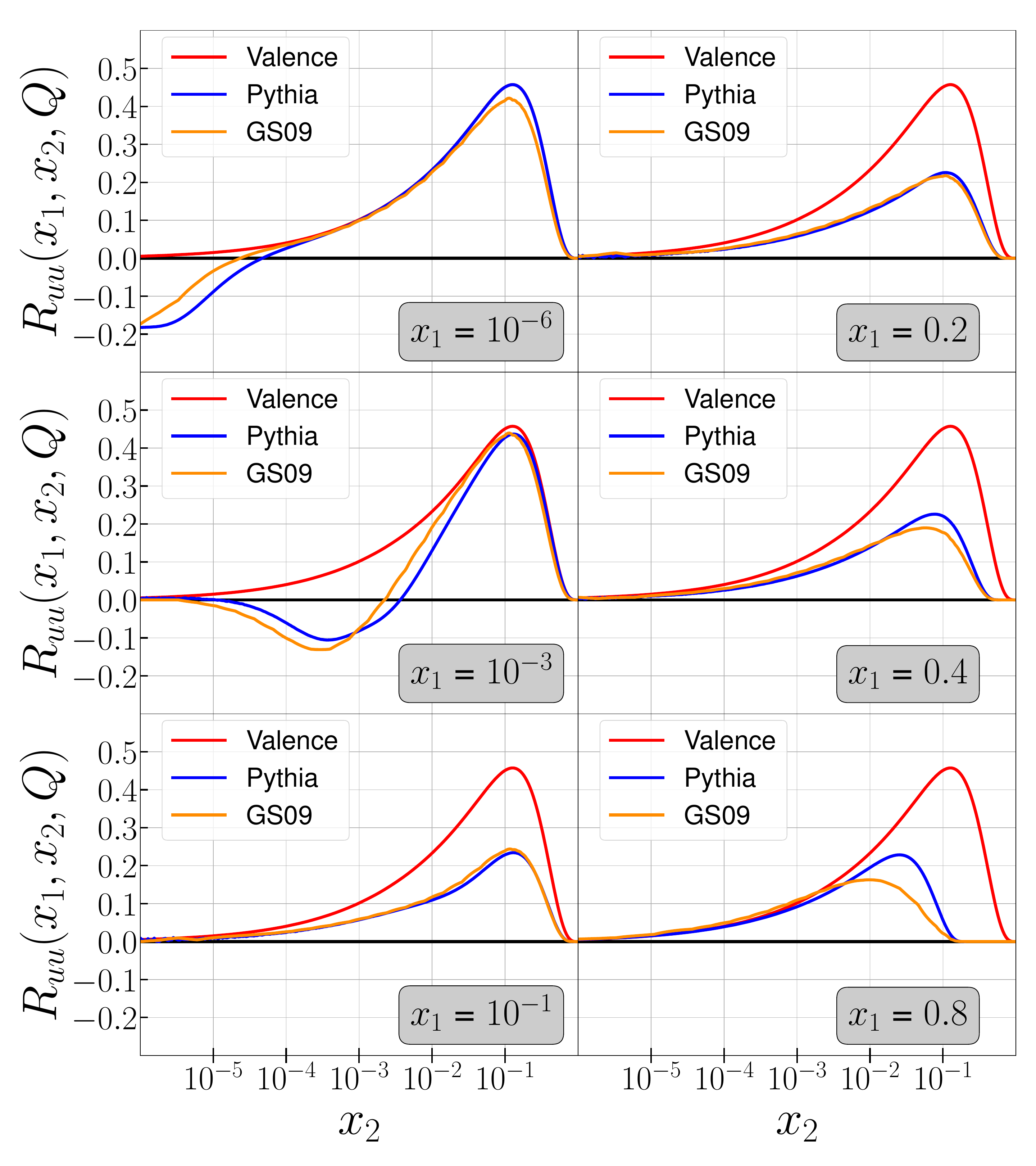}} \\a)
\end{minipage}
\hfill
\begin{minipage}[h]{0.49\linewidth}
\center{\includegraphics[width=1.0\linewidth]{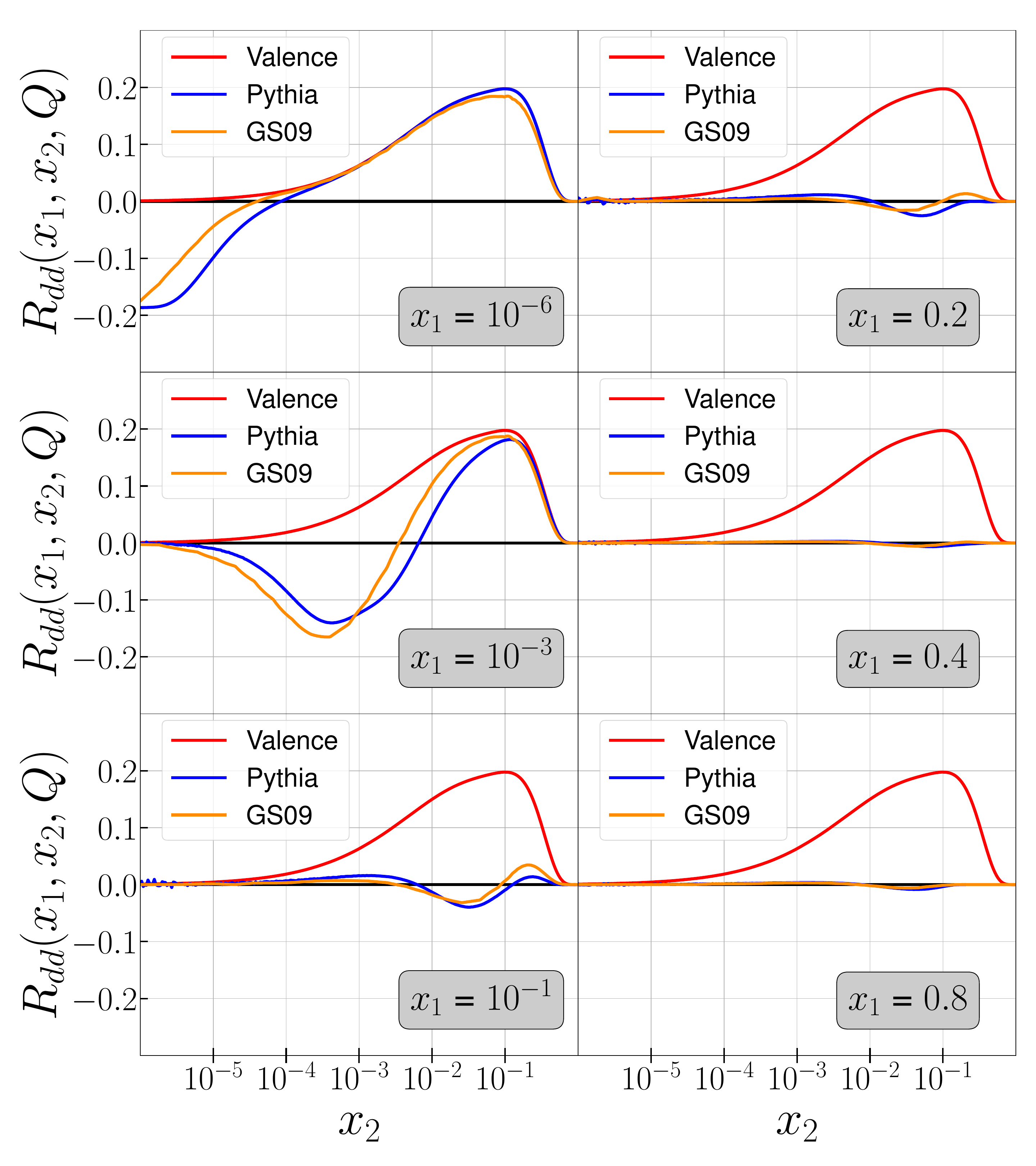}} \\b)
\end{minipage}
\cprotect\caption{The responses  of the valence quarks as functions of $x_2$ at fixed $x_1$. The factorization scale $Q = 91$ GeV. Comparison between the GS09 and \pythia dPDFs: \mbox{a) $R_{u u}(x_1, x_2, Q)$} response function, b) $R_{d d}(x_1, x_2, Q)$ response function. 
}
\label{fig:Pythia_sum_rules_uval_dval}
\end{figure}

Now let us check how well \pythia and GS09 models satisfy the number rule given by Eq.~\eqref{eq:gs_number_rule_inv}  taking as an example the $R_{u u}(x_1, x_2, Q)$ and $R_{d d}(x_1, x_2, Q)$ response functions. 
In Fig.~\ref{fig:Pythia_sum_rules_uval_dval} we compare  response functions evaluated with the \pythia and GS09 dPDFs against raw  valence $u$- and  $d$-quark  sPDFs.  
For our checks we consider six different fixed values of $x_1$, namely $10^{-6}, \, 10^{-3}, \, 10^{-1}, \, 0.2, \, 0.4$ and $0.8$. 
One can see how the $1\rightarrow2$ splitting, simulated according to Eq.~\eqref{eq:companion_quark}, leads to negative values of  the  response functions at small values of $x_2$. 
For example, if in the first interaction we probe a  $u$-quark  at $x_1 = 10^{-6}$ it is most likely a sea quark  which was created due to a  perturbative  splitting of a gluon into a $u\bar{u}$ pair. 
Therefore, the remaining companion $\bar{u}$-quark has to be included into a $\bar{u}$-quark sPDF which means that 
$f_{\bar{u}}^{r}(x_2, Q) \, < \, f_{\bar{u}}^{m\leftarrow u, x_1}(x_2, Q)$ for $x_2 \sim 10^{-6}$.
It  implies that $R_{u u}(x_1, x_2, Q)$ becomes negative if $x_2 \sim 10^{-6}$.  
In Fig.~\ref{fig:Pythia_sum_rules_uval_dval} one can see how the minimum of $R_{u u}(x_1, x_2, Q)$ propagates with the value of $x_1$ (see $x_1 = 10^{-6}$ and $x_1 = 10^{-3}$   cases). 
The same considerations, obviously, apply to the $g\rightarrow d\bar{d}$ splitting. 
By contrast, when $x_1 \geq 10^{-1}$ it becomes more likely to pick a valence  quark and, therefore, \pythia starts to use  Eq.~\eqref{eq:pythia_val_quark_reweight} more frequently than Eq.~\eqref{eq:companion_quark}, which results in a decrease of the peak of the valence distributions. 
We also would like to note that in the case of the $R_{uu}$ response function we observe almost a factor of two reduction in the height of the peak of the modified valence quark distribution comparing to the raw valence distribution if $x_1 = 0.1$ and that in case of  the $R_{dd}$ response function the peak of the modified valence quark distribution almost disappears completely (as one would naively expect).

The results given in Fig.~\ref{fig:Pythia_sum_rules_uval_dval} demonstrate that \pythia and GS09 models give similar predictions for the response of the valence $u$- and  $d$-quarks. 
Now let us check numerically how  well both models satisfy the number sum rule. In Table~\ref{tab:pythia_GS09_uval} and Table~\ref{tab:pythia_GS09_dval} we provide a numerical check of Eq.~\eqref{eq:gs_number_rule_inv} for the $u$- and $d$-quarks correspondingly. 
We see that both models satisfy  Eq.~\eqref{eq:gs_number_rule_inv}  at the  accuracy  level of a few percent for most values of $x_1$. 
The violation of  Eq.~\eqref{eq:gs_number_rule_inv} by GS09 dPDFs for the case $x_1 = 10^{-6}$ is a boundary effect due to the lower limit of the GS09 grids we use. 
One can see in Fig.~\ref{fig:Pythia_sum_rules_uval_dval} that if $x_1 = 10^{-6}$, a significant part of the response function ``leaks'' outside of the plotting range (compare $x_1 = 10^{-6}$ and $x_1 = 10^{-3}$ cases).
When working with the \pythia dPDFs we set the lower integration limit to $10^{-8}$ in order to avoid the aforementioned boundary effects (the lowest accessible $x$ value in \pythia is equal to $10^{-10}$). 
Comparing results of numerical integration  obtained with \pythia and GS09 dPDFs for the $x_1 = 10^{-6}$ case shown in Tables~\ref{tab:pythia_GS09_uval} and \ref{tab:pythia_GS09_dval} we see  that by setting a lower integration limit to $10^{-8}$  in the \pythia code one can avoid violation of the number sum rule due to the boundary effects.

\begin{table}
\begin{center}
	\begin{tabular}{ | c | c | c | c |}
    \hline
        $x_1$ 		& $N_{u_{v1}}$ \pythia 		& $N_{u_{v1}}$ GS09 & $N_{u_{v1}}$ Naive\\ \hline
		
		$10^{-6}$ & 1.022 & 1.526 & 2.006 \\ \hline
        $10^{-3}$ & 1.009 & 1.000 & 2.006 \\ \hline
        $10^{-1}$ & 1.003 & 1.036 & 2.005 \\ \hline
        0.2       & 1.002 & 1.016 & 2.005 \\ \hline
        0.4       & 1.001 & 0.949 & 1.997 \\ \hline
        0.8       & 0.999 & 0.862 & 1.708 \\ \hline

    \end{tabular}
	\captionof{table}{Numerical integration with respect to $x_2$ over $R_{uu}(x_1, x_2, Q)$ response function at fixed $x_1$. Here $N_{u_{v1}}$ stands for remaining number of valence $u$-quarks after the first hard interaction. In the ideal situation when the GS sum rules are perfectly satisfied $N_{u_{v1}} = 1$. 
	}
	\label{tab:pythia_GS09_uval}
\end{center}
\end{table}
\begin{table}
\begin{center}
	\begin{tabular}{ | c | c | c | c |}
    \hline
        $x_1$ 	& $N_{d_{v1}}$ \pythia 	& $N_{d_{v1}}$ GS09 & $N_{d_{v1}}$ Naive\\ \hline
		
		$10^{-6}$ & 0.015  & 0.633 & 0.999 \\ \hline
        $10^{-3}$ & 0.001  & 0.029 & 0.999 \\ \hline
        $10^{-1}$ & -0.002 & 0.004 & 0.999 \\ \hline
        0.2       & -0.001 & 0.006 & 0.999 \\ \hline
        0.4       & -0.001 & 0.002 & 0.998 \\ \hline
        0.8       & -0.001 & 0.000 & 0.897 \\ \hline
		
	\end{tabular}
	\captionof{table}{Same as in Table~\ref{tab:pythia_GS09_uval} but for the $R_{dd}(x_1, x_2, Q)$ response function.  In the ideal situation when the GS sum rules are perfectly satisfied $N_{d_{v1}} = 0$. 
	}
	\label{tab:pythia_GS09_dval}
\end{center}
\end{table}

Now let us check numerically how well the GS09 and  asymmetric \pythia dPDFs satisfy the momentum sum rule. 
In order to do that we write Eq.~\eqref{eq:gs_momentum_rule_inv} and Eq.~\eqref{eq:pythia_momentum_rule_intermediate} as 
\begin{eqnarray}
	&&\frac{1}{1 - x_1}\int\limits^{1 - x_1}_0 dx_2 \, x_2 \, \sum\limits_{j_2}  \, \frac{ D_{j_1 j_2} (x_1, x_2, Q)}{f^r_{j_1} (x_1, Q)} \overset{?}{=} 1,
	\label{eq:schematic_form_mr1}\\
	&&\frac{1}{1 - x_1}\int\limits^{1 - x_1}_0 dx_2 \, x_2 \,
	\sum\limits_j f_j^{m\leftarrow j_1, x_1}(x_2, Q) \overset{?}{=} 1,
	\label{eq:schematic_form_mr2}
\end{eqnarray}
and check how well the LHS of Eq.~\eqref{eq:schematic_form_mr1}  and  Eq.~\eqref{eq:schematic_form_mr2} match unity when we insert the GS09 dPDFs and the modified \pythia  sPDFs respectively. 

The results of the numerical checks are given in Table~\ref{tab:test_momentum_rule_Pythia_GS}.  
We see that, as expected from Eq.~\eqref{eq:pythia_momentum_rule_final}, \pythia dPDFs obey the momentum rule at the permille accuracy level  and that  GS09 dPDFs  satisfy the momentum rule at a few percent accuracy level.
We also note that the naive dPDFs obey the momentum sum rule at the few permille accuracy level if $x_1$ is in the range $10^{-6} - 10^{-3}$, but show a strong disagreement if $x_1 \geq 0.1$. 
The reason for this becomes clear if one substitutes Eq.~\eqref{eq:naive_dPDFs} in Eq.~\eqref{eq:gs_momentum_rule_inv} which leads to
\begin{eqnarray}
	\frac{1}{1 - x_1}\int\limits^{1 - x_1}_0 dx_2 \, x_2 \,\sum\limits_{j_2}  f^{r}_{j_2} (x_2, Q) \overset{?}{=} 1,
	\label{eq:schematic_form_mr3}
\end{eqnarray}
which holds only in the $x_1 \rightarrow 0$ limit.
\begin{center}
    \begin{tabular}{ | c | c | c | c | c |}
    \hline
        $x_1$ & $j_1$ & \pythia dPDFs. & GS09  dPDFs. & Naive dPDFs. \\ \hline
                		
	    $10^{-6}$	& $d$  & 0.996 & 0.931 & 0.996 \\ \hline
        $10^{-3}$	& $d$  & 0.996 & 1.009 & 0.997 \\ \hline
        $10^{-1}$	& $d$  & 1.003 & 1.026 & 1.106 \\ \hline
        0.2     	& $d$  & 1.007 & 1.006 & 1.244 \\ \hline
        0.4     	& $d$  & 1.009 & 0.980 & 1.649 \\ \hline
        0.8     	& $d$  & 1.009 & 0.854 & 4.131 \\ \hline

   \end{tabular}
   \captionof{table}{Check of the momentum sum rule for the \pythia, GS09 and naive models of dPDFs. 
   In the ideal situation when the momentum sum rule is perfectly satisfied the number in each cell should be equal to unity.
   } 
   \label{tab:test_momentum_rule_Pythia_GS}
\end{center} 

\begin{figure}
\begin{minipage}[h]{0.49\linewidth}
\center{\includegraphics[width=1.0\linewidth]{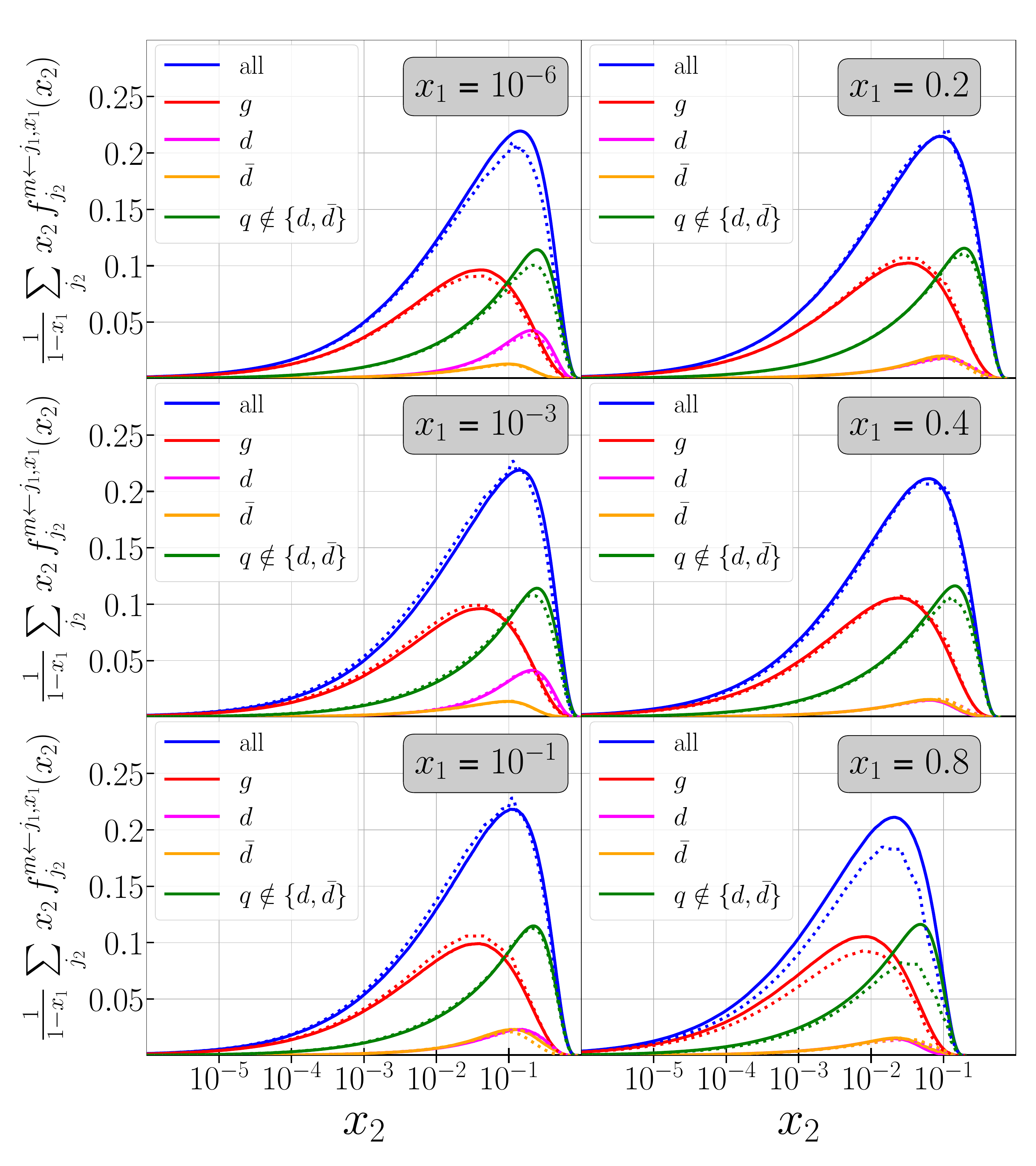}}\\a)
\end{minipage}
\hfill
\begin{minipage}[h]{0.49\linewidth}
\center{\includegraphics[width=1.0\linewidth]{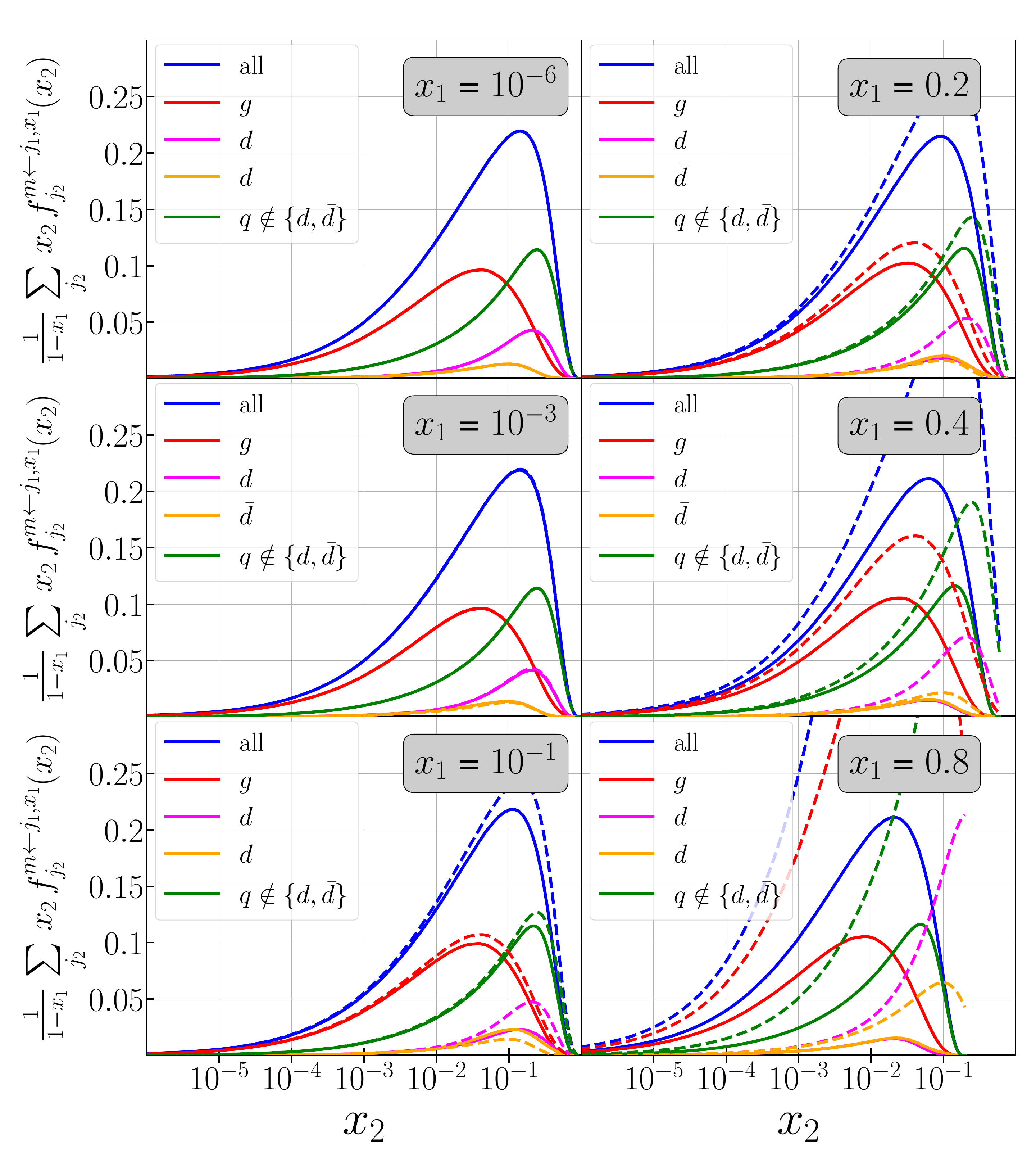}}\\b)
\end{minipage}
\cprotect\caption{Behaviour of the momentum sum rule response function, as defined in Eq.~\eqref{eq:momentum_sum}, for the dPDF models considered. Solid, dotted and dashed lines correspond to the \pythia dPDFs, GS09 and naive dPDFs correspondingly.}
\label{fig:Pythia_momentum_rule}
\end{figure}

Similarly to the case of the number sum rule, it is interesting to visualize the behaviour of the integrands in Eqs.~\eqref{eq:schematic_form_mr1}, \eqref{eq:schematic_form_mr2}.
In Fig.~\ref{fig:Pythia_momentum_rule} we plot the response function for the momentum sum rule:
\begin{eqnarray}
	\mathcal{R}_{j_1} \equiv \frac{1}{1 - x_1}  \sum_{j_2} \frac{ D_{j_1 j_2} (x_1, x_2, Q)}{f^r_{j_1} (x_1, Q)} \overset{\text{\pythia}}{=}  \frac{1}{1 - x_1} \,
	\sum\limits_{j_2} f_{j_2}^{m\leftarrow j_1, x_1}(x_2, Q)
	\label{eq:momentum_sum}
\end{eqnarray}
as a function of $x_2$ for six fixed values of $x_1$, and for the case in which $j_1 = d$. 
In Fig.~\ref{fig:Pythia_momentum_rule} a) we compare results obtained with \pythia dPDFs (solid lines) and GS09 dPDFs (dotted lines) whereas in Fig.~\ref{fig:Pythia_momentum_rule} b) we compare \pythia predictions against  results obtained with naive  dPDFs (dashed lines).
The blue lines in Fig.~\ref{fig:Pythia_momentum_rule} correspond to the sum over all flavours in Eq.~\eqref{eq:momentum_sum} whereas red, magenta and  orange lines stand for partial contributions from gluon, $d$- and $\bar{d}$-quarks correspondingly. 
Finally, the green lines are given by contribution from all quarks except $d$- and $\bar{d}$-flavours.

By comparing results obtained with \pythia and GS09 dPDFs shown in Fig.~\ref{fig:Pythia_momentum_rule} a) we see that both models give similar predictions for all partial contributions except in the $x_1 = 10^{-6}$ and $x_1 = 0.8$ cases where the GS09 curves lie consistently below the \pythia ones.
It is also noteworthy that, for all $x_1$ values probed, the bulk of area under the curves lies above $x_2 = 10^{-6}$  (in contrast to the corresponding plots for the number sum rule in Fig.~\ref{fig:Pythia_sum_rules_uval_dval}, where we find that for $x_1 = 10^{-6}$ an important part of the curve lies below $x_2 = 10^{-6}$).
This means that the value of 0.931 at $x_1=10^{-6}$ in Table~\ref{tab:test_momentum_rule_Pythia_GS} cannot be ascribed to the fact that the GS09 grids are limited to $x_i>10^{-6}$, but is instead linked to the GS09 curve being somewhat ``too low'' at larger $x_2$ values $\sim 10^{-2}-10^{-1}$ for $x_1=10^{-6}$. 
This deficiency in the extent to which the GS09 set satisfies the momentum sum rule at the lowest $x_1$ values only appears after evolution -- when one constructs Table~\ref{tab:test_momentum_rule_Pythia_GS} at $Q=1$ GeV, one finds a value of 0.997 at $x_1=10^{-6}$, and the sum rule integrand curve for GS09 lies almost exactly on top of the \pythia one for this $x_1$ value. 
We were not able to isolate a simple reason why this deficiency appears after evolution -- it is presumably due to deficiencies at larger $x_i$ values migrating downwards, but the region $x_1=10^{-6}, x_2 = 10^{-2} - 10^{-1}$ is fed by various $x$ values and flavours during evolution, making it very difficult to identify what features of the initial conditions are driving this effect. 
Since the main goal of this paper  is in any case to study dPDFs and tPDFs generated using the \pythia model, and the GS09 results are used as a baseline, we defer further investigation of this phenomenon to future studies.

Now let us discuss the momentum response functions for the naive dPDFs, as shown in  Fig.~\ref{fig:Pythia_momentum_rule} b). 
We see that if $x_1$ is equal to $10^{-6}$ or $10^{-3}$ there is no visible difference between the naive and \pythia response functions (solid and dashed lines practically coincide) which is in agreement with the results of the numerical integration shown in Table~\ref{tab:test_momentum_rule_Pythia_GS}.
We also see that starting from $x_1 = 10^{-1}$ the  naive model starts to consistently overestimate the sum rule integrand.
One would expect such an overestimate from the fact that the 
naive momentum sum rule curves do not go to zero at $x_2 = 1 - x_1$ as they should, but their trajectories are such that they would go to zero at $x_2 = 1$, if there were not the $\theta$-function cutting them off. 
This effect is hardly noticeable for very small $x_1 \lesssim 10^{-1}$, but becomes more and more significant at large $x_1$ -- see \textit{e.g.} the $x_1 = 0.8$ curve of Fig.~\ref{fig:Pythia_momentum_rule} b).
This increasing overshoot results in the naive model momentum sum rule integral deviating more and more from $1$ as $x_1$ increases, as we observed in Table~\ref{tab:test_momentum_rule_Pythia_GS}. 
%


Before moving on, let us briefly comment on the case of unequal scale dPDFs in \pythia. Using the machinery discussed in Appendix~\ref{appendix:how_to_call_modified_PDFs}, we can access asymmetric unequal-scale dPDFs, with either $Q_1<Q_2$ or $Q_1>Q_2$. The sPDF of the ``second'' parton is always modified as in Section~\ref{ss:pythia_model_of_MPI}. For such dPDFs it is clear that, when integrating over the second parton, the momentum and number and sum rules will hold for the unequal scale \pythia dPDFs to a very similar extent as they held for the equal scale \pythia dPDFs just discussed\footnote{We also checked this directly for the  $Q_1 = 45$, $Q_2 = 91$ GeV and  $Q_2 = 45$, $Q_1 = 91$ GeV cases.}. It is worth remarking that it is not necessarily expected theoretically that the dPDFs should satisfy the GS09 sum rules for the case of $Q_1 \neq Q_2$. If one follows the procedure discussed in Section~\ref{ss:gs09_intro} to obtain the unequal scale dPDFs from the equal scale dPDFs (and it is the case that those equal scale dPDFs satisfy the sum rules) then the unequal scale dPDFs will satisfy the sum rules when integrating over the parton with higher $Q$, but not when integrating over the parton with lower $Q$.

\subsection{Symmetrized double parton distribution functions}
\label{ss:results_dPDFs_sym}
In Section~\ref{ss:results_dPDFs_asym} we demonstrated that the asymmetric dPDFs constructed as in Eq.~\eqref{eq:pythia_dPDFs} satisfy the GS sum rules in the second argument at a few percent accuracy level  in the wide range of Bjorken-$x$'s from $10^{-6}$ up to $0.8$. 
We also found that both \pythia and GS09 dPDFs have similar behaviour of the response functions as shown in Fig.~\ref{fig:Pythia_sum_rules_uval_dval} and Fig.~\ref{fig:Pythia_momentum_rule}.
However, as we have mentioned before, QCD requires dPDFs to be symmetric under simultaneous interchange of parton flavours, Bjorken-$x$'es and factorization scales. 
Obviously, the dPDFs  defined by Eq.~\eqref{eq:pythia_dPDFs} do not satisfy this requirement;
in this section we check how well the symmetrized dPDFs defined in as  Eq.~\eqref{eq:pythia_dPDFs_sym} satisfy the GS sum rules. 
We start with the momentum sum rule. 
By substituting Eq.~\eqref{eq:pythia_dPDFs_sym} in Eq.~\eqref{eq:gs_momentum_rule_inv} we get
\begin{eqnarray}
	\int\limits^{1 - x_1}_0 dx_2 \, x_2 \, \mathcal{R}^{\ast}_{j_1}
	\overset{?}{=}  1,
	 \label{eq:mom_rule_sym_tmp}
\end{eqnarray}
where, similarly to Eq.~\eqref{eq:momentum_sum}, we defined
\begin{eqnarray}
    \mathcal{R}^{\ast}_{j_1} \equiv \frac{1}{1 - x_1}  \sum_{j_2} \frac{ D^{\rm sym}_{j_1 j_2} (x_1, x_2, Q)}{f^r_{j_1} (x_1, Q)}.
\end{eqnarray}

In Table~\ref{tab:test_momentum_rule_Pythia_SYM_GS} we check  Eq.~\eqref{eq:mom_rule_sym_tmp} for the symmetrized \pythia dPDFs for the case where  $j_1 = u$. 
As a reference we also add the check of Eq.~\eqref{eq:pythia_momentum_rule_intermediate} for asymmetric \pythia dPDFs as well as for the GS09 and naive dPDFs. 
We see that if $x_1 \ = 10^{-6}-10^{-1}$ the symmetric, asymmetric, and naive dPDFs satisfy the GS momentum sum rule at the 2$\%$ accuracy level; the GS09 dPDFs satisfy the sum rules to a similar extent except at $x_1 = 10^{-6}$, as discussed in the previous section.
We also note that the difference between the considered models increases with the value of $x_1$. 
For example, if $x_1 = 0.8$ the asymmetric \pythia dPDFs  and symmetric GS09 dPDFs satisfy the momentum sum rule at the  1$\%$ and 13$\%$ accuracy level correspondingly, whereas the symmetrized \pythia dPDFs and naive dPDFs are demonstrating strong violations of momentum conservation.

\begin{center}
    \begin{tabular}{ | c | c | c | c | c | c |}
    \hline
        $x_1$ 		& 	$j_1$	&  \pythia dPDFs. &  \pythia sym. dPDFs. &  GS09  dPDFs. &  Naive  dPDFs. \\ \hline
   
   		$10^{-6}$ & $u$ & 0.996 & 0.979 & 0.931 & 0.996 \\ \hline
        $10^{-3}$ & $u$ & 0.996 & 0.980 & 1.007 & 0.997 \\ \hline
        $10^{-1}$ & $u$ & 1.008 & 1.014 & 0.994 & 1.106 \\ \hline
        0.2       & $u$ & 1.010 & 1.047 & 0.978 & 1.244 \\ \hline
        0.4       & $u$ & 1.011 & 1.133 & 0.948 & 1.649 \\ \hline
        0.8       & $u$ & 1.011 & 1.679 & 0.882 & 4.131 \\ \hline

   \end{tabular}
   \captionof{table}{Test of the momentum sum rule for the symmetrized \pythia dPDFs. } 
   \label{tab:test_momentum_rule_Pythia_SYM_GS}
\end{center} 

The additional terms introduced by the symmetrization procedure also lead to violation of the number sum rule given by  Eq.~\eqref{eq:gs_number_rule_inv}. 
Let us illustrate this statement by considering $u\bar{u}$ and $\bar{u}u$ flavour combinations as the most prominent examples.
Similarly to Eq.~\eqref{eq:responce_function_def} we define a response function for the symmetrized \pythia dPDFs as
\begin{eqnarray}
	R^\ast_{j_1 j_2}(x_1, x_2, Q) \equiv x_2 \, 
	\frac{ D^{\rm sym}_{j_{1} j_2}(x_1, x_2, Q) - D^{\rm sym}_{j_1 \bar{j_2}}(x_1, x_2, Q) }{f_{j_1}^{r}(x_1, Q)},
\end{eqnarray}
which can be written as 
\begin{eqnarray}
	R^\ast_{j_1 j_2}(x_1, x_2, Q) = 
	\frac{x_2}{2 \, f^r_{j_1}(x_1) }	
	\left[
	\tilde{R}_{j_1 j_2}(x_1, x_2) - \tilde{R}_{j_1\bar{j}_2}(x_1, x_2)
	\right],
	\label{eq:r_sym_two_term}
\end{eqnarray}
where, omitting factorization scale labels, we defined
\begin{eqnarray}
	\tilde{R}_{j_1 j_2}(x_1, x_2) &\equiv& 
		f^r_{j_1}(x_1) f^{m \leftarrow j_1, x_1}_{j_2}(x_2) + 
		f^r_{j_2}(x_2) f^{m \leftarrow j_2, x_2}_{j_1}(x_1).
	\label{eq:r_tild_exp_1}
\end{eqnarray}

\begin{figure}
\begin{minipage}[h]{0.48\linewidth}
\center{\includegraphics[width=1\linewidth]{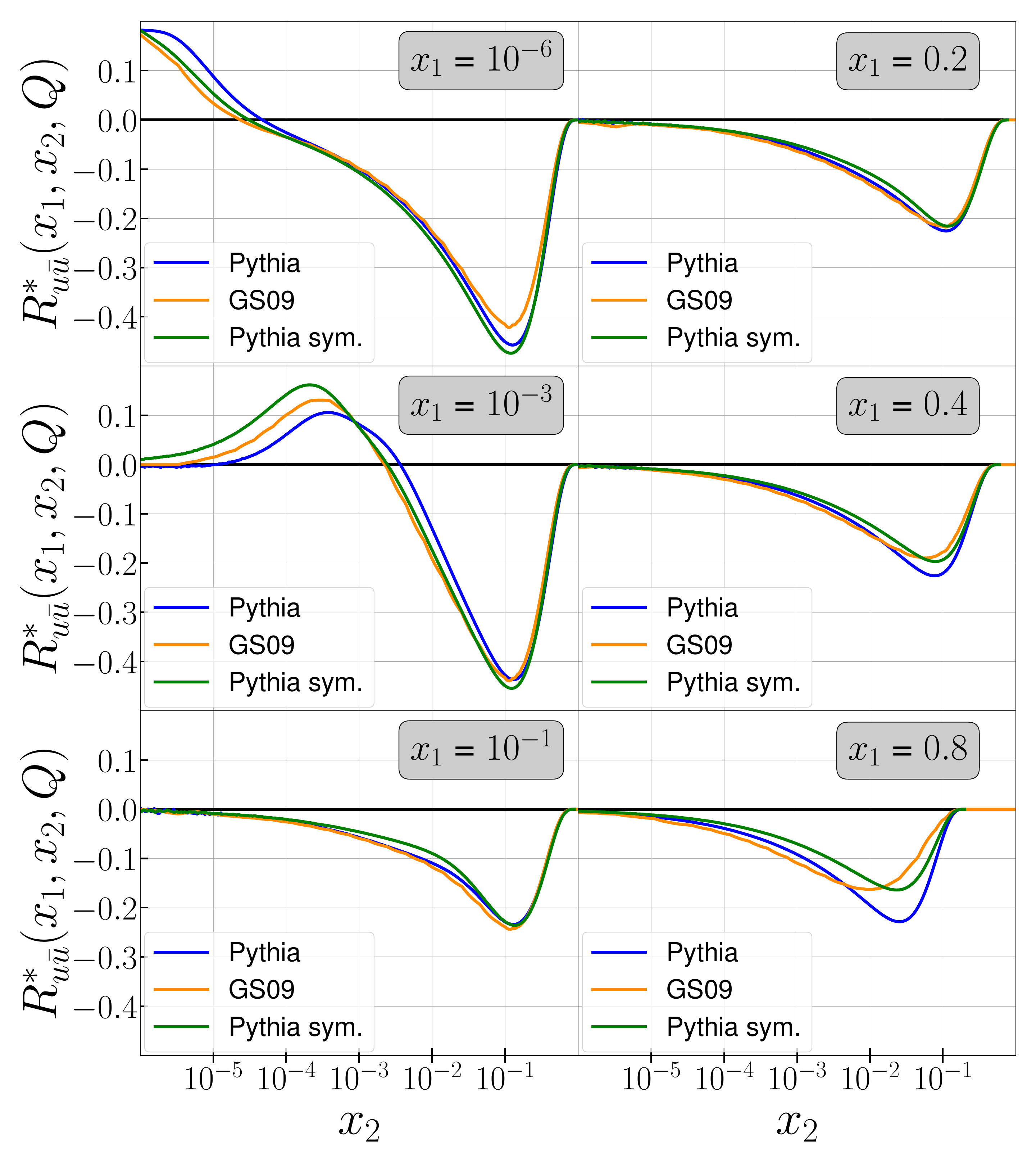}}\\a)
\end{minipage}
\hfill
\begin{minipage}[h]{0.48\linewidth}
\center{\includegraphics[width=1\linewidth]{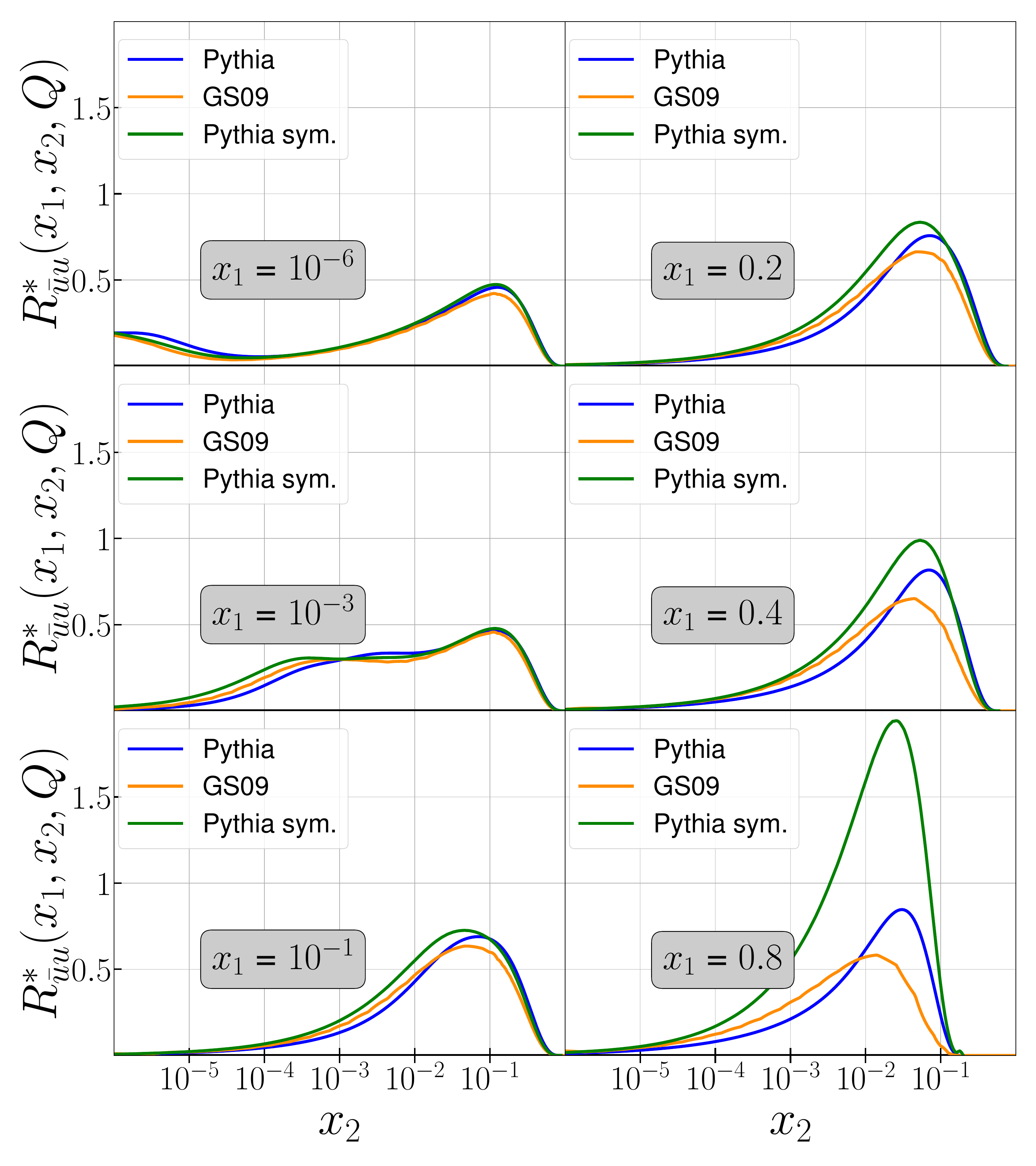}}\\b)
\end{minipage}
\cprotect\caption{Same as in Fig.~\ref{fig:Pythia_sum_rules_uval_dval} but for the $u\bar{u}$ and $\bar{u}u$ flavour combinations: a) $R^\ast_{u \bar{u}}$ response function, b) $R^\ast_{\bar{u} u}$ response function. 
}
\label{fig:Pythia_SYM_sum_rules_uubar_ubaru}
\end{figure}

The second term  on the right hand side of Eq.~\eqref{eq:r_tild_exp_1} is new and leads to the violation of the number sum rule by the symmetrized \pythia dPDFs. 
Let us consider first the $u\bar{u}$ flavour combination. 
The $R^\ast_{u\bar{u}}(x_1, x_2, Q)$ response functions  are given in Fig.~\ref{fig:Pythia_SYM_sum_rules_uubar_ubaru} a) and the corresponding  checks of the number sum rule are provided in Table~\ref{tab:test_numer_rule_Pythia_SYM_GS_uubar}. 
The checks for the  $\bar{u}u$ flavour combinations are given in  Fig.~\ref{fig:Pythia_SYM_sum_rules_uubar_ubaru} b) and Table~\ref{tab:test_numer_rule_Pythia_SYM_GS_ubaru}. 
We see that in the $u\bar{u}$ case the deviation of the number sum rule integral from the expected value of $N_{\bar{u}_{v1}} \equiv -N_{u_{v1}} = -1$ is about $10 - 25\%$ depending on the value of $x_1$.  
However, in the $\bar{u}u$ case  the deviations become stronger. 
In particular, the sum rule integral shows a deviation from the expected value of $3$ by around 16$\%$  for moderate ($[10^{-3}, 10^{-1}]$)  values of $x_1$ and a large deviation for the valence region (about  130$\%$ deviation if $x_1 = 0.8$). 

\begin{center}
    \begin{tabular}{ | c | c | c | c | c |}
    \hline
    $x_1$ 	&  $N_{\bar{u}_{v1}}$ \pythia  & $N_{\bar{u}_{v1}}$ \pythia sym.  & $N_{\bar{u}_{v1}}$ GS09 & $N_{u_{v1}}$ Naive \\ \hline
   	
		$10^{-6}$ 	 & -1.022 & -1.227 & -1.526 & -2.006 \\ \hline
        $10^{-3}$ 	 & -1.007 & -0.847 & -1.000 & -2.006 \\ \hline
        $10^{-1}$ 	 & -1.004 & -0.925 & -1.036 & -2.005 \\ \hline
        0.2      	 & -1.003 & -0.928 & -1.016 & -2.005 \\ \hline
        0.4          & -1.002 & -0.884 & -0.949 & -1.997 \\ \hline
        0.8          & -0.998 & -0.740 & -0.862 & -1.708 \\ \hline

   \end{tabular}
   \captionof{table}{Numerical integration with respect to $x_2$ over $R^\ast_{u\bar{u}}(x_1, x_2, Q)$ response function at fixed $x_1$. In the ideal situation when the GS sum rules are perfectly satisfied $N_{\bar{u}_{v1}} = -1$. 
   } 
   \label{tab:test_numer_rule_Pythia_SYM_GS_uubar}
\end{center} 

\begin{center}
    \begin{tabular}{ | c | c | c | c | c |}
    \hline
    $x_1$ 	&  $N_{u_{v1}}$ \pythia  
    			& $N_{u_{v1}}$ \pythia sym.  
    			& $N_{u_{v1}}$ GS09 
    			& $N_{u_{v1}}$ Naive \\ \hline
   
        $10^{-6}$  & 2.989 & 2.961 & 2.163 & 2.006 \\ \hline
        $10^{-3}$  & 3.003 & 3.351 & 2.982 & 2.006 \\ \hline
        $10^{-1}$  & 3.002 & 3.491 & 2.968 & 2.005 \\ \hline
        
        0.2 & 3.002 & 3.580 & 2.912 & 2.005 \\ \hline
        0.4 & 3.000 & 3.858 & 2.833 & 1.997 \\ \hline
        0.8 & 2.994 & 7.048 & 2.667 & 1.708 \\ \hline

   \end{tabular}
   \captionof{table}{Numerical integration with respect to $x_2$ over $R_{\bar{u}u}(x_1, x_2, Q)$ response function at fixed $x_1$.  In the ideal situation when the GS sum rules are perfectly satisfied $N_{u_{v1}} = 3$. 
   } 
   \label{tab:test_numer_rule_Pythia_SYM_GS_ubaru}
\end{center} 

\begin{figure}
\begin{minipage}[h]{0.49\linewidth}
\center{\includegraphics[width=1.0\linewidth]{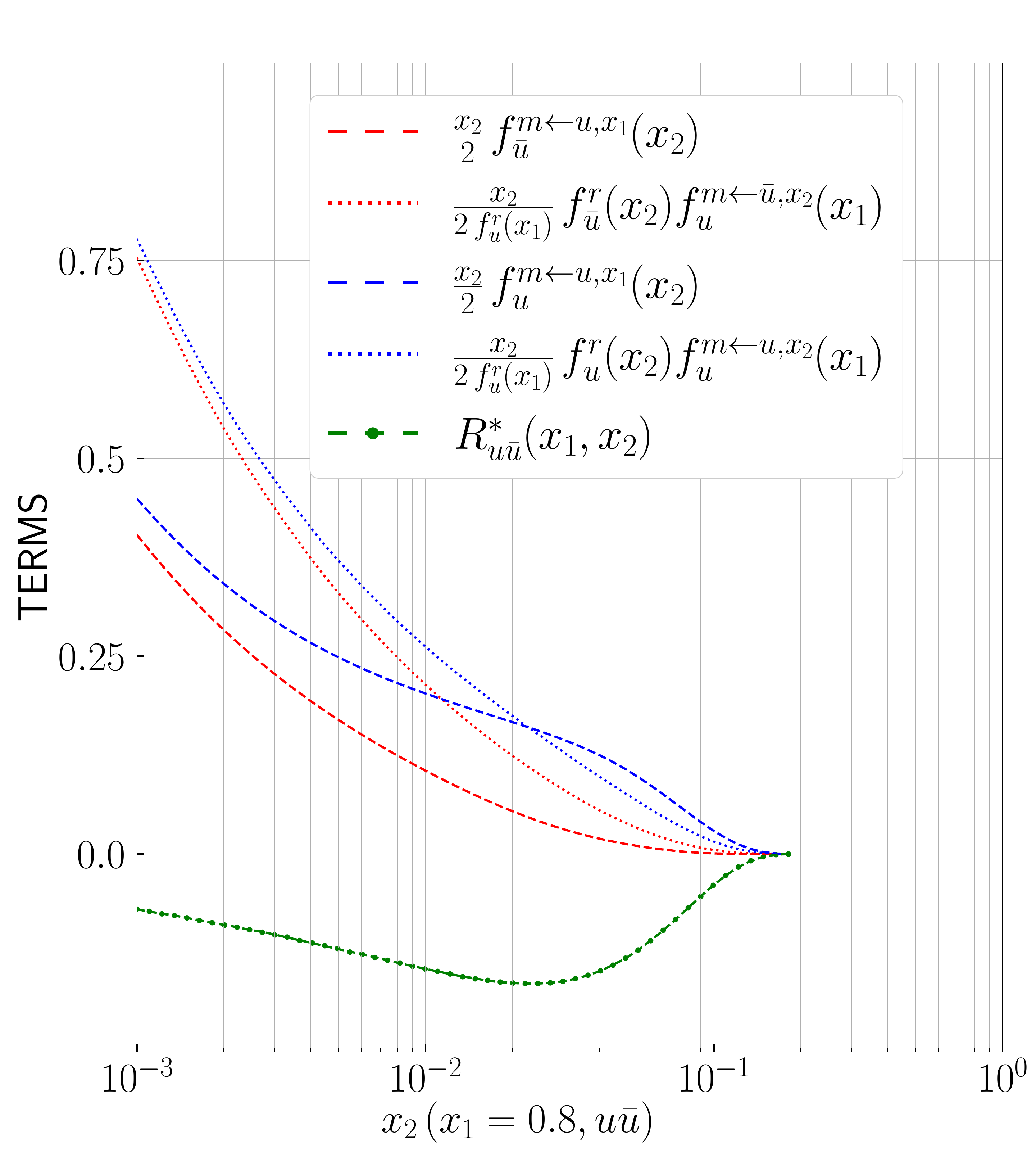}} a)
\end{minipage}
\hfill
\begin{minipage}[h]{0.49\linewidth}
\center{\includegraphics[width=1.0\linewidth]{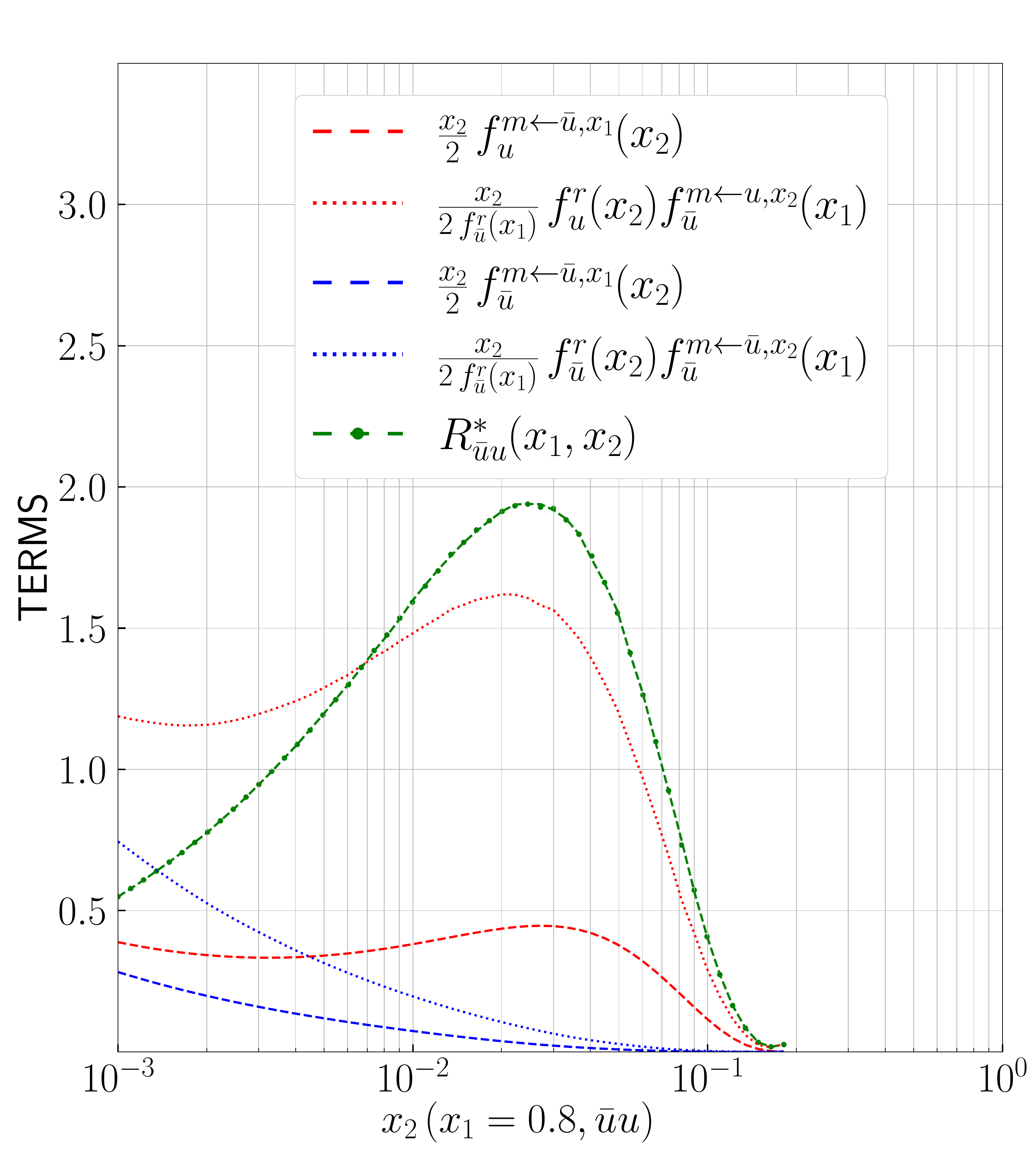}} b)
\end{minipage}
\cprotect\caption{Different terms contributing to a) $R^\ast_{u \bar{u}}(x_1, x_2)$ and  b) $R^\ast_{\bar{u} u}(x_1, x_2)$.}
\label{fig:Pythia_sum_terms}
\end{figure}

In order to find out the origin of the peak of $R^\ast_{\bar{u} u}$ at $x_1 = 0.8$ that is responsible for this large deviation, we study in more detail the terms on the right hand side of Eq.~\eqref{eq:r_sym_two_term}. 
Let us start with the $u\bar{u}$ flavour combination.  
The corresponding terms are given by 
\begin{eqnarray}
	\frac{x_2}{2 \, f^r_{u}(x_1)} \,	
	\tilde{R}_{u \bar{u}}(x_1, x_2) &=&
	\frac{x_2}{2} 
	\left[
		f^{m \leftarrow  u, x_1}_{\bar{u}}(x_2) + 
		\frac{f^r_{\bar{u}}(x_2) f^{m \leftarrow \bar{u}, x_2}_{u}(x_1)}{f^r_{u}(x_1)}
	\right],\label{eq:u_ubar_1}\\
	\frac{x_2}{2 \, f^r_{u}(x_1)} \,	
	\tilde{R}_{u u}(x_1, x_2) &=&
	\frac{x_2}{2} 
	\left[
		f^{m \leftarrow u, x_1}_{u}(x_2) + 
		\frac{f^r_{u}(x_2) f^{m \leftarrow u, x_2}_{u}(x_1)}{f^r_{u}(x_1)}
	\right].\label{eq:u_ubar_2}
\end{eqnarray}
In Fig.~\ref{fig:Pythia_sum_terms} a) we plot the four different terms given in Eqs.~\eqref{eq:u_ubar_1} -- \eqref{eq:u_ubar_2} (red and blue curves). 
The difference between Eq.~\eqref{eq:u_ubar_1} and Eq.~\eqref{eq:u_ubar_2} (which gives us $R^\ast_{u\bar{u}}$) is given by the green dash-dotted line.
We see that all four terms have rather similar behaviour. 
Now let us consider the $\bar{u}u$ case which yields
\begin{eqnarray}
	\frac{x_2}{2 \, f^r_{\bar{u}}(x_1)} \,
	\tilde{R}_{\bar{u} u}(x_1, x_2) &=& 
	\frac{x_2}{2} 
	\left[
		f^{m \leftarrow \bar{u}, x_1}_{u}(x_2) + 
		\frac{f^r_{u}(x_2) f^{m \leftarrow u, x_2}_{\bar{u}}(x_1)}{f^r_{\bar{u}}(x_1)}
	\right],
	\label{eq:ubar_u_3}\\
	\frac{x_2}{2 \, f^r_{\bar{u}}(x_1)} \,
	\tilde{R}_{\bar{u} \bar{u}}(x_1, x_2) &=&
	\frac{x_2}{2} 
	\left[
		f^{m \leftarrow \bar{u}, x_1}_{\bar{u}}(x_2) + 
		\frac{f^r_{\bar{u}}(x_2) f^{m \leftarrow \bar{u}, x_2}_{\bar{u}}(x_1)}{f^r_{\bar{u}}(x_1)}
	\right].
	\label{eq:ubar_u_4}
\end{eqnarray}

\begin{figure}
\begin{minipage}[h]{0.49\linewidth}
\center{\includegraphics[width=1.0\linewidth]{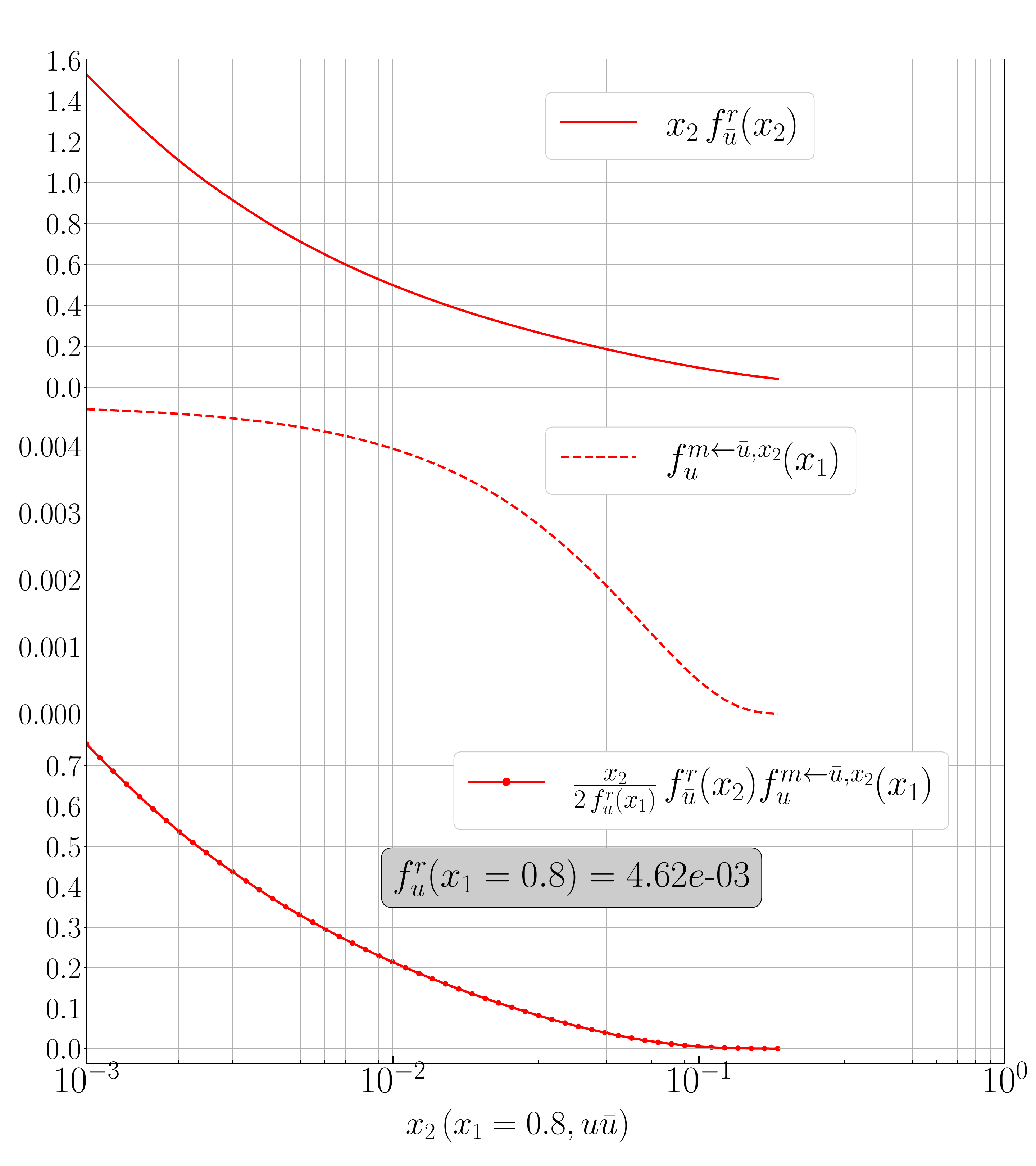}} a)
\end{minipage}
\hfill
\begin{minipage}[h]{0.49\linewidth}
\center{\includegraphics[width=1.0\linewidth]{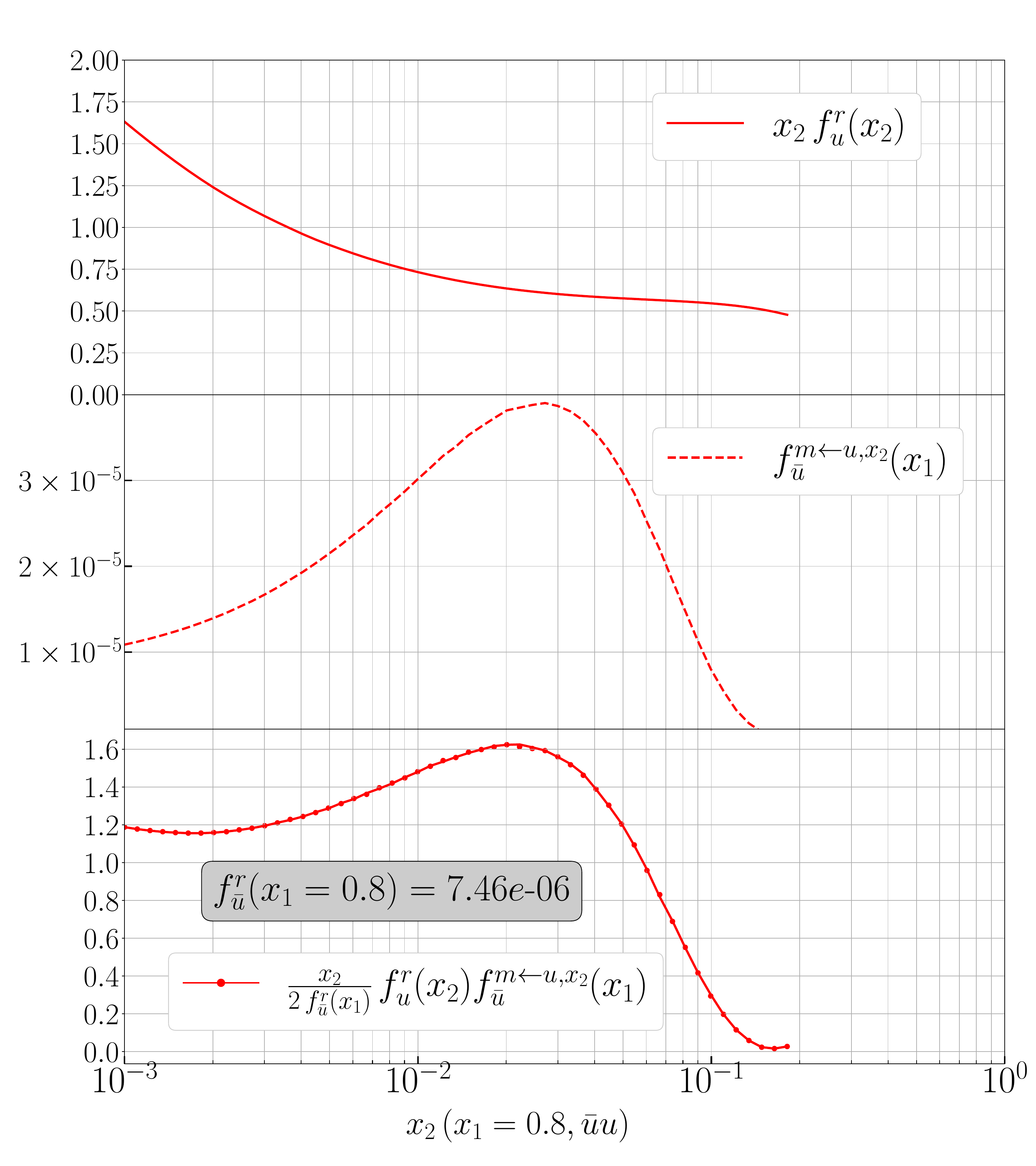}} b)
\end{minipage}
\cprotect\caption{Decomposition into factors of  a) $f^r_{\bar{u}}(x_2) f^{m \leftarrow \bar{u}, x_2}_{u}(x_1) / f^r_{u}(x_1)$ and 
\mbox{b) $f^r_{u}(x_2) f^{m \leftarrow u, x_2}_{\bar{u}}(x_1) / f^r_{\bar{u}}(x_1)$} terms.}
\label{fig:main_peak_par}
\end{figure}

\begin{figure}[!h]
\center{\includegraphics[width=0.65\linewidth]{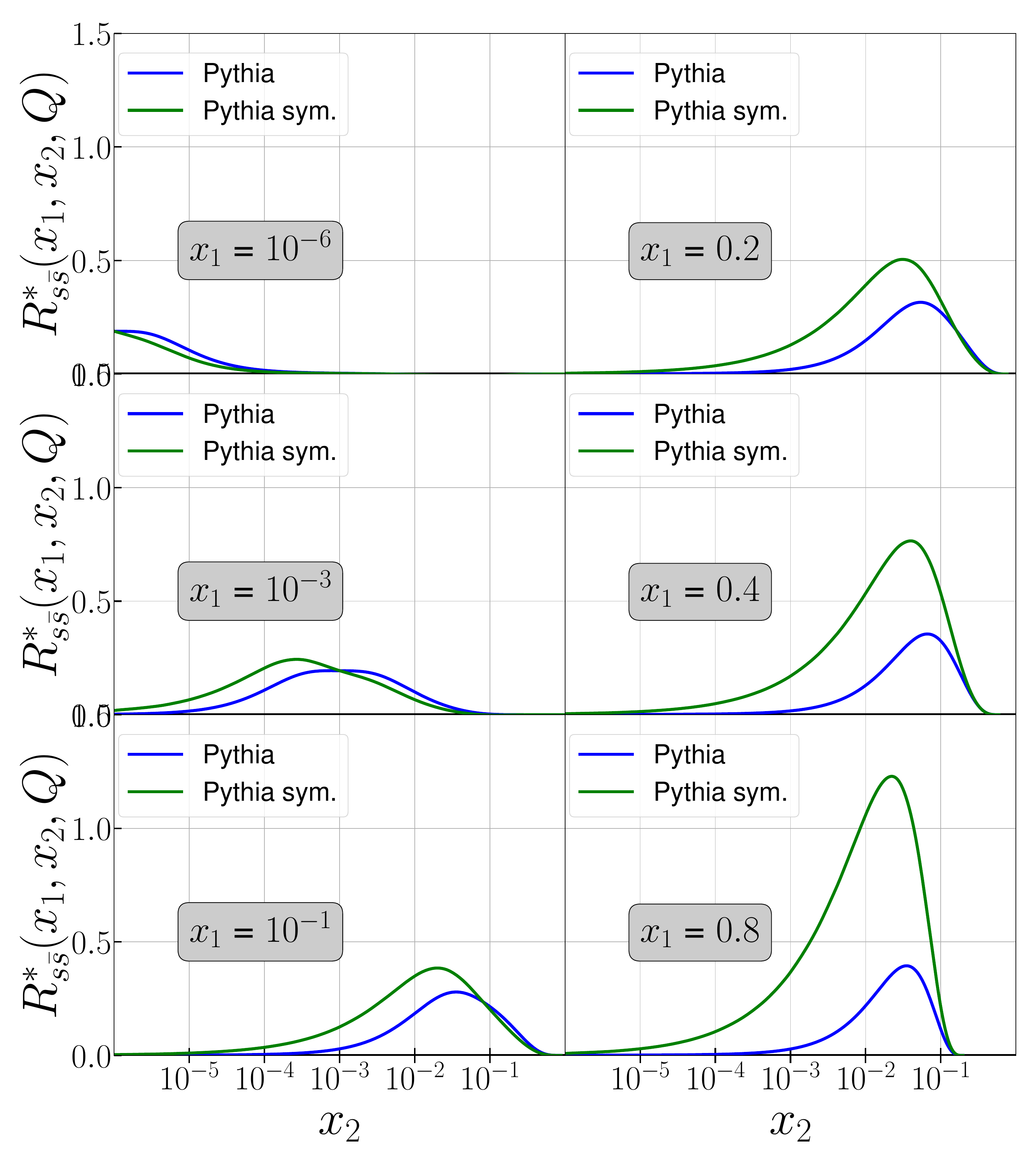}}
\cprotect\caption{Same as in Fig.~\ref{fig:Pythia_SYM_sum_rules_uubar_ubaru} but for the $R^\ast_{s\bar{s}}$ response function.}
\label{fig:Pythia_sum_rules_Rssbar_sym}
\end{figure}

In Fig.~\ref{fig:Pythia_sum_terms} b) we plot the four different terms given in Eqs.~\eqref{eq:ubar_u_3}--\eqref{eq:ubar_u_4}.  
Unlike the $u\bar{u}$ case we see that now different terms have a different behaviour. 
Namely, we see that the term proportional to $f^r_{u}(x_2) f^{m \leftarrow u, x_2}_{\bar{u}}(x_1) / f^r_{\bar{u}}(x_1)$ (red dotted line) has the biggest contribution and is largely responsible for the large peak given by the green
dash-dotted line.
Let us study  the aforementioned ratio in more details. 
In  Fig.~\ref{fig:main_peak_par}  a) we plot a decomposition of the
$f^r_{\bar{u}}(x_2) f^{m \leftarrow \bar{u}, x_2}_{u}(x_1) / f^r_{u}(x_1)$ term and in Fig.~\ref{fig:main_peak_par}  b) we plot its analogue  for the
$\bar{u}u$  flavour combination  $f^r_{u}(x_2) f^{m \leftarrow u, x_2}_{\bar{u}}(x_1) / f^r_{\bar{u}}(x_1)$. 
By comparing the middle plots in Fig.~\ref{fig:main_peak_par} we see that  $f^{m \leftarrow \bar{u}, x_2}_{u}(x_1)$  rapidly drops down in the region between $10^{-2}$ and $10^{-1}$  whereas $f^{m \leftarrow u, x_2}_{\bar{u}}(x_1)$ becomes peaked. 
The reason for  such a different behaviour is as follows. In the case of $f^{m\leftarrow\bar{u},x_2}_u(x_1 = 0.8)$ we are looking at how finding a $\bar{u}$ quark with momentum $x_2$ affects the distribution of $u$ quarks with momentum $x_1 = 0.8$. Since there is a sizeable distribution of $u$ quarks at large $x$ (arising from the valence contribution), the boost to this distribution from the companion quark effects is negligible, and the only visible effect from the $\bar{u}$ is just the ``squeezing'' caused by momentum conservation. 
That is, according to Section~\ref{ss:pythia_model_of_MPI},  for the middle plot in Fig.~\ref{fig:main_peak_par}  a) at a given factorization scale  we have 
\begin{eqnarray}
	f^{m\leftarrow\bar{u},x_2}_u(x_1) \approx   
	\frac{1}{1 - x_2} f^{r}_u\left(\frac{x_1}{1 - x_2}\right).
\end{eqnarray}
However, in case of the middle plot in Fig.~\ref{fig:main_peak_par}  b) the situation changes. 
The $\bar{u}$-quark sPDF, having no valence component, is much smaller at large $x$ than the  $u$-quark sPDF\footnote{At $Q = 91$ GeV, we have  $f^r_u(x_1 = 0.8) \sim 10^{-3}$ whereas \mbox{$f^r_{\bar{u}}(x_1 = 0.8) \sim 10^{-5}$}. 
}. 
In this case, the companion quark effects are noticeable, particularly when $x_2$ is close in magnitude with $x_1$. This results in  a peak in the middle plot in \mbox{Fig.~\ref{fig:main_peak_par} b)}.
Additionally, the comparison between upper plots in Fig.~\ref{fig:main_peak_par} shows us that the $x_2 f^r_u(x_2)$ term stays roughly the same in the region of interest (changes from 0.75 to 0.5) whereas $x_2 f^r_{\bar{u}}(x_2)$ drops from 0.6 down to zero.
Combining upper and middle plots in Fig.~\ref{fig:main_peak_par} together and dividing by $2 \, f^r_u(x_1 = 0.8)$ or $2 \, f^r_{\bar{u}}(x_1 = 0.8)$  we get a smoothly decreasing function in the lower plot in Fig.~\ref{fig:main_peak_par} a) and a peaked  function in the lower plot in Fig.~\ref{fig:main_peak_par} b) correspondingly.
Therefore, we conclude that the large peak in the response functions for the symmetrized dPDFs shown in Fig.~\ref{fig:Pythia_SYM_sum_rules_uubar_ubaru} b) is due to the companion quark contribution to the $\bar{u}$ sPDF $f^{m\leftarrow u, x_2}_{\bar{u}}(x_1)$.
We also shall note that similar large peaks occur for other flavour combinations where a sea quark sPDF receives a large contribution from a $g\rightarrow q\bar{q}$ splitting, \textit{e.g.} for the $s\bar{s}$ flavour combination as shown in Fig.~\ref{fig:Pythia_sum_rules_Rssbar_sym}.

Thus, we see that the symmetric dPDFs constructed using the MPI model of \pythia do not satisfy the sum rules perfectly.
Nevertheless, we shall note that the largest violation of the GS sum rules by dPDFs constructed as in Eq.~\eqref{eq:pythia_dPDFs_sym} occur at very large values of Bjorken-$x$'es (if $x \geq 0.4$) where the absolute values of dPDFs are strongly suppressed. 
For the rest of the considered Bjorken-$x$'es the symmetric dPDFs obey the GS sum rules at $10 - 25\%$ accuracy level. 
It would be interesting to modify the \pythia algorithm to suppress the large peaks in Figs.~\ref{fig:Pythia_SYM_sum_rules_uubar_ubaru}  and \ref{fig:Pythia_sum_rules_Rssbar_sym} such that we obtain symmetric dPDFs which obey the GS sum rules at a better accuracy level (for example, by suppressing the probability to find a companion quark accompanying a quark, when both partons have large $x$ fractions).
However, such a study is beyond the scope of this paper.

\subsection{Double Drell-Yan production}
\label{ss:double_DY_pheno}

In Section~\ref{ss:results_dPDFs_asym} and Section~\ref{ss:results_dPDFs_sym} we studied how  well asymmetric  and symmetric \pythia dPDFs satisfy the GS sum rules, and compared these to the GS09 dPDFs. Notably, we saw that the response functions defined by Eqs.~\eqref{eq:responce_function_def} and \eqref{eq:momentum_sum} match rather closely between the GS09 and \pythia dPDFs.

Given this close match, an interesting question is to what extent the \pythia and GS09 dPDFs are similar overall. 
Here we study this question via phenomenological MC simulations using both sets of dPDFs. We also include predictions obtained using the naive dPDFs defined in Eq.~\eqref{eq:naive_dPDFs}, as a benchmark.
It is most interesting to study a DPS process where the effects of the GS sum rules can potentially play an important role - \textit{i.e.} one which receives contributions associated with two quarks of the same (anti)flavour belonging to the same hadron. 
Therefore, we concentrate on a four-lepton production through the \textit{double Drell-Yan} (dDY) process~\cite{Goebel:1979mi, Mekhfi:1983az, Halzen:1986ue, Kom:2011nu, Myska:2011ji, Kasemets:2012pr, Krasny:2013aca}, defined as $\left(q\bar{q} \rightarrow \gamma^\ast/Z^\ast \rightarrow 2l\right) \otimes \left(q\bar{q} \rightarrow \gamma^\ast/Z^\ast \rightarrow 2l\right)$, where $l$  labels either an electron or muon final state. 

Before presenting our results, let us note that the older versions of  \pythia (version $\leq$ 8.235) were using asymmetric DPS luminosities  
\begin{eqnarray}
	\mathcal{L}_{j_1, j_2, j_3, j_4}(x_1, x_2, x_3, x_4) = 
	D_{j_1, j_2}(x_1, x_2) \,
	D_{j_3, j_4}(x_3, x_4)
	\label{eq:DPS_lumi_gen}
\end{eqnarray}
to simulate DPS production, whereas starting from  
version 8.240 \pythia uses  the symmetric ones\cprotect\footnote{More precisely,  staring from \pythia version 8.240 Eq.~\eqref{eq:DPS_lumi_gen_sym} is used when the option  \verb|SecondHard| is switched on. 
However, if the event is produced through the standard MPI machinery then a simple product of sPDFs as  in Eq.~\eqref{eq:DPS_lumi_gen} is used.}
\begin{align}
	\mathcal{L}^{\rm sym}_{j_1, j_2, j_3, j_4}(x_1, x_2, x_3, x_4) &= 
	\frac{1}{2} \,
	\mathcal{L}_{j_1, j_2, j_3, j_4}(x_1, x_2, x_3, x_4) \nonumber\\
	&\phantomrel{=}+\frac{1}{2} \, 
	\mathcal{L}_{j_2, j_1, j_4, j_3}(x_2, x_1, x_4, x_3),
	\label{eq:DPS_lumi_gen_sym}
\end{align}
where for the sake of simplicity we skipped factorization scale labels. 
However, if the number of generated events is high enough then MC simulations with Eq.~\eqref{eq:DPS_lumi_gen} and Eq.~\eqref{eq:DPS_lumi_gen_sym}  lead to the same results, since both  equivalent  ways to produce the $A\,B$ final state ($j_1 \, j_3 \,\rightarrow\, A \,\otimes\, j_2 \, j_4 \,\rightarrow\, B$  and  $j_2 \, j_4 \,\rightarrow\, A \,\otimes\, j_1 \, j_3 \,\rightarrow\, B$)  will be used approximately equal number of times.
To check this statement we simulated the DPS distributions presented in this section with  \pythia 8.235 and   \pythia 8.240, and found basically no  difference between both programs.
Therefore, throughout this section we use the symmetrized DPS luminosities  Eq.~\eqref{eq:DPS_lumi_gen_sym} only and omit the ``$\rm sym$'' labels  to keep our notation simple\footnote{Bear in mind that the usage of the  symmetic DPS luminosities as in Eq.~\eqref{eq:DPS_lumi_gen_sym} does not correspond to working with symmetric dPDFs. 
The luminosities obtained using symmetric dPDFs are given by:
\begin{align} 
	D^{\rm sym}_{j_1, j_2}(x_1, x_2) \,
	D^{\rm sym}_{j_3, j_4}(x_3, x_4) &=
	\frac{1}{2} \,
	\mathcal{L}^{\rm sym}_{j_1, j_2, j_3, j_4}(x_1, x_2, x_3, x_4)\nonumber\\
	&\phantomrel{=}+ 
	\frac{1}{2} \,
	\mathcal{L}^{\rm sym}_{j_1, j_2, j_4, j_3}(x_1, x_2, x_4, x_3).
	\label{eq:DPS_lumi_all_permut}
\end{align}
By comparing Eq.~\eqref{eq:DPS_lumi_all_permut} with Eq.~\eqref{eq:DPS_lumi_gen_sym} we see that the symmetrization procedure as in \pythia 8.240 does not take into account the second term on the right hand side of Eq.~\eqref{eq:DPS_lumi_all_permut}.}.

To obtain the predictions with the GS09 and naive dPDFs,  we change the weight of each \pythia event by multiplying it  by a corresponding ratio of DPS luminosities:\cprotect\footnote{To perform the reweighting procedure one has to either use the method we describe  in Appendix~\ref{appendix:how_to_call_modified_PDFs} to compute 
$\mathcal{L}^{\rm Pythia}$ or to be able to access DPS luminosities used by \pythia for a given DPS event. 
The latter can be achieved, for example, by adding new members to the \verb|Event|  class to store the  information about the sPDFs being used to simulate the second interaction. 
This additional information has to be  added to the event record in the  \verb|ProcessLevel::nextOne(Event& process)|  method of the \verb|ProcessLevel| class after the successful generation of the second hard interaction. 
The information  assigned in such a way is then accessible together with the other standard information stored in the event record and can be used to evaluate $\omega^{\rm GS09}$ and $\omega^{\rm Naive}$  during the generation procedure. Also note that by default in pp collisions \pythia assign  a weight equal to one to each event.}
\begin{eqnarray}
	\omega^{\rm GS09}  &=& 
	\frac
	{
		\mathcal{L}^{\rm GS09}
	}
	{
		\mathcal{L}^{\rm Pythia}
	},
	\label{eq:GS09_Pythia_W}\\
	\omega^{\rm Naive}  &=& 
	\frac
	{
		\mathcal{L}^{\rm Naive}
	}
	{
		\mathcal{L}^{\rm Pythia}
	}.
	\label{eq:Naive_Pythia_W}
\end{eqnarray}

Before discussing our results we would also like to note that  according to the \pythia model the total DPS cross section is given by 
\begin{eqnarray}
	\sigma_{\rm DPS} = \frac{\sigma_{\rm ND}}{2} \left(\frac{\sigma_1}{\sigma_{\rm ND}}\right)^2,
	\label{eq:dps_pythia_sigma_ND}
\end{eqnarray}
where $\sigma_{\rm ND}$ is a total non-diffractive cross section~\cite{Sjostrand:2004pf}.  
This differs by a factor of $\sigma_{\rm eff} / \sigma_{\rm ND}$
from the usual ``pocket formula'' result: $\sigma_{\rm DPS} = \sigma^2_1 / 2 \sigma_{\rm eff}$. 
The transition from Eq.~\eqref{eq:dps_pythia_sigma_ND} to the pocket formula expression in \pythia, however, 
is used only in the MPI machinery and is not performed if the flag \verb|SecondHard| is turned on.  
This implies that the DPS cross sections in our simulations are evaluated according 
to \mbox{Eq.~\eqref{eq:dps_pythia_sigma_ND}} with $\sigma_{\rm ND}$ instead of $\sigma_{\rm eff}$. 
However,  since we are interested only in the relative difference between distributions of DPS events produced with different models of dPDFs, the aforementioned difference in the value of the DPS cross section does not affect our analysis. 

\begin{figure}
\begin{minipage}[h]{0.49\linewidth}
\center{\includegraphics[width=1.0\linewidth]{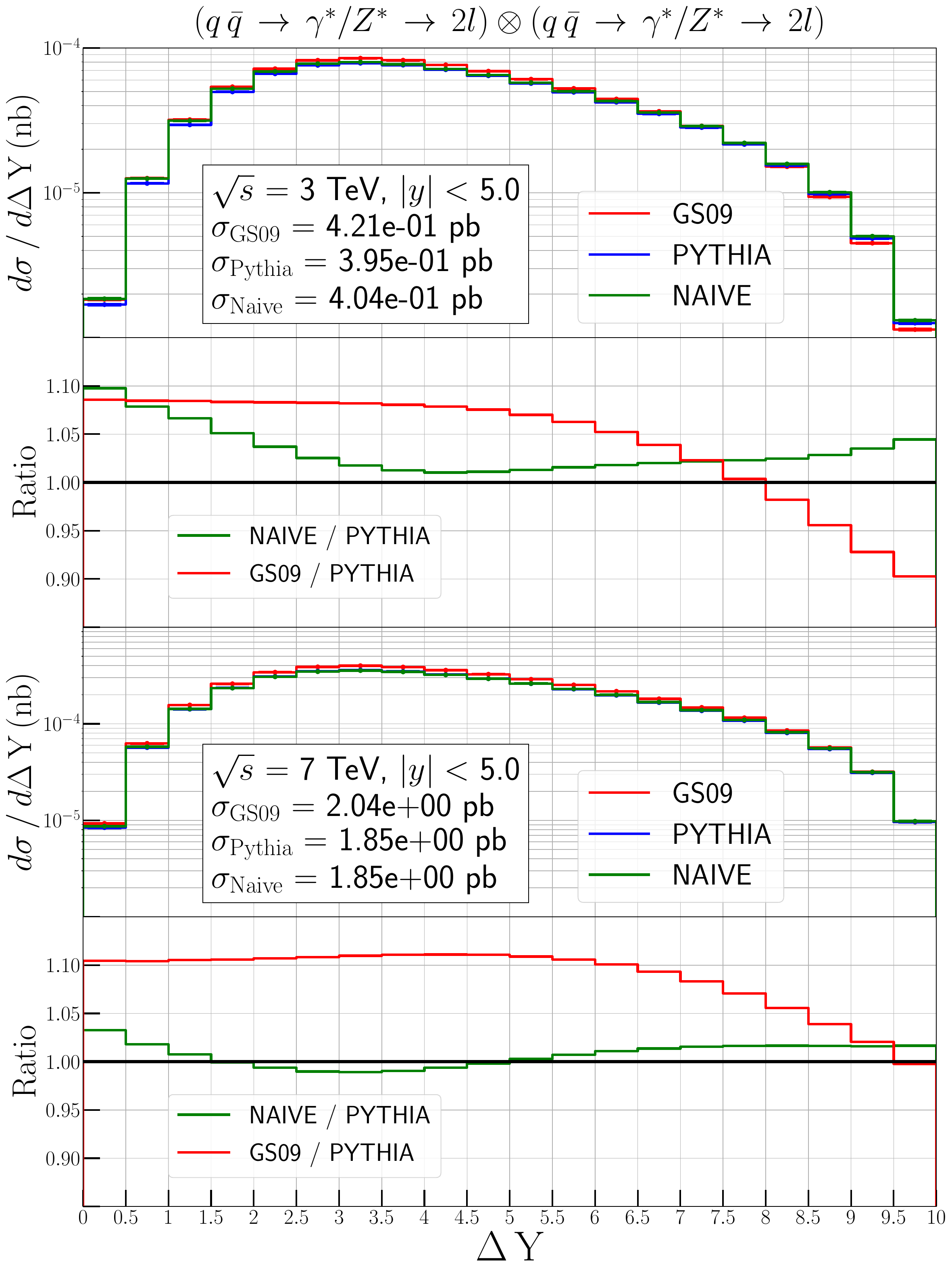}}
\end{minipage}
\hfill
\begin{minipage}[h]{0.49\linewidth}
\center{\includegraphics[width=1.0\linewidth]{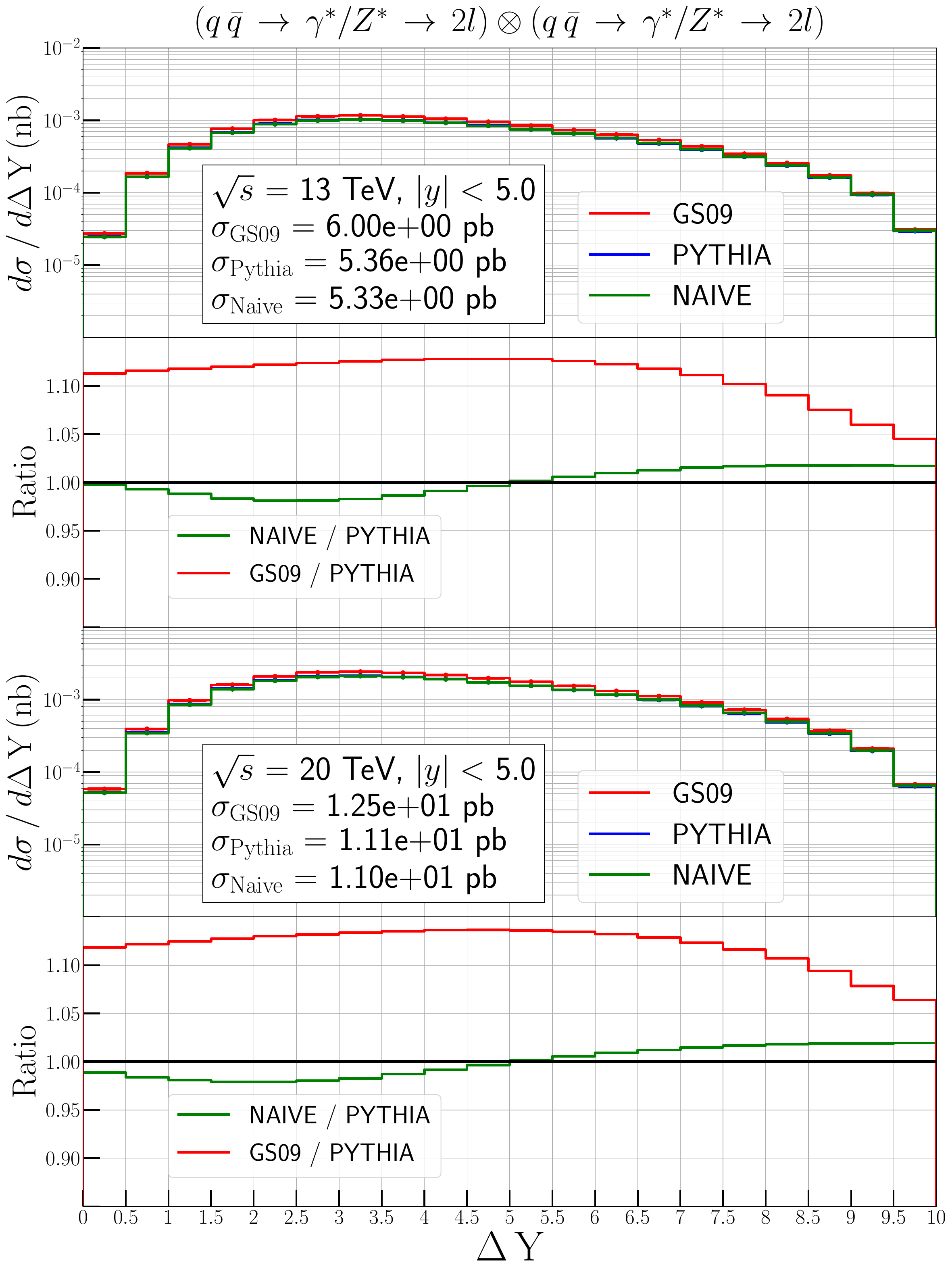}}
\end{minipage}
\cprotect\caption{Comparison between ${\rm \Delta Y} = {\rm max}|y_i - y_j|$ DPS distributions generated with the GS09, \pythia and naive dPDFs. 
The collision energy $\sqrt{S}$ is equal to $3$, $7$, $13$ and $20$ TeV. 
}
\label{fig:dY_dist_all}
\end{figure}

The differential distributions generated with the naive, \pythia  and GS09 dPDFs are given in Fig.~\ref{fig:dY_dist_all}.  
We perform binning in terms of the maximal rapidity difference ${\rm \Delta Y} \equiv {\rm max}|y_i - y_j|$ of the produced leptons~\cite{Kom:2011bd}.  
The absolute values of rapidities of all produced leptons were requested to be $\le 5$. 
The complete \pythia setup we use is given in Appendix \ref{appendix:set_up_we_use}.
Since we are essentially interested in comparison between the different models of dPDFs we do not impose any additional cuts on the final state leptons.
Instead, in order to have better control of the  probed values of the Bjorken-$x$'es, we consider different values of the collision energy, namely 3, 7, 13 and \mbox{20 TeV}.  
We fix the renormalization scale equal to the factorization scale, and set the factorization scale of both interactions to \mbox{91 GeV ( $\sim M_Z$)}. 

We see that the ratio $d\sigma_{\rm{Naive}}/d\sigma_{\rm{Pythia}}$ is fairly flat in ${\rm \Delta Y}$, and remains rather close to $1$ for the whole range of ${\rm \Delta Y}$ (except at the 3 TeV collision energy where we see some increase at small and large values of ${\rm \Delta Y}$), whereas $d\sigma_{\rm{GS09}}/d\sigma_{\rm{Pythia}}$ decreases as one approaches the largest values of ${\rm \Delta Y}$. 
We see also that the ratio $d\sigma_{\rm{GS09}}/d\sigma_{\rm{Pythia}}$ remains more or less constant at small and moderate ${\rm \Delta Y}$ as $\sqrt{S}$ is varied, whilst there is a noticeable decrease in this ratio at large ${\rm \Delta Y}$ as $\sqrt{S}$ decreases.

\begin{figure}
\begin{minipage}[h!]{0.59\linewidth}
\center{\includegraphics[width=1.\linewidth]{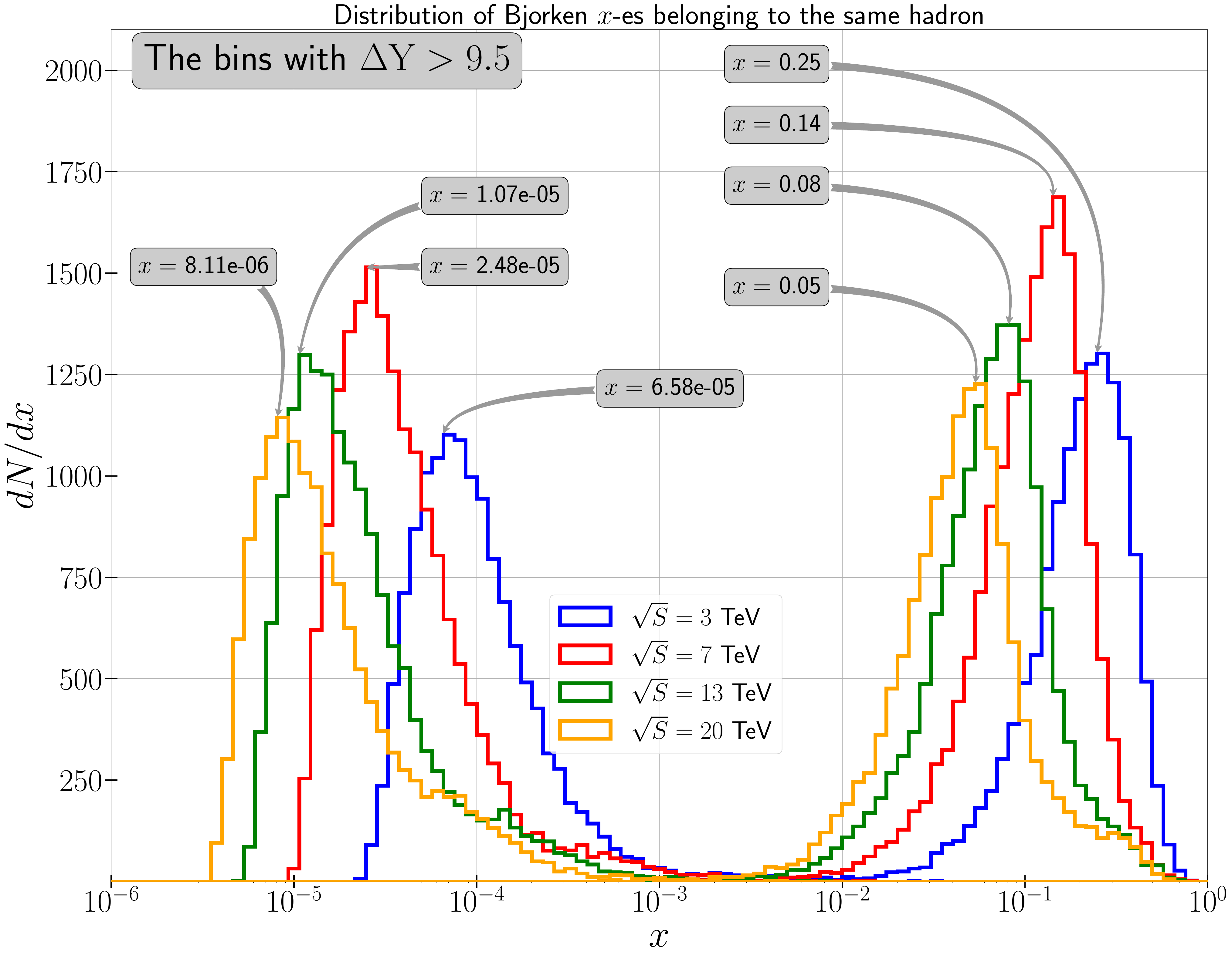} \\a)}
\end{minipage}
\hfill
\begin{minipage}[h!]{0.4\linewidth}
\center{\includegraphics[width=0.85\linewidth]{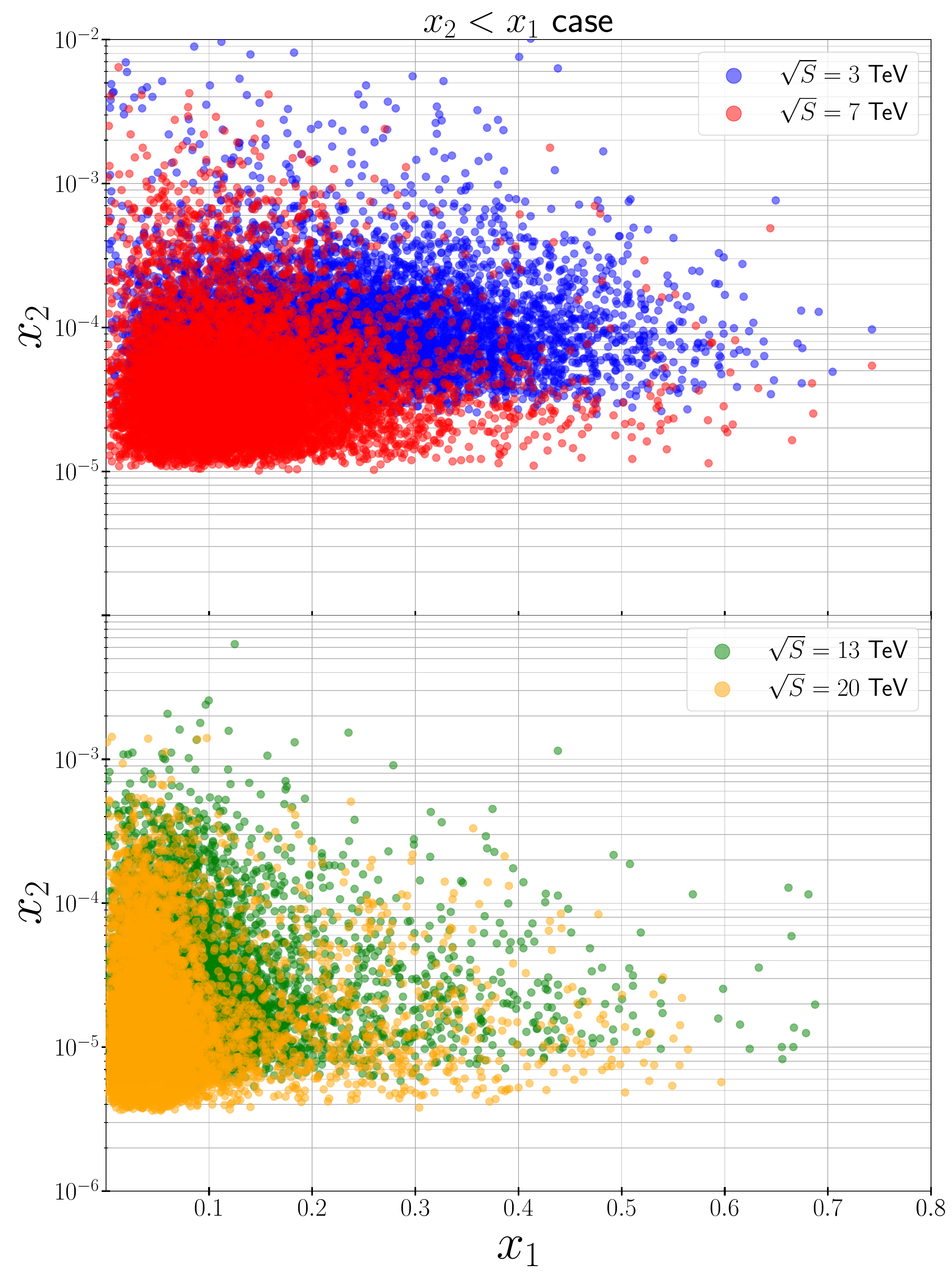}\\b)}
\end{minipage}
\cprotect\caption{Distribution of Bjorken $x$-es belonging to the  same hadron for the bin with  ${\rm \Delta Y} \in [9.5, 10]$. Left plot: distribution of the  minimal and maximal Bjorken $x$-es at 3, 7, 13 and 20 TeV collision energies.  Right plot: distribution of the DPS events contributing to the ${\rm \Delta Y} \in [9.5, 10]$ bin in the $x_1$ - $x_2$ plane for the $x_2 < x_1$ case.}
\label{fig:x_dist}
\end{figure}

Let us now try to explain all of these features, starting with the very large ${\rm \Delta Y}$ region (and in particular the last bin with ${\rm \Delta Y} \in [9.5, 10]$). We expect this ${\rm \Delta Y}$ bin to probe a very asymmetric configuration of $x$ fractions in the dPDF, with the values of $x$ increasing as $\sqrt{S}$ is lowered. This is confirmed in the plots of Fig.~\ref{fig:x_dist}, which also shows the values of $x$ probed at each value of $\sqrt{S}$.

\begin{figure}
\begin{minipage}[h!]{1.0\linewidth}
\center{\includegraphics[width=0.75\linewidth]{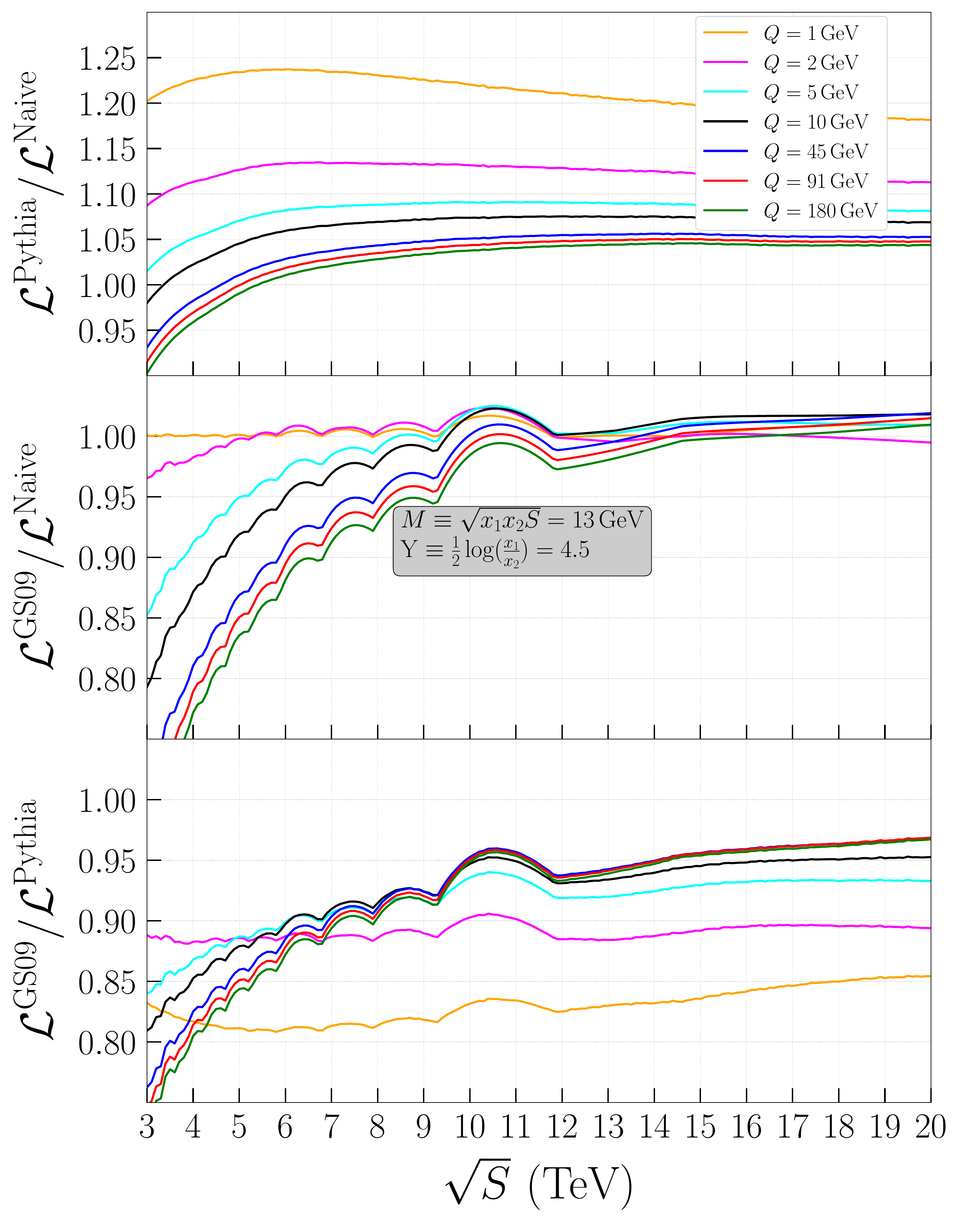}}
\end{minipage}
\cprotect\caption{Ratios of the DPS luminosities at  different values of factorization scale $Q$ as  functions of collision energy  $\sqrt{S}$ for the $u$-$\bar{u}$ flavour combination.}
\label{fig:x_ratio_PDS_lumi}
\end{figure}

Why does the ratio $d\sigma_{\rm{GS09}}/d\sigma_{\rm{Pythia}}$ decrease as we decrease $\sqrt{S}$ and probe larger $x$ values, whilst $d\sigma_{\rm{Naive}}/d\sigma_{\rm{Pythia}}$ stays fairly close to $1$? To try to understand this, we use a simplified setup, in which we consider only one partonic contribution;  the one in which a $u$ and a $\bar{u}$ from one proton collide with a $\bar{u}$ and a $u$ in the other. This sort of contribution should be dominant at large $\Delta{\rm Y}$, as we can have a right-moving valence quark in the first subprocess, and a left-moving valence quark in the second. We fix the $x$ values in the dPDFs to correspond to the production of two heavy states of mass \mbox{$M = \sqrt{x_1\, x_2\, S} = 13$ GeV}, with the first being produced with rapidity ${\rm Y} = \log(x_1 / x_2) / 2 = 4.5$, and the other being produced with rapidity ${\rm Y}=-4.5$ (yielding ${\rm \Delta Y} = 9$ overall), and for simplicity we consider the luminosity rather than the cross section.
The corresponding results are given in Fig.~\ref{fig:x_ratio_PDS_lumi} where we plot ratios of DPS luminosities  as a function of collision energy $\sqrt{S}$.
As we mentioned before, Fig.~\ref{fig:dY_dist_all} is generated using the particular choice of  factorization scale \mbox{$Q = 91$ GeV}.
However, in Fig.~\ref{fig:x_ratio_PDS_lumi} we choose to generate the ratios for a range of factorization scales in between $1$ and $180$ GeV, as this helps us explain the observed behaviour at \mbox{$Q = 91$ GeV}.

Let us first study the $Q$-dependence of the $\mathcal{L}^{\rm Pythia} / \mathcal{L}^{\rm Naive}$ ratio which is driven by the modification of the sPDFs affecting the $\mathcal{L}^{\rm Pythia}$ luminosity. We see in the upper plot of Fig.~\ref{fig:x_ratio_PDS_lumi} that at $Q=91$ GeV the ratio $\mathcal{L}^{\rm Pythia} / \mathcal{L}^{\rm Naive}$ changes if $\sqrt{S} \in [3, 9]$ TeV and becomes almost constant if $\sqrt{S} > 9$ TeV, which is consistent with our results shown  in Fig.~\ref{fig:dY_dist_all}\footnote{Note  that here we consider a given flavour combination and fixed $x$-values. Therefore, we expect only qualitative agreement between Figs.~\ref{fig:dY_dist_all} and \ref{fig:x_ratio_PDS_lumi}.}. We also see that the
ratio $\mathcal{L}^{\rm Pythia} / \mathcal{L}^{\rm Naive}$ demonstrates a considerable change as the factorization scale increases from $1$ to $180$ GeV (compare orange and green curves).
The $Q$-dependence becomes milder as $Q$ increases, and is very gradual for $Q \geq 45$ GeV.
To investigate this behaviour let us note that for a given configuration 
($x_1 = x_3,\, x_2 = x_4$, $j_1 = j_3,\, j_2 = j_4$) $\mathcal{L}^{\rm Pythia} / \mathcal{L}^{\rm Naive}$ is given by 
\begin{equation}
	\frac
	{
	    D^{\rm Pythia}_{u\bar{u}}(x_1, x_2)\,
	    D^{\rm Pythia}_{\bar{u}u}(x_2, x_1)
	}
	{
	    D^{\rm Naive}_{u\bar{u}}(x_1, x_2)\, 
	    D^{\rm Naive}_{\bar{u}u}(x_2, x_1)
	} 
	= 
	\left[
		\frac{f^{m \leftarrow u, x_1}_{\bar{u}}(x_2, Q)}{f^r_{\bar{u}}(x_2, Q)}
	\right]
	\left[
	    \frac{f^{m \leftarrow \bar{u}, x_2}_{u}(x_1, Q)}{f^r_{u}(x_1, Q)}
	\right].
	\label{eq:pythia_naive_lumi}
\end{equation}
The kinematics we consider implies that $x_1 \in [10^{-2}, 1]$ whereas $x_2$ is well between $10^{-4}$ and $10^{-6}$ (for instance, if $\sqrt{S} = 3$ TeV $x_1 \approx 0.4$ and $x_2 \approx 4.81 \times 10^{-5}$).
We found that for the Bjorken-$x$'es corresponding to Fig.~\ref{fig:x_ratio_PDS_lumi} the second   factor on the right hand side of   Eq.~\eqref{eq:pythia_naive_lumi} is very close to unity whereas the first one is responsible for the observed differences.
Since the first factor  on the right hand side of Eq.~\eqref{eq:pythia_naive_lumi} implies that in the first interaction we probe a valence $u$-quark and a sea $\bar{u}$-quark in the second one, then, according to the \pythia model, we deal with two modification mechanisms: first we squeeze all sPDFs according to Eqs.~\eqref{eq:pythia_squezing_1}, \eqref{eq:pythia_val_quark_reweight} and then rescale  the sea sPDFs by multiplying them by the $a$-factor as in  Eqs.~\eqref{eq:pythia_sea_gluon_rescale_1}, \eqref{eq:pythia_sea_gluon_rescale_2} to preserve overall momentum conservation. 
We found that the rescaling  of the $\bar{u}$ sPDFs according to  Eq.~\eqref{eq:pythia_sea_gluon_rescale_1}  increases the value of $f^{m \leftarrow u, x_1}_{\bar{u}}$ whereas the squeezing according to Eq.~\eqref{eq:pythia_squezing_1}  tends to make it smaller.   
The effect of the squeezing depends on the rate at which the $\bar{u}$ PDF grows at small $x$ -- if the $\bar{u}$ PDF at small $x$ is proportional to $x^{-b}$, then taking only the squeezing effect into account we would have
\begin{align}
    \frac{f^{m \leftarrow u, x_1}_{\bar{u}}(x_2, Q)}{f^r_{\bar{u}}(x_2, Q)} = (1-x_1)^{b-1}.
\end{align}
For the MSTW set we use, $b>1$ and grows rapidly at small $Q \sim 1-2$ GeV before reaching an approximately stable value of $b \sim 1.5$ at larger $Q$ values. Thus, the squeezing effect corresponds to a suppression of the ratio $f^{m \leftarrow u, x_1}_{\bar{u}}/f^r_{\bar{u}}$, which increases at low $Q$ but stabilises at larger $Q$ values. This suppression also increases as $x_1$ increases, or $\sqrt{S}$ decreases.

The ``$a$-rescaling'' effect is initially quite large -- we find $a \sim 1.2-1.3$ at $Q = 1$ GeV, which explains the (perhaps unexpected!) fact that $\mathcal{L}^{\rm Pythia} / \mathcal{L}^{\rm Naive}$ significantly exceeds $1$ at this $Q$ value. This enhancement rapidly reduces to $\sim 1.1$ at $Q \sim 5$ GeV, and thereafter changes rather slowly. At $Q = 1$ GeV, there is a noticeable decrease in $a$ as $\sqrt{S}$ increases, but at higher $Q$ values the behaviour of $a$ is rather flat in $\sqrt{S}$.

Thus, we see that the reduction of $\mathcal{L}^{\rm Pythia} / \mathcal{L}^{\rm Naive}$ as $Q$ increases is caused both by the squeezing suppression getting stronger and the $a$-rescaling enhancement getting weaker. At large $Q$ values such as $Q=91$ GeV the downward trend in the ratio as $\sqrt{S}$ decreases is mainly driven by the squeezing effect getting stronger, as the $a$-rescaling enhancement is roughly constant with $\sqrt{S}$.

Now let us  consider the ratio $\mathcal{L}^{\rm GS09} / \mathcal{L}^{\rm Naive}$ shown in the middle plot of  Fig.~\ref{fig:x_ratio_PDS_lumi}\footnote{Note that the oscillation of the ratios involving the $\mathcal{L}^{\rm GS09}$ luminosities in  Fig.~\ref{fig:x_ratio_PDS_lumi} are numerical artifacts of the linear interpolation used in the GS09 grids for the Bjorken-$x$'es $\geq 0.1$. }. 
At $Q = 1$ GeV this ratio is very close to unity, 
which is essentially by design -- when the $x$ value of one parton is very small (as we have here), the GS09 inputs are constructed to match closely with the naive dPDFs, as this will ensure the momentum sum rule is well-satisfied when integrating over the other parton (\textit{c.f.} Eq.~\eqref{eq:schematic_form_mr3}). At $Q = 1$ GeV the differences between the GS09 and naive dPDFs predominantly manifest themselves in the region where both $x$ values in the dPDF are large  (see Ref.~\cite{Gaunt:2009re}). 
However, as $Q$ starts to increase we observe some important deviations of the ratio $\mathcal{L}^{\rm GS09} / \mathcal{L}^{\rm Naive}$ from 1 in Fig.~\ref{fig:x_ratio_PDS_lumi} caused by the dDGLAP evolution.
We see that if $\sqrt{S}$ is approximately in  the $[3, 10]$ TeV interval the ratio  $\mathcal{L}^{\rm GS09} / \mathcal{L}^{\rm Naive}$ is smaller than one, whereas at large $\sqrt{S}$ we observe the opposite behaviour.
This observation is consistent with the picture of the dDGLAP evolution from Ref.~\cite{Gaunt:2009re}; at the starting value of the evolution scale $Q = 1$ GeV the differences between the GS09 and naive dPDFs are located in the region where both $x_i$ are large, however, as $Q$ grows the differences propagate down to smaller values of $x_1, x_2$ leading to a decrease of the GS09 $D_{u\bar{u}}$ dPDFs relative to the naive ones. 
Taking into account the fact that in our analysis  the decrease in $\sqrt{S}$ leads to an increase of both $x_1$ and $x_2$ (see the distributions in Fig.~\ref{fig:x_dist}) we conclude that if \mbox{$\sqrt{S} \in [3, 10]$ TeV} the $x_i$ values enter the region significantly affected by the aforementioned propagation of the differences between two models down to the smaller values of momentum fractions. 
However, as $\sqrt{S}$ increases both $x_1$ and $x_2$ become smaller which lead us to the region of small $x$'es where the $1\rightarrow2$ splitting contributions due to the last term on the right hand side of Eq.~\eqref{eq:double_dglap} become important.
Since the $1\rightarrow2$ feeding term causes an increase of the parton densities we conclude that the change of the behaviour of the $\mathcal{L}^{\rm GS09} / \mathcal{L}^{\rm Naive}$  ratio at large values of $\sqrt{S}$ is due to the positive contribution from $1\rightarrow2$ splittings.

Having now studied and understood the behaviour of the $\mathcal{L}^{\rm Pythia} / \mathcal{L}^{\rm Naive}$ and  $\mathcal{L}^{\rm GS09} / \mathcal{L}^{\rm Naive}$ ratios, we can now understand the behaviour of $\mathcal{L}^{\rm GS09} / \mathcal{L}^{\rm Pythia}$ as this can simply be constructed by taking the quotient of the ratios already studied. In particular, we note that at very low $Q$'s the ratio $\mathcal{L}^{\rm GS09} / \mathcal{L}^{\rm Pythia}$ lies significantly below $1$ predominantly due to the $a$-rescaling enhancement in \pythia. At large $Q$, the trend in $\sqrt{S}$ is mainly driven by momentum/number sum rule suppression in the GS09 dPDFs, which has started to propagate downwards into the region where one $x_i$ is large and the other small.
As one can see the behaviour of the $\mathcal{L}^{\rm GS09} / \mathcal{L}^{\rm Pythia}$ at $Q=91$ GeV is consistent with the distributions shown in Fig.~\ref{fig:dY_dist_all}.

Now let us focus on the small $\Delta \rm Y$ region in Fig.~\ref{fig:dY_dist_all}, and initially restrict our attention to the behaviour of $d\sigma_{\rm{Naive}}/d\sigma_{\rm{Pythia}}$. We saw in Fig.~\ref{fig:dY_dist_all} that this ratio stays within 5\% of $1$ for the whole range of $\Delta \rm Y$ at 7, 13 and 20 TeV collision energies. However, if $\sqrt{S} = 3$ TeV we observe about 10\% deviation of the aforementioned ratio from $1$ at small $\Delta \rm Y \in [0, 0.5]$.

\begin{figure}
\begin{minipage}[h!]{0.49\linewidth}
\center{\includegraphics[width=1.\linewidth]{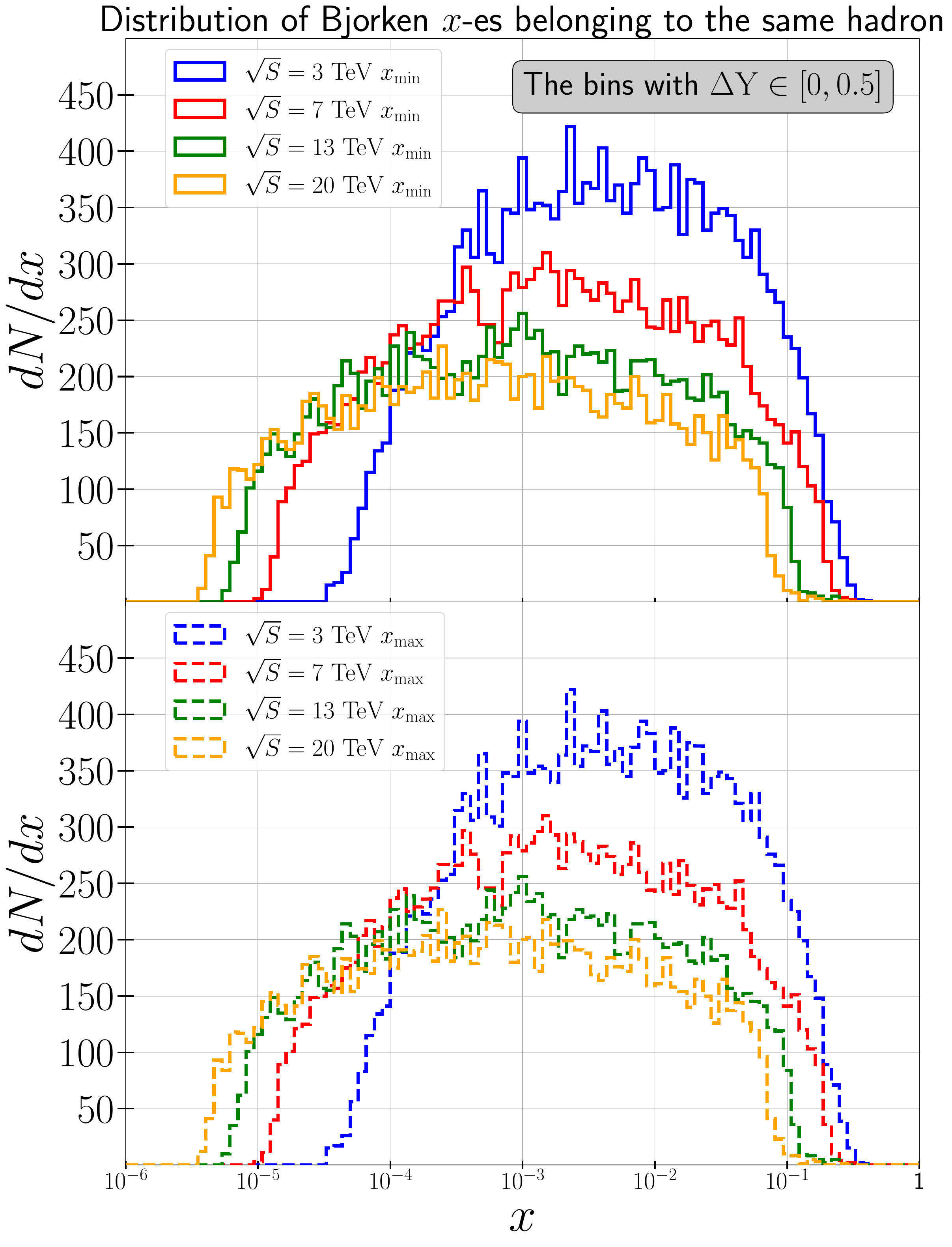}\\a)}
\end{minipage}
\hfill
\begin{minipage}[h!]{0.49\linewidth}
\center{\includegraphics[width=1.\linewidth]{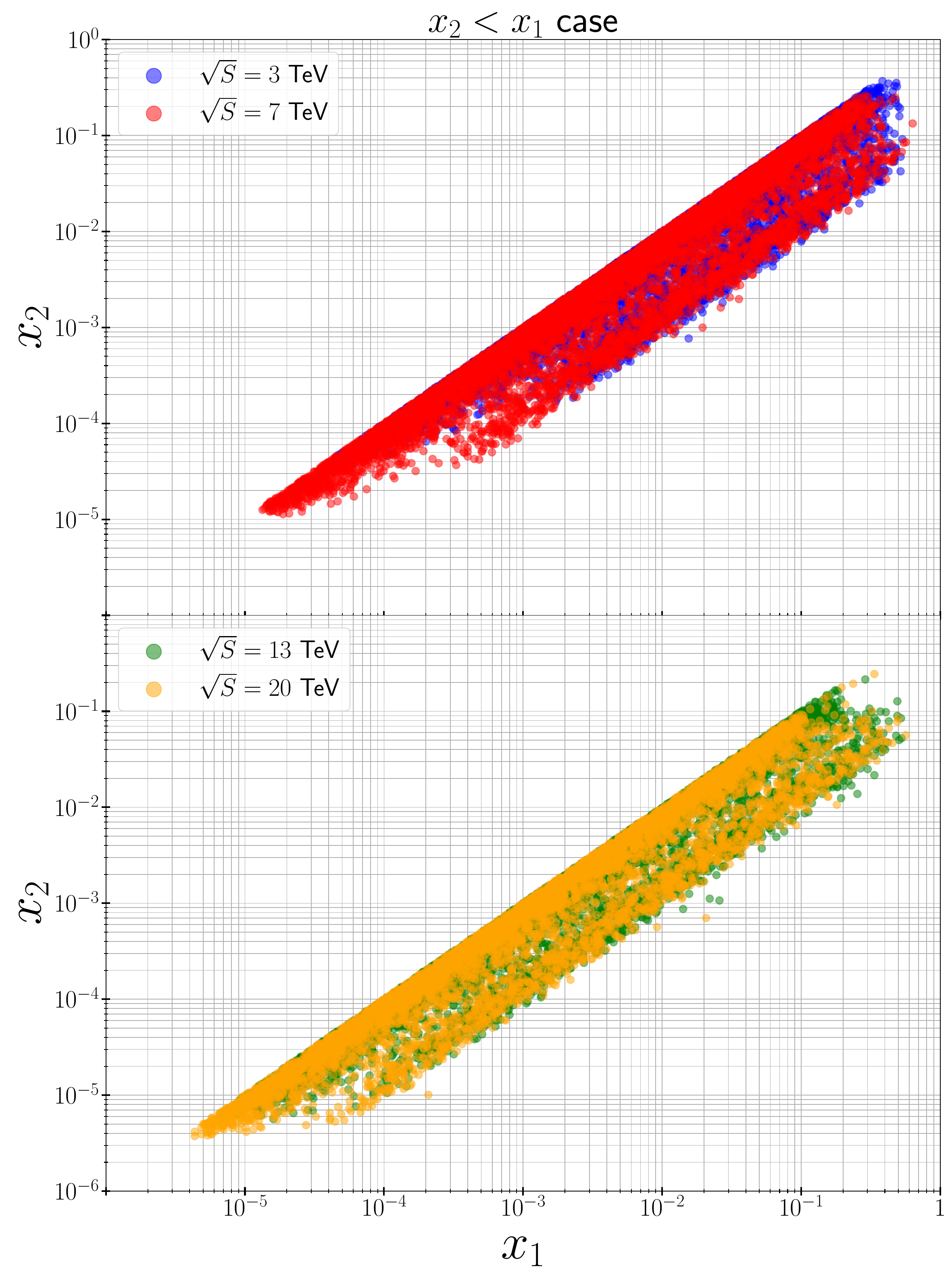}\\b)}
\end{minipage}
\cprotect\caption{Distribution of Bjorken-$x$'es belonging to the  proton with the  $D_{j_1 j_2}(x_1, x_2, Q)$ dPDF for the bin with  ${\rm \Delta Y} \in [0, 0.5]$. Left plot: distribution of the  minimal and maximal Bjorken-$x$'es for 3, 7, 13 and 20 TeV collision energies.  Right plot: distribution of the DPS events contributing to the ${\rm \Delta Y} \in [0, 0.5]$ bin in the $x_1$ - $x_2$ plane for the $x_2 < x_1$ case.}
\label{fig:x_dist_first}
\end{figure}

The $\Delta \rm Y \in [0, 0.5]$ bin should probe dPDFs where the two $x_i$ values are of the same magnitude -- this is confirmed in Fig.~\ref{fig:x_dist_first} where we plot 
the distribution of Bjorken-$x$'es for the  \pythia events contributing to this bin. We also see that the distribution of events in this common $x$ value is approximately flat with the bulk of events having $x \in [10^{-5}, 10^{-1}]$. Furthermore, unlike the $\Delta \rm Y \in [9.5, 10]$ bin which predominantly probes dPDFs with flavour structure $q_i\bar{q_j}$ ($i, j  = \{u, d, s, c, b\}$), the events in the $\Delta \rm Y \in [0, 0.5]$ bin arise from an approximately even mix of events where a $q_iq_j$ in one proton collide with a $\bar{q}_i\bar{q}_j$ in the other, and events where we have a $q_i\bar{q}_j$ vs. $\bar{q}_iq_j$ collision -- see Table~\ref{tab:dY_qq_qqbar}. Finally, we found that amongst the five (anti-)quark flavours we consider, the combinations of $\bar{u}$, $\bar{d}$, $u$ and $d$ flavours occurs most frequently.

\begin{figure}
\begin{minipage}[h!]{1.0\linewidth}
\center{\includegraphics[width=1.\linewidth]{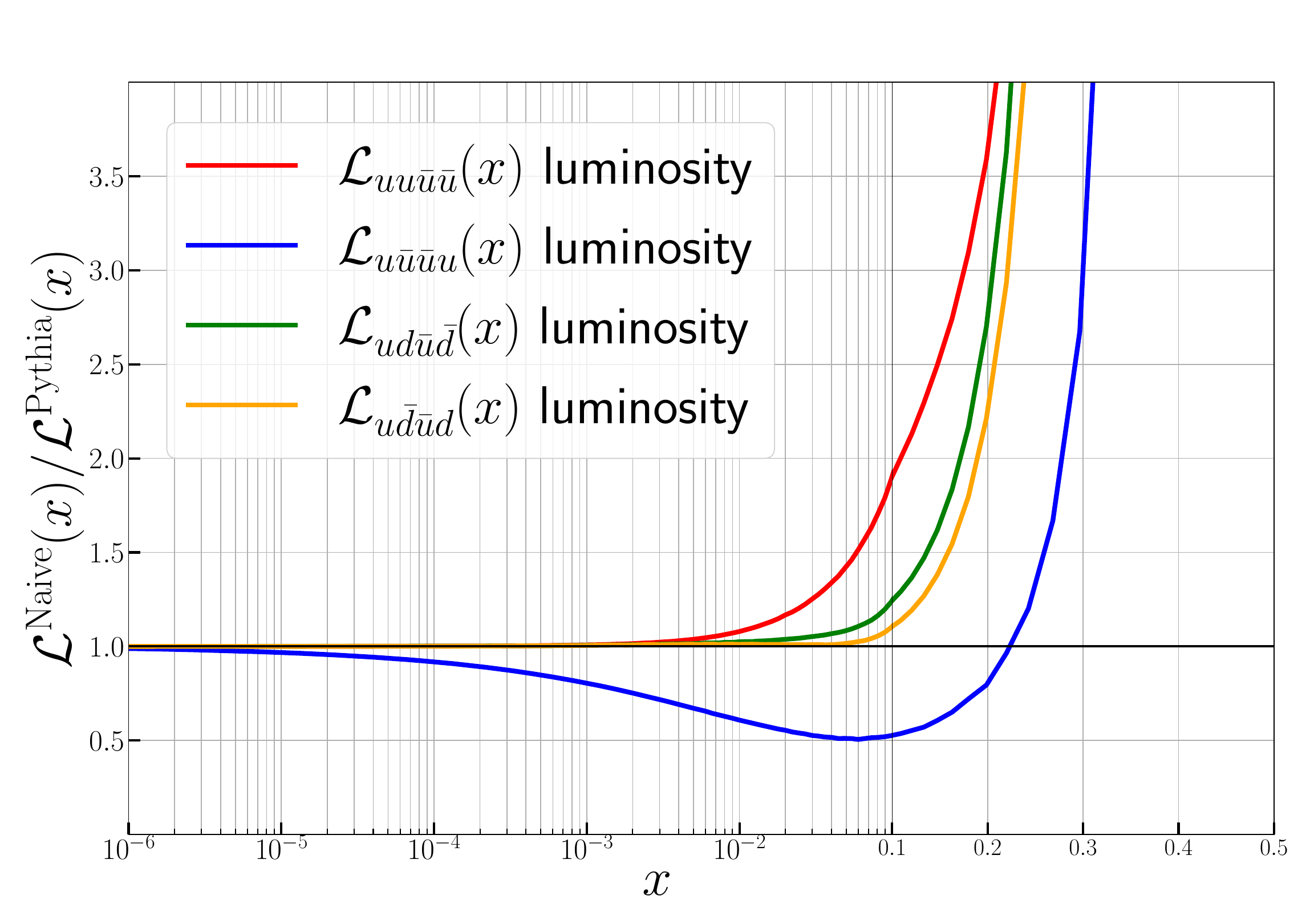}}
\end{minipage}
\cprotect\caption{Ratio of the DPS luminosities $\mathcal{L}^{\mathrm{Naive}} / \mathcal{L}^{\mathrm{Pythia}}$ as a function of $x$ where $x_1 = x_2 = x_3 = x_4 = x$. 
}
\label{fig:x_ratio_PDS_lumi_same_x}
\end{figure}

\begin{table}
\begin{center}
	\begin{tabular}{ | c | c | c |}
    \hline
        $\sqrt{S}$  & $\Delta \rm Y \in [0, 0.5]$ & $\Delta \rm Y \in [9.5, 10]$\\ \hline
		
		$\sqrt{S} = 3$ TeV,  $q_i \, q_j$ & 53.3\% 	& 19.2\%  \\ \hline
		$\sqrt{S} = 7$ TeV,  $q_i \, q_j$ & 51.3\% 	& 29.9\%  \\ \hline
		$\sqrt{S} = 13$ TeV, $q_i \, q_j$ & 50.7\% 	& 38.5\%  \\ \hline
		$\sqrt{S} = 20$ TeV, $q_i \, q_j$ & 50.1\% 	& 42.7\%  \\ \hline
	
   	\end{tabular}
	\captionof{table}{Flavour composition   of the dPDFs belonging to the same proton and contributing to the first and last $\Delta \rm Y$ bins ($q_i \, q_j \,+\, q_i \, \bar{q}_j = 100\%$). }
	\label{tab:dY_qq_qqbar}
\end{center}
\end{table}
In order to get a better understanding of $d\sigma_{\rm{Naive}}/d\sigma_{\rm{Pythia}}$ in the region $\Delta \rm Y \in [0, 0.5]$, we plot the ratio $\mathcal{L}^{\rm Naive} / \mathcal{L}^{\rm Pythia}$ where we set all Bjorken-$x$'es to the same value in Fig.~\ref{fig:x_ratio_PDS_lumi_same_x}.
In this figure we consider combinations of $u$ and $d$ flavours having different behaviour, namely $\mathcal{L}_{uu\bar{u}\bar{u}}$, $\mathcal{L}_{u\bar{u}\bar{u}u}$, $\mathcal{L}_{ud\bar{u}\bar{d}}$ and $\mathcal{L}_{u\bar{d}\bar{u}d}$ (here we skip ${dd\bar{d}\bar{d}}$ and ${d\bar{d}\bar{d}d}$ flavour combinations since they have behaviour similar to the analogous combinations of $u$ and $\bar{u}$ quarks). 
We see that for all considered flavour combinations, except for the $u\bar{u}\bar{u}u$ case,  the ratio is bigger than unity for the whole range of $x$. 
The decrease of the ratio  $\mathcal{L}^{\rm Naive} / \mathcal{L}^{\rm Pythia}$ for the $u\bar{u}\bar{u}u$ flavour combination below $1$ is due to the companion quark contribution from $g \rightarrow u \bar{u}$ splitting in the \pythia case.
We also see that the ratio corresponding to the $uu\bar{u}\bar{u}$ flavour combination increases faster compared to the $ud\bar{u}\bar{d}$ and  $u\bar{d}\bar{u}d$ cases.
This behaviour can be explained by the fact that for the $uu$ dPDF we invoke not only momentum squeezing mechanism in \pythia, as in Eq.~\eqref{eq:pythia_squezing}, but also have a suppression effect due to the change in remaining number of valence quarks, as in Eq.~\eqref{eq:pythia_val_quark_reweight}.
It is also interesting to note that the effects due to the momentum and number sum rule constraints can penetrate down to rather small Bjorken-$x$ values $\sim 10^{-2} - 10^{-1}$.
Since at $\sqrt{S} = 3$ TeV collision energy we probe Bjorken-$x$'es at larger values  than at 7 - 20 TeV (see blue distributions in  Fig.~\ref{fig:x_dist_first} a)), we conclude that the increase in the ratio of naive to \pythia distributions at small values of $\Delta \rm Y$ at 3 TeV collision energy, as shown in Fig.~\ref{fig:dY_dist_all}, is due to the number and momentum sum rule effects  which manifest themselves at the larger values of Bjorken-$x$'es accessible at the 3 TeV collision energy.

\begin{figure}
\begin{minipage}[h]{0.49\linewidth}
\center{\includegraphics[width=1.0\linewidth]{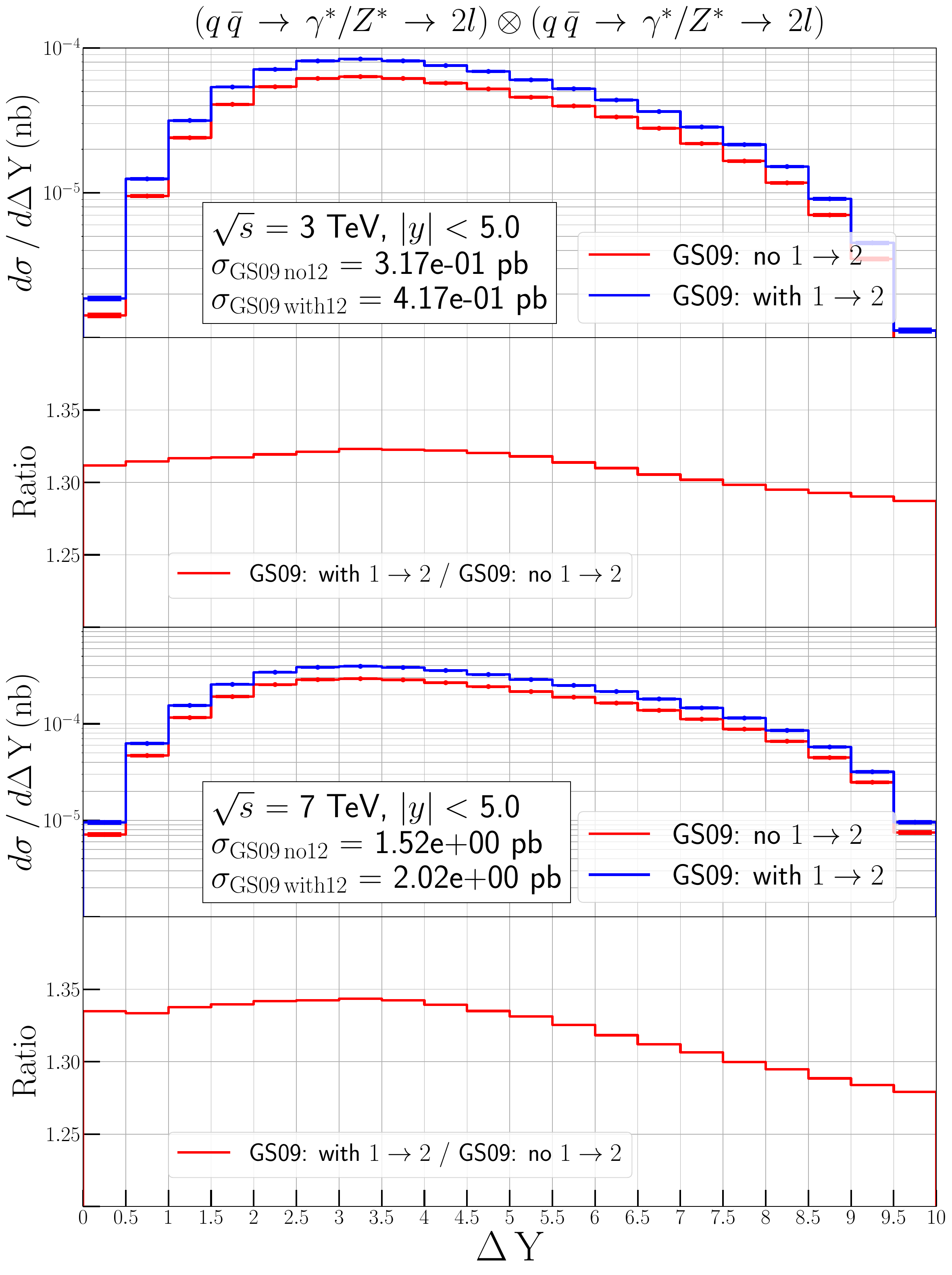}}
\end{minipage}
\hfill
\begin{minipage}[h]{0.49\linewidth}
\center{\includegraphics[width=1.0\linewidth]{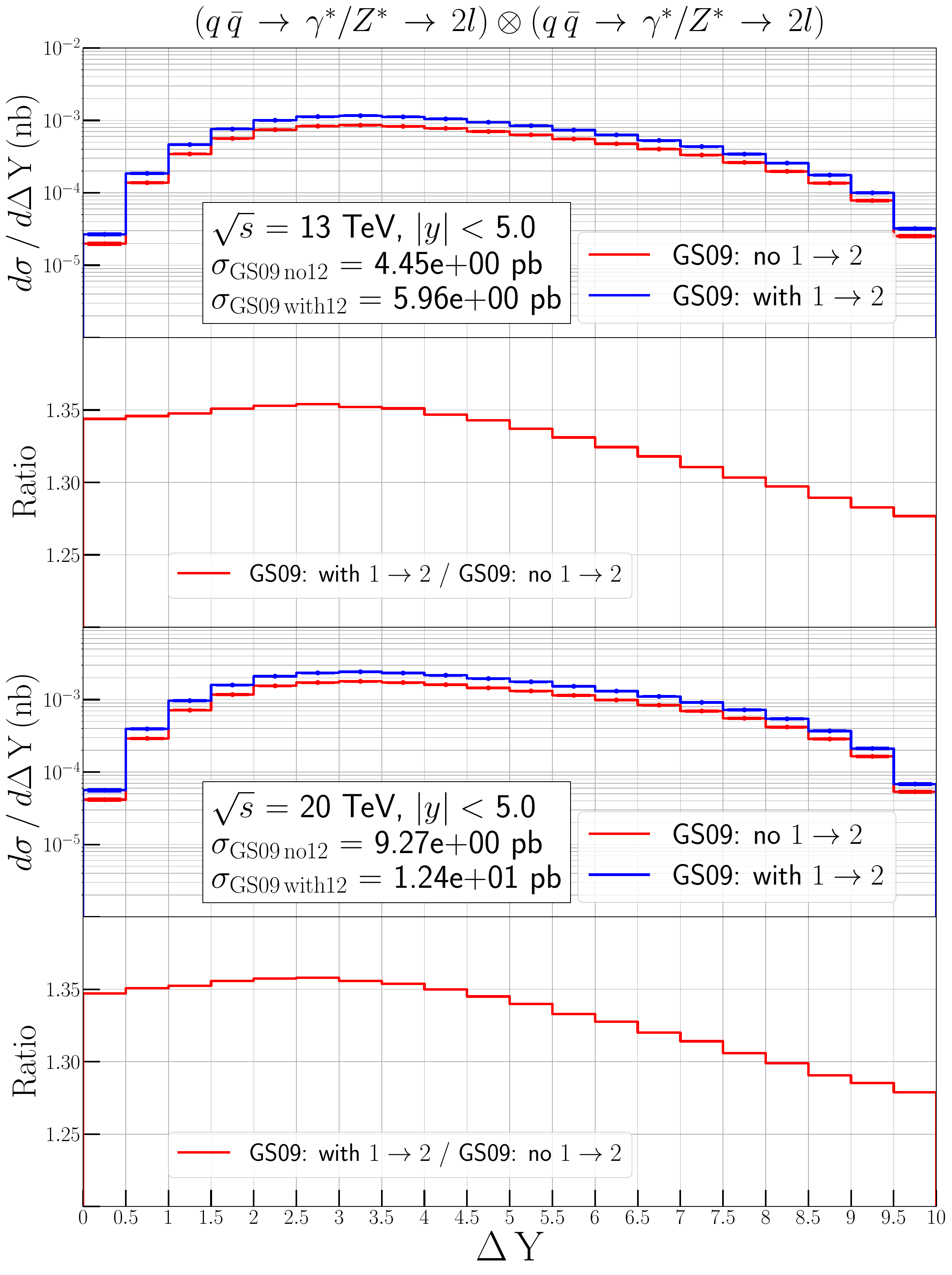}}
\end{minipage}
\cprotect\caption{Comparison between ${\rm \Delta Y} = {\rm max}|y_i - y_j|$ DPS distributions generated with the GS09 dPDFs produced by solving dDGLAP evolution equation with and without $1\rightarrow2$ splitting. The collision energy $\sqrt{S}$ is equal to $3$, $7$, $13$ and $20$ TeV.
}
\label{fig:dY_dist_12}
\end{figure}

Let us finally turn to considering the ratio $d\sigma_{\rm{GS09}}/d\sigma_{\rm{Pythia}}$ at small to moderate values of $\rm \Delta Y \in [0, 5]$. We see in Fig.~\ref{fig:dY_dist_all} that this ratio is approximately constant for $\rm \Delta Y \in [0, 5]$, and is $\sim 1.1$ across all values of $\sqrt{S}$. This enhancement of GS09 over \pythia is linked to $1 \to 2$ splitting contributions to the GS09 dPDFs during dDGLAP evolution, see the the last term on the right hand side of Eq.~\eqref{eq:double_dglap}. To illustrate this, in Fig.~\ref{fig:dY_dist_12} we plot the ratio of the $\Delta \rm Y$ distributions obtained by solving dDGLAP evolution equation with and without the $1 \rightarrow 2$ splitting contribution. 
As one can see the evolution effects due to the $1\rightarrow2$ splittings increase the $\Delta \rm Y$ distributions reasonably uniformly in the whole range of binning but somewhat more at the small values of $\Delta \rm Y$ than at the large values of $\Delta \rm Y$.
One notes that the enhancement factor when including the $1 \to 2$ splitting effects is $\sim 1.30-1.35$ at small $\rm \Delta Y$, which is rather larger than the factor $\sim 1.1$ enhancement of GS09 over \pythia that we see in Fig.~\ref{fig:dY_dist_all}. This implies that there is some suppression in GS09 from momentum/number sum rule suppression effects migrating to lower $x$ values during evolution, which partially compensates the $1 \to 2$ splitting enhancement.

To finish this section, we would like to comment that despite the similarities between the response functions computed with \pythia and GS09 dPDFs, as shown in Figs.~\ref{fig:Pythia_sum_rules_uval_dval} and \ref{fig:Pythia_momentum_rule}, the $\Delta \rm Y$ distributions generated by these models of dPDFs differ from one another, especially at small and large values of rapidity separation. This implies some important differences between the dPDF sets themselves.
We also note that the $\Delta \rm Y$ distributions generated with naive and \pythia  dPDFs also show some important differences not only at large values of rapidity separation (something one would expect, since large values of $\Delta \rm Y$ allow to probe dPDFs with one large $x$ value, where momentum sum rule effects play a significant role) but also at small values of $\Delta \rm Y$ at the 3 and 7 TeV collision energies.

\section{Sum rules for triple PDFs}
\label{s:results_tPDFs}
In the previous sections we studied how well the asymmetric and symmetric dPDFs constructed with \pythia code obey the GS sum rules.
Now we would like to extend the analysis performed in Sections~\ref{ss:results_dPDFs_asym}--\ref{ss:results_dPDFs_sym} to the case of triple PDFs. 
However, since the sum rules for tPDFs do not exist in the literature, we first need to establish what these sum rules actually are; in Section~\ref{ss:bare_tPDFs_sum_rules} and Section~\ref{ss:renorm_tPDFs_sum_rules}  we derive the GS sum rules for the bare and renormalized tPDFs correspondingly. Then, in Section~\ref{ss:results_tPDFs_check}, we investigate how well the \pythia tPDFs satisfy these sum rules.

\subsection{Sum rule for bare tPDFs}
\label{ss:bare_tPDFs_sum_rules}

We start by writing down the GS sum rules generalized to the case of bare tPDFs.
The momentum sum rule  is given by
\begin{eqnarray}
	\sum\limits_{j_3}\int\limits^{1 - x_1 - x_2}_0 dx_3 ~x_3 
	~T^B_{j_1 j_2 j_3}(x_1, x_2, x_3) = (1 - x_1 - x_2) D^B_{j_1 j_2}(x_1, x_2),
	\label{eq:triple_sum_rule_momentum}
\end{eqnarray}
where $T^B_{j_1 j_2 j_3}$  and $D^B_{j_1 j_2}(x_1, x_2)$ are the bare triple and double parton distribution functions respectively. 
To prove this relation we  apply the same method as was used to 
show that GS sum rules hold for bare dPDFs in Ref.~\cite{Gaunt:2012tfk}.
First of all let us write down the light-cone representation~\cite{Harindranath:1998pd, Zhang:1993is, Zhang:1993dd, Harindranath:1993de, Harindranath:1998pc} of dPDFs and tPDFs:
\begin{align}
	D^B_{j_1 j_2}(x_1, x_2) &= \sum\limits_{N, \{\beta_i\} }\int [dz]_N ~[d^2{\bm{k}}]_N 
	                          ~ \left| \Phi_N\left( \{\beta_i, z_i,  {\bm{k}}_i\} \right) \right|^2  						  	 
						 	   ~ \sum\limits^N_k ~\delta(x_1 - z_k) ~\delta_{j_1 p_k} \nonumber\\
						&\phantomrel{=}\times
						  \sum\limits^N_{l \neq k}  ~\delta(x_2 - z_l) ~\delta_{j_2 p_l},
	\label{eq:dPDF_LC}\\
	T^B_{j_1 j_2 j_3}(x_1, x_2, x_3) &= \sum\limits_{N, \{\beta_i\} }\int [dz]_N ~[d^2{\bm{k}}]_N 
	                              ~ \left| \Phi_N\left( \{\beta_i, z_i,  {\bm{k}}_i\} \right) \right|^2
						  		   ~ \sum\limits^N_k ~\delta(x_1 - z_k) ~\delta_{j_1 p_k}	  \nonumber\\
								&\phantomrel{=}\times	 						  
						  			\sum\limits^N_{l \neq k}  ~\delta(x_2 - z_l) ~\delta_{j_2 p_l}
						  			~\sum\limits^N_{m \neq k,l}  ~\delta(x_3 - z_m) ~\delta_{j_3 p_m},
	\label{eq:tPDF_LC}
\end{align}
where
\begin{eqnarray}
	[dz]_N &\equiv& \prod\limits^{N}_{i = 1} ~dz_i ~\delta\left(1 - \sum\limits^N_i ~z_i\right), \\
	\left[d^2{\bm{k}}\right]_N &\equiv& \prod\limits^{N}_{i = 1} ~d^2{\bm{k}}_i  ~\delta^2\left( \sum\limits^N_i ~{\bm{k}}_i  \right),
\end{eqnarray}
and $\Phi_N\left( \{\beta_i, z_i,  {\bm{k}}_i\} \right)$  is the bare light-cone wave function to find the hadron being composed from $N$ partons with longitudinal momentum fractions $\{z_i\}$, transverse momenta $\{{\bm{k}}_i\}$ and other quantum numbers $\{\beta_i\}$ including flavour indices $\{p_i\}$ ($i = 1...N$). 
The sum $\sum_{N, \{\beta_i\} }$ in Eq.~\eqref{eq:dPDF_LC} and Eq.~\eqref{eq:tPDF_LC} runs over all distinguishable Fock states which reproduce appropriate hadronic quantum numbers.

By substituting Eq.~\eqref{eq:tPDF_LC} in Eq.~\eqref{eq:triple_sum_rule_momentum} we get
\begin{align}
	&\sum\limits_{j_3}\int\limits^{1 - x_1 - x_2}_0 dx_3 ~x_3 ~T^B_{j_1 j_2 j_3}(x_1, x_2, x_3) =
			\sum\limits_{N, \{\beta_i\} }\int [dz]_N ~[d^2{\bm{k}}]_N 
	       ~ \left| \Phi_N\left( \{\beta_i, z_i,  {\bm{k}}_i\} \right) \right|^2 \nonumber\\
	       &\phantomrel{=}\times
	       \sum\limits_{j_3}\int\limits^{1 - x_1 - x_2}_0 dx_3 ~ x_3
		   ~\sum\limits^N_k ~\delta(x_1 - z_k) ~\delta_{j_1 p_k}
		   ~\sum\limits^N_{l \neq k}  ~\delta(x_2 - z_l) ~\delta_{j_2 p_l}
		   ~\sum\limits^N_{m \neq k,l}  ~\delta(x_3 - z_m) ~\delta_{j_3 p_m} \nonumber\\
	&=\sum\limits_{N, \{\beta_i\} }\int [dz]_N ~[d^2{\bm{k}}]_N 
	       ~ \left| \Phi_N\left( \{\beta_i, z_i,  {\bm{k}}_i\} \right) \right|^2 
		   ~\sum\limits^N_k ~\delta(x_1 - z_k) ~\delta_{j_1 p_k} 
   		   ~\sum\limits^N_{l \neq k}  ~\delta(x_2 - z_l) ~\delta_{j_2 p_l} \nonumber\\
  &\phantomrel{=}\times \sum\limits_{j_3}\sum\limits^N_{m \neq k,l}  z_m ~\delta_{j_3 p_m}.
  \label{eq:triple_proof_1}
\end{align}
The last sum in Eq.~\eqref{eq:triple_proof_1} can be written as 
\begin{eqnarray}
	\sum\limits_{j_3}\sum\limits^N_{m \neq k,l}  z_m ~\delta_{j_3 p_m} = \sum\limits^N_{m}  z_m  - z_k - z_l = 1 - x_1 - x_2,
\end{eqnarray}
where we used the fact that $\sum^N_m  z_m = 1$ and that $z_k = x_1$ and $z_l = x_2$.  
Therefore, Eq.~\eqref{eq:triple_proof_1} turns into
\begin{align}
	 \sum\limits_{j_3}\int\limits^{1 - x_1 - x_2}_0 & dx_3 ~x_3 ~T^B_{j_1 j_2 j_3}(x_1, x_2, x_3) \nonumber
	\\ = & \left(1 - x_1 - x_2\right) 
		   \sum\limits_{N, \{\beta_i\} }\int [dz]_N ~[d^2{\bm{k}}]_N 
	       ~ \left| \Phi_N\left( \{\beta_i, z_i,  {\bm{k}}_i\} \right) \right|^2 \nonumber \\ & \times \sum\limits^N_k ~\delta(x_1 - z_k) ~\delta_{j_1 p_k} 
   		   ~\sum\limits^N_{l \neq k}  ~\delta(x_2 - z_l) ~\delta_{j_2 p_l} \nonumber\\
   	= & \left(1 - x_1 - x_2\right) D^B_{j_1 j_2}(x_1, x_2).
  \label{eq:triple_proof_2}
\end{align}

Now let us concentrate on the generalization of the number rule Eq.~\eqref{eq:gs_number_rule_inv} for the case of bare tPDFs. 
First of all let us note that 
\begin{align}
	\int\limits^{1 - x_1 - x_2}_0 dx_3 
	~T^B_{j_1 j_2 j_3}(x_1, x_2, x_3) &= \sum\limits_{N, \{\beta_i\} }\int [dz]_N ~[d^2{\bm{k}}]_N 
	       ~ \left| \Phi_N\left( \{\beta_i, z_i,  {\bm{k}}_i\} \right) \right|^2 
		   \\ \nonumber
   		   &\phantomrel{=}
   		   \times\sum\limits^N_k ~\delta(x_1 - z_k) ~\delta_{j_1 p_k} ~ \sum\limits^N_{l \neq k}  ~\delta(x_2 - z_l) ~\delta_{j_2 p_l} 
           ~\sum\limits^N_{m \neq k,l}  ~\delta_{j_3 p_m},
\end{align}
and, therefore,
\begin{align}
	&\int\limits^{1 - x_1 - x_2}_0 dx_3 ~T^B_{j_1 j_2 j_{3v}}(x_1, x_2, x_3) \equiv
	\int\limits^{1 - x_1 - x_2}_0 dx_3 
	~\left[T^B_{j_1 j_2 j_3}(x_1, x_2, x_3) - T^B_{j_1 j_2 \bar{j}_3}(x_1, x_2, x_3)\right]  \nonumber\\
	&=\sum\limits_{N, \{\beta_i\} }\int [dz]_N ~[d^2{\bm{k}}]_N 
	~\left| \Phi_N\left( \{\beta_i, z_i,  \bm{k}_i\} \right) \right|^2
	\sum\limits^N_k ~\delta(x_1 - z_k) ~\delta_{j_1 p_k} 
   	~\sum\limits^N_{l \neq k}  ~\delta(x_2 - z_l) ~\delta_{j_2 p_l} \nonumber\\
    &\phantomrel{=}\times
    \left( \sum\limits^N_{m \neq k,l}  ~\delta_{j_3 p_m}  - \sum\limits^N_{m \neq k,l}  ~\delta_{\bar{j}_3 p_m} \right).
    \label{eq:triple_proof_3}
\end{align}
The two sums in the parentheses in Eq.~\eqref{eq:triple_proof_3} can be written as 
\begin{eqnarray}
	\sum\limits^N_{m \neq k,l}~\delta_{j_3 p_m} &=& \sum\limits^N_{m}~\delta_{j_3 p_m} - \delta_{j_3 j_1} - \delta_{j_3 j_2}
	= N_{j_3 | \{ \beta_i\}} - \delta_{j_3 j_1} - \delta_{j_3 j_2},\nonumber\\
	\sum\limits^N_{m \neq k,l}~\delta_{\bar{j}_3 p_m} &=& \sum\limits^N_{m}~\delta_{\bar{j}_3 p_m} - \delta_{\bar{j}_3 j_1} - \delta_{\bar{j}_3 j_2}
	= N_{\bar{j}_3 | \{ \beta_i\}} - \delta_{\bar{j}_3 j_1} - \delta_{\bar{j}_3 j_2},
\label{eq:triple_proof_4}
\end{eqnarray}
where $N_{j_3 | \{ \beta_i\}}$ ($N_{\bar{j}_3 | \{ \beta_i\}}$) is the number of $j_3$ ($\bar{j}_3$) partons in state $\{ \beta_i\}$. 
The difference between $N_{j_3 | \{ \beta_i\}}$ and $N_{\bar{j}_3 | \{ \beta_i\}}$ is the number of valence quarks, which is independent of $\{ \beta_i\}$:
\begin{eqnarray}
    N_{j_{3v}} = N_{j_3 | \{ \beta_i\}} - N_{\bar{j}_3 | \{ \beta_i\}}.
    \label{eq:valencedef}
\end{eqnarray}

Inserting Eqs.~\eqref{eq:triple_proof_4} and \eqref{eq:valencedef} into Eq.~\eqref{eq:triple_proof_3}, we obtain
\begin{align}
	\int\limits^{1 - x_1 - x_2}_0 dx_3 
	~T^B_{j_1 j_2 j_{3v}}(x_1, x_2, x_3)  
	&= \left(N_{j_{3v}} - \delta_{j_3 j_1} - \delta_{j_3 j_2}  + \delta_{\bar{j}_3 j_1} + \delta_{\bar{j}_3 j_2}\right) \nonumber\\ 
	&\phantomrel{=}\times D^B_{j_1 j_2}(x_1, x_2),
    \label{eq:tiple_sum_rule_number}
\end{align}
which is  Eq.~\eqref{eq:gs_number_rule_inv} generalized for the case of bare tPDFs.

\subsection{Sum rules for renormalized tPDFs}
\label{ss:renorm_tPDFs_sum_rules}
We have shown  that Eq.~\eqref{eq:triple_sum_rule_momentum}, \eqref{eq:tiple_sum_rule_number}, hold for the bare distributions. 
Now let us sketch the all order proof for  the renormalized tPDFs following the method similar to the one used in Ref.~\cite{Diehl:2018kgr}. Using notation of Refs.~\cite{Diehl:2011yj, Diehl:2018kgr}, the bare sPDF, dPDF and tPDF are defined as 
\begin{align}
	f^B_{j_1}(x_1) &= (x_1 p^+)^{-n_1}
	\int \, \frac{dz^{-}_1}{2\pi} \, e^{\di x_1 z^{-}_1 p^+} \,
	\langle p |
		\mathcal{O}_{j_1}(0, z_1)
	| p \rangle	
	\biggr\rvert_{z^+_1 = 0, \, \bm{z}_1 = \bm{0}}, 
	\label{eq:bare_sPDF_def_1}	\\
	D^B_{j_1 j_2}(x_1, x_2)  
	&=(x_1 p^+)^{-n_1} (x_2 p^+)^{-n_2} 2p^+
	\int \, \frac{dz^{-}_1}{2\pi} \, \frac{dz^{-}_2}{2\pi} \, dy^- \, d^{2-2\varepsilon}\bm{y}
	\, 
	e^{\di (\sum_{i=1}^2x_i z^-_i)p^+}	 
	\nonumber\\
	&\phantomrel{=}\times	
	\langle p |
		\mathcal{O}_{j_1}(y, z_1)
		\mathcal{O}_{j_2}(0, z_2)
	| p \rangle	
	\biggr\rvert_{z^+_i = y^+ = 0, \, \bm{z}_i = \bm{0}}, 
	\label{eq:bare_sPDF_def_2}\\
	T^B_{j_1 j_2 j_3}(x_1, x_2, x_3) 
	&= (x_1 p^+)^{-n_1} (x_2 p^+)^{-n_2} (x_3 p^+)^{-n_3} (2p^+)^2 \nonumber\\
	&\phantomrel{=}\times\int \, \frac{dz^{-}_1}{2\pi} \, \frac{dz^{-}_2}{2\pi} \, \frac{dz^{-}_3}{2\pi}\, dy_1^- dy_2^- \, d^{2-2\varepsilon}\bm{y}_1 \, d^{2-2\varepsilon}\bm{y}_2
	\,
	e^{\di (\sum_{i=1}^3x_i z^-_i)p^+} \nonumber\\
	&\phantomrel{=}\times	
	\langle p |
		\mathcal{O}_{j_1}(y_1, z_1)
		\mathcal{O}_{j_2}(y_2, z_2)
		\mathcal{O}_{j_3}(0, z_3)
	| p \rangle	
	\biggr\rvert_{z^+_i = y_i^+ = 0, \, \bm{z}_i = \bm{0}},	
	\label{eq:bare_sPDF_def_3}
\end{align}
where we use the light-cone representation for four-vectors:  $v^\pm = (v^0 \pm v^3) / \sqrt{2}$, $\bm{v} = (v^1, v^2)$. The quantity $n_j$ is $0$ if $j$ is a quark or antiquark, and $1$ if $j$ is a gluon.
The twist-two operators  $\mathcal{O}$ in Eqs.~\eqref{eq:bare_sPDF_def_1}-\eqref{eq:bare_sPDF_def_3} are bilinear in partonic fields and are defined as 
\begin{eqnarray}
	\mathcal{O}_q (y,z) &=& 
	\frac{1}{2} \, 
	\bar{q}\left(y - \frac{z}{2}\right) \, \gamma^+ \,
	q\left(y + \frac{z}{2}\right),
	\label{eq:op_def_1} \\
	\mathcal{O}_{\bar{q}} (y,z) &=& 
	-\frac{1}{2} \, 
	\bar{q}\left(y + \frac{z}{2}\right) \, \gamma^+ \,
	q\left(y - \frac{z}{2}\right), 
	\label{eq:op_def_2} \\
	\mathcal{O}_{g} (y,z) &=&  
	G^{+i}\left(y - \frac{z}{2}\right) \,
	G^{+i}\left(y + \frac{z}{2}\right), 
	\label{eq:op_def_3}
\end{eqnarray}
where $q,$ and $\bar{q}$ are the quark and antiquark fields, $G$ is the gluon field strength, and $i$ is a transverse component\footnote{Note that these definitions are strictly speaking only appropriate when we are using light-cone gauge $A^+=0$. In other gauges Wilson line factors appear in the definitions of the operators $\mathcal{O}_j$.}.

The quantity $\bm{y}$ in Eq.~\eqref{eq:bare_sPDF_def_2} corresponds to the transverse separation between the two partons probed. Similarly, in Eq.~\eqref{eq:bare_sPDF_def_3}, $\bm{y}_1$ and $\bm{y}_2$ correspond to the transverse displacements of partons $1$ and $2$ from parton $3$ (with the transverse displacement between partons $1$ and $2$ then being given by $\bm{y}_1 - \bm{y}_2$).
As was noted in Ref.~\cite{Diehl:2011yj}, there is a UV singularity
in Eq.~\eqref{eq:bare_sPDF_def_2} when $\bm{y} = 0$ and in Eq.~\eqref{eq:bare_sPDF_def_3} when $\bm{y}_i = 0$ or $\bm{y}_1 - \bm{y}_2 = 0$. 
In Section~\ref{s:gs09_pythia} a cut-off on $\bm{y}$ was used to regulate the integral for the dPDF, but as was also mentioned there, the dPDF sum rules only hold exactly when we use dimensional regularization and $\overline{\rm MS}$  renormalization -- thus, here we will follow the latter approach.
For the bare dPDF and tPDF in Eqs.~\eqref{eq:bare_sPDF_def_2}  and \eqref{eq:bare_sPDF_def_3}  we should change the integrations with respect to $d^2\bm{y}_{(i)}$ to  $d^{2 - 2\varepsilon}\bm{y}_{(i)}$ (in fact in the previous section, integrations over transverse momenta $\bm{k}_{i}$ should be similarly altered to regularize the UV divergences, although this does not affect the overall argument presented there).

Generalizing the renormalization equation for the dPDFs, we write down the following equation for the renormalization of the tPDFs:
\begin{align}
	T_{j_1 j_2 j_3}(x_1, x_2, x_3, Q) 
	&= 
	\sum\limits_{j^\prime_1 j^\prime_2 j^\prime_3} 
	Z_{j_1  j^\prime_1}(Q) \otimes_1 
	Z_{j_2  j^\prime_2}(Q) \otimes_2 
	Z_{j_3  j^\prime_3}(Q) \otimes_3  
	T^B_{j^\prime_1 j^\prime_2 j^\prime_3}(z_1, z_2, z_3)\nonumber\\
	&\phantomrel{=}+
	\sum\limits_{j^\prime_1 j^\prime_2} 
	Z_{j_1  j^\prime_1}(Q) \otimes_1 
	Z_{j_2 j_3, j^\prime_2}(Q) \otimes_{23} 
	D^B_{j^\prime_1 j^\prime_2}(z_1, z_2) + 
	\mathrm{(permutations)} \nonumber\\
	&\phantomrel{=}+
	\sum\limits_{j^\prime_1} 
	Z_{j_1 j_2 j_3, j^\prime_1}(Q) \otimes_{123} 
	f^B_{j^\prime_1}(z_1),
	\label{eq:tPDF_RGE_direct}
\end{align}
where we use the same notation as in Ref.~\cite{Diehl:2018kgr}. The 
definitions of the convolution symbols in this expression are given by:
\begin{eqnarray}
	A \otimes_1 B &=& 
	\int\frac{dz}{z} A\left(\frac{x_1}{z}\right) B(z),\\
	A \otimes_{12} B &=& 
	\int\frac{dz}{z^2} A\left(\frac{x_1}{z}, \frac{x_2}{z} \right) B(z),\\
	A \otimes_{123} B &=& 
	\int\frac{dz}{z^3} A\left(\frac{x_1}{z}, \frac{x_2}{z}, \frac{x_3}{z} \right) B(z).
\end{eqnarray}
The terms on the first two lines of Eq.~\eqref{eq:tPDF_RGE_direct} are analogous to the two terms in the renormalization of the dPDFs. The quantity $Z_{ij}$ is the renormalization factor that appears in the renormalization of the sPDFs, and $Z_{ij,k}$ is a renormalization factor linked with the $1 \to 2$ splitting singularity. The term ``(permutations)'' in Eq.~\ref{eq:tPDF_RGE_direct} refers to two additional terms which have the same structure as the first term on the second line of that equation, but which have $\{j_1,j_2,j_3\}$ replaced by $\{j_2,j_1,j_3\}$ or $\{j_3,j_1,j_2\}$. The final term in Eq.~\eqref{eq:tPDF_RGE_direct} is new and is associated with a $1 \to 3$ splitting process.

Let us use the particular implementation of the $\overline{\rm MS}$ scheme discussed in Eqs.~(79) and (80) of Ref.~\cite{Diehl:2018kgr}. 
Then, the structures of the renormalization factors are as follows:
\begin{align}
    Z_{ij} &= \delta_{ij} \delta(1-x) + \sum_{n=1}^{\infty}\alpha_s^n \sum_{m \ge 1} \dfrac{Z_{nm;ij}}{\varepsilon^m},
    \label{eq:zij_div}\\
    Z_{jk,i} &= \sum_{n=1}^{\infty}\alpha_s^n \sum_{m \ge 1} \dfrac{Z_{nm;jk,i}}{\varepsilon^m},
    \label{eq:zijk_div}\\
    Z_{jkl, i} &= \sum_{n=2}^{\infty}\alpha_s^n \sum_{m \ge 1} \dfrac{Z_{nm;jk,i}}{\varepsilon^m}.
    \label{eq:zjkli_div}
\end{align}

Note that $Z_{jk,i}$ begins at order $\alpha_s$, whilst $Z_{jk,i}$ begins at order $\alpha_s^2$ (since they are related to the $1\to 2$ and $1\to3$ processes respectively). This then means that, in the $\overline{\rm MS}$ scheme, they contain no finite pieces, only pole terms.

\paragraph{Notation.} In the below, we shall make use of the notation introduced in Section 6.1 of \cite{Diehl:2018kgr}; very briefly, we use the following shorthand for integrals:
\begin{equation}
    \int_m T = \int dx_m T(x_1,...,x_m,...)
\end{equation}
We also introduce operators $X_m^n$ that act on a function by multiplying by $x_m^n$:
\begin{equation}
    X_m^n T = x_m^n T(x_1,...,x_m,...)
\end{equation}

\paragraph{Momentum sum rule.} To prove the momentum sum rule for the renormalized tPDFs, let us consider:
\begin{align} \label{eq:DelMtm}
    \Delta^{j_1j_2}_{\text{mtm}} \equiv \sum_{j_3} \int_3 X_3 T_{j_1 j_2 j_3} - (1-X_1-X_2)D_{j_1j_2},
\end{align}
For the momentum sum rule to hold, $\Delta^{j_1j_2}_{\text{mtm}}$ must be zero. The first term on the right hand side of Eq.~\eqref{eq:DelMtm}, after inserting the expression for $T_{j_1 j_2 j_3}$ from Eq.~\eqref{eq:tPDF_RGE_direct}, becomes:
\begin{align} \label{eq:DelMtmFirst}
	& \sum\limits_{j^\prime_1 j^\prime_2 j^\prime_3} 
	Z_{j_1  j^\prime_1} \otimes_1 
	Z_{j_2  j^\prime_2} \otimes_2 
	\sum_{j_3}\int_3 X_3 Z_{j_3  j^\prime_3} \otimes_3  
	T^B_{j^\prime_1 j^\prime_2 j^\prime_3}\nonumber\\
	+&
	\sum\limits_{j^\prime_1 j^\prime_2} 
	Z_{j_1 j_2, j^\prime_1} \otimes_{12} 
		\sum_{j_3}\int_3 X_3 Z_{j_3  j^\prime_2} \otimes_3 
	D^B_{j^\prime_1 j^\prime_2} + 
	\mathrm{(permutations)}  \nonumber\\
	+&
	\sum\limits_{j^\prime_1} 
	\sum_{j_3}\int_3 X_3 Z_{j_1 j_2 j_3, j^\prime_1} \otimes_{123} 
	f^B_{j^\prime_1}.
\end{align}
We now simplify each of the terms in this expression. For the first term in Eq.~\eqref{eq:DelMtmFirst}, we have:
\begin{align} \nonumber
	&\sum\limits_{j^\prime_1 j^\prime_2 j^\prime_3} 
	Z_{j_1  j^\prime_1} \otimes_1 
	Z_{j_2  j^\prime_2} \otimes_2 
	\sum_{j_3}\int_3 X_3 Z_{j_3  j^\prime_3} \otimes_3  
	T^B_{j^\prime_1 j^\prime_2 j^\prime_3} \\ \nonumber
	=& \sum\limits_{j^\prime_1 j^\prime_2 j^\prime_3} 
	Z_{j_1  j^\prime_1} \otimes_1 
	Z_{j_2  j^\prime_2} \otimes_2 
	\left( \sum_{j_3}\int X Z_{j_3  j^\prime_3} \right)  
	\int_3 X_3 T^B_{j^\prime_1 j^\prime_2 j^\prime_3} \\
	=& \sum\limits_{j^\prime_1 j^\prime_2} 
	Z_{j_1  j^\prime_1} \otimes_1 
	Z_{j_2  j^\prime_2} \otimes_2 
	 (\colorbox{red!50}{{$\displaystyle 1$}}-\colorbox{yellow!50}{{$\displaystyle X_1$}}-\colorbox{green!50}{{$\displaystyle X_2$}}) D^B_{j^\prime_1 j^\prime_2 }.
\end{align}
Going from the first line to the second line, we have used Eq.~(64) from Ref.~\cite{Diehl:2018kgr},
\begin{equation}
X^n (A \otimes B) = (X^n A) \otimes (X^n B) \quad \quad \int X^n (A \otimes B) = \int X^n A \int X^n B 
\end{equation}
where $A$ and $B$ are arbitrary single argument functions (specifically we have used the latter relation). Then, going from the second to the third line, we have used the momentum sum rule for the bare tPDFs, plus the momentum sum rule for the $Z_{ij}$: $\sum_{j_3}\int X Z_{j_3  j^\prime_3}=1$ (see Eq.~(97) in Ref.~\cite{Diehl:2018kgr}).

For the second term in Eq.~\eqref{eq:DelMtmFirst}, we have:
\begin{align} \nonumber
    & \sum\limits_{j^\prime_1 j^\prime_2} 
	Z_{j_1 j_2, j^\prime_1} \otimes_{12} 
		\sum_{j_3}\int_3 X_3 Z_{j_3  j^\prime_2} \otimes_3 
	D^B_{j^\prime_1 j^\prime_2} \\ \nonumber
	=& \sum\limits_{j^\prime_1 j^\prime_2} 
	Z_{j_1 j_2, j^\prime_1} \otimes_{12} 
		\left(\sum_{j_3}\int X Z_{j_3  j^\prime_2} \right) \int_2 X_2 
	D^B_{j^\prime_1 j^\prime_2} \\
	=& \sum\limits_{j_1^\prime} 
	Z_{j_1 j_2, j_1^\prime} \otimes_{12} 
		 (1-X)f^B_{j_1^\prime}.
\end{align}
Once again we use Eq.~(64) from Ref.~\cite{Diehl:2018kgr} to go from the first to the second line, and then we use the momentum sum rule for the $Z_{ij}$ along with the momentum sum rule for the bare dPDFs (as proven in Ref.~\cite{Diehl:2018kgr}).

In one of the terms denoted by ``(permutations)'', we have:
\begin{align} \nonumber
    & \sum\limits_{j^\prime_1 j^\prime_2} 
	Z_{j_1,j^\prime_1} \otimes_1 \sum_{j_3}\int_3 X_3 Z_{j_2 j_3, j^\prime_2} \otimes_{23} 
	D^B_{j^\prime_1 j^\prime_2} \\ \nonumber
	=& \sum\limits_{j^\prime_1 j^\prime_2} 
	Z_{j_1,j^\prime_1} \otimes_1 \sum_{j_3}\left(\int_3 X_3 Z_{j_2 j_3, j^\prime_2} \right) \otimes_{2} 
	\left( X_2 D^B_{j^\prime_1 j^\prime_2} \right) \\
	=& \sum\limits_{j^\prime_1 j^\prime_2} 
	Z_{j_1,j^\prime_1} \otimes_1 \left[ (\colorbox{green!50}{{$\displaystyle 1$}}-\colorbox{blue!50}{{$\displaystyle X_2$}})Z_{j_2j_2^\prime} \right] \otimes_{2} 
	\left( X_2 D^B_{j^\prime_1 j^\prime_2} \right).
\end{align}
Here we use Eq.~(66) from Ref.~\cite{Diehl:2018kgr} to go from the first to the second line, 
\begin{equation}
\int_2 X_2^n (D \otimes_{12} A) = \left( \int_2 X_2^n D \right) \otimes_{1} (X^nA)
\end{equation}
with $D$ an arbitrary function of two arguments and $A$ a function of one. Then to get to the third line, we use the momentum sum rule for the $Z_{ij,k}$ factors as given in Eq.~(103) of Ref.~\cite{Diehl:2018kgr}. The other term in the ``(permutations)'' has the same structure but with $1 \leftrightarrow 2$.

Let us now consider the second term in Eq.~\eqref{eq:DelMtm}. Expressing $D_{j_1j_2}$ in terms of bare quantities, using Eq.~(70) from Ref.~\cite{Diehl:2018kgr}, we have:
\begin{align}
    -& (1-X_1-X_2)D_{j_1j_2} \nonumber\\
    =&  (\colorbox{yellow!50}{{$\displaystyle X_1$}}+\colorbox{blue!50}{{$\displaystyle X_2$}}-\colorbox{red!50}{1}) \sum_{j^\prime_1j^\prime_2} Z_{j_1  j^\prime_1} \otimes_1 
	Z_{j_2  j^\prime_2} \otimes_2  
	D^B_{j^\prime_1 j^\prime_2} + 
	(X_1+X_2-1) \sum\limits_{j^\prime} 
	Z_{j_1 j_2, j^\prime} \otimes_{12} 
	f^B_{j^\prime}. 
\end{align}

Now many of the terms on the right hand side of the expression for $\Delta_{\text{mtm}}$ cancel. The terms in the \colorbox{red!50}{{$\phantom{1}$}}, \colorbox{green!50}{{$\phantom{1}$}}, \colorbox{blue!50}{{$\phantom{1}$}} boxes cancel against each other directly, whilst the terms in the \colorbox{yellow!50}{{$\phantom{1}$}} boxes cancel against the other ``(permutations)'' term that we have not written out explicitly. All the terms containing $D^B$ on the right hand side then cancel out, and we are left with:
\begin{align} \label{eq:DelMtmFinal} \nonumber
\Delta^{j_1j_2}_{\text{mtm}} =& \sum\limits_{j^\prime} 
	\sum_{j_3}\int_3 X_3 Z_{j_1 j_2 j_3, j^\prime} \otimes_{123} 
	f^B_{j^\prime}
	+ \sum\limits_{j^\prime} 
	Z_{j_1 j_2, j^\prime} \otimes_{12} 
		 (1-X)f^B_{j^\prime} \\ \nonumber
	&+ (X_1+X_2-1) \sum\limits_{j^\prime} 
	Z_{j_1 j_2, j^\prime} \otimes_{12} 
	f^B_{j^\prime} \\
	=& \sum_{j^\prime k^\prime} \left( \sum_{j_3} \int_3 X_3 Z_{j_1 j_2 j_3, j^\prime} - (1-X_1-X_2) Z_{j_1j_2,j^\prime} \right) \otimes_{12} \left( XZ^{-1}_{j^\prime k^\prime}\right) \otimes \left( Xf_{k^\prime}\right).
\end{align}
To reach the last line we have used:
\begin{align} \label{eq:123simplify}
    \int_3 X_3^n T \otimes_{123} 
	A = \left( \int_3 X_3^n T \right) \otimes_{12} X^n A
\end{align}
and
\begin{align}
    (X_1+X_2) \left( D \otimes_{12} A \right) = \left( [X_1+X_2] D \right) \otimes_{12}  \left( X A\right).
\end{align}

The quantity $\Delta^{j_1j_2}_{\text{mtm}}$ must be finite, as it is expressed in terms of renormalized quantities. However, in Eq.~\eqref{eq:DelMtmFinal} we see that it can be expressed in terms of a pure pole factor convolved with a finite quantity ($f_{k'}$). The only way this can be true is if both left and right hand side are zero - thus $\Delta^{j_1j_2}_{\text{mtm}} = 0$ and the momentum sum rule for tPDFs holds at the renormalized level.

Given that $\Delta^{j_1j_2}_{\text{mtm}}=0$, we obtain from Eq.~\eqref{eq:DelMtmFinal} a sum rule for the $Z_{ijk,l}$ factors:
\begin{align}
    \sum_{j_3} \int_3 X_3 Z_{j_1 j_2 j_3, j^\prime} = (1-X_1-X_2) Z_{j_1j_2,j^\prime}.
\end{align}

\paragraph{Number sum rule.} The proof of the number sum rule for renormalized tPDFs proceeds in an analogous fashion. Consider:
\begin{align} \label{eq:NumSum}
    \Delta^{j_1j_2}_{\text{num}} \equiv \int_3 T_{j_1 j_2 j_{3v}} - (N_{j_3v}+\delta_{j_3\bar{j}_1}+\delta_{j_3\bar{j}_2}-\delta_{j_3j_1}+\delta_{j_3j_2}) D_{j_1j_2}.
\end{align}
If this is zero, the number sum rule holds. We consider \mbox{$\int_3 T_{j_1 j_2 j_{3v}}$}, and express it in terms of bare quantities using Eq.~\eqref{eq:tPDF_RGE_direct}:
\begin{align} \label{eq:NumSumFirst}
	& \sum\limits_{j^\prime_1 j^\prime_2 j^\prime_3} 
	Z_{j_1  j^\prime_1} \otimes_1 
	Z_{j_2  j^\prime_2} \otimes_2 
	\int_3 Z_{j_{3v}  j^\prime_3} \otimes_3  
	T^B_{j^\prime_1 j^\prime_2 j^\prime_3}\nonumber\\
	+&
	\sum\limits_{j^\prime_1 j^\prime_2} 
	Z_{j_1 j_2, j^\prime_1} \otimes_{12} 
		\int_3 Z_{j_{3v}  j^\prime_2} \otimes_3 
	D^B_{j^\prime_1 j^\prime_2} + 
	\mathrm{(permutations)}  \nonumber\\
	+&
	\sum\limits_{j^\prime_1} 
	\int_3 Z_{j_1 j_2 j_{3v}, j^\prime_1} \otimes_{123} 
	f^B_{j^\prime_1}.
\end{align}
Using Eq.~(64) from Ref.~\cite{Diehl:2018kgr}, the number sum rule for the $Z_{ij}$ (see Eq.~(90) from Ref.~\cite{Diehl:2018kgr}), and the number sum rule for the bare tPDFs, we may express the first term of Eq.~\eqref{eq:NumSumFirst} as: 
\begin{align} \nonumber
	& \sum\limits_{j^\prime_1 j^\prime_2 j^\prime_3} 
	Z_{j_1  j^\prime_1} \otimes_1 
	Z_{j_2  j^\prime_2} \otimes_2 
	\int_3 Z_{j_{3v}  j^\prime_3} \otimes_3  
	T^B_{j^\prime_1 j^\prime_2 j^\prime_3} \\ \nonumber
	=& \sum\limits_{j^\prime_1 j^\prime_2 j^\prime_3} 
	Z_{j_1  j^\prime_1} \otimes_1 
	Z_{j_2  j^\prime_2} \otimes_2 
	\left( \delta_{j_3j^\prime_3} - \delta_{j_3\bar{j}^\prime_3} \right)  
	\int_3 T^B_{j^\prime_1 j^\prime_2 j^\prime_3} \\
	=& \sum\limits_{j^\prime_1 j^\prime_2} 
	Z_{j_1  j^\prime_1} \otimes_1 
	Z_{j_2  j^\prime_2} \otimes_2 
	 (\colorbox{red!50}{{$\displaystyle N_{j_3v}$}}+\colorbox{yellow!50}{{$\displaystyle \delta_{j_3\bar{j}'_1}$}}+\colorbox{green!50}{{$\displaystyle \delta_{j_3\bar{j}'_2}$}}-\colorbox{yellow!50}{{$\displaystyle \delta_{j_3j'_1}$}}-\colorbox{orange!50}{{$\displaystyle \delta_{j_3j_2'}$}}) D^B_{j^\prime_1 j^\prime_2 }.
\end{align}
For the second term, we use Eq.~(64) from Ref.~\cite{Diehl:2018kgr}, the number sum rule for the $Z_{ij}$, and the number sum rule for the bare dPDFs, to obtain:
\begin{align} \nonumber
    & \sum\limits_{j^\prime_1 j^\prime_2} 
	Z_{j_1 j_2, j^\prime_1} \otimes_{12} 
		\int_3 Z_{j_{3v}  j^\prime_2} \otimes_3 
	D^B_{j^\prime_1 j^\prime_2} \\ \nonumber
	=& \sum\limits_{j^\prime_1 j^\prime_2} 
	Z_{j_1 j_2, j^\prime_1} \otimes_{12} 
		\left(\delta_{j_3j_2^\prime} - \delta_{j_3\bar{j}_2^\prime}\right) \int_2
	D^B_{j^\prime_1 j^\prime_2} \\
	=& \sum\limits_{j_1^\prime} 
	Z_{j_1 j_2, j_1^\prime} \otimes_{12} 
		 (N_{j_3v}+\delta_{j_3\bar{j}'_1}-\delta_{j_3j'_1})f^B_{j_1^\prime}. 
\end{align}
In one of the terms in the ``(permutations)'', we can use Eq.~(66) from Ref.~\cite{Diehl:2018kgr} and the number sum rule for the $Z_{ij,k}$ (as given in Eq.~(94) of Ref.~\cite{Diehl:2018kgr}):
\begin{align} \nonumber
    & \sum\limits_{j^\prime_1 j^\prime_2} 
	Z_{j_1,j^\prime_1} \otimes_1 \int_3 Z_{j_2 j_{3v}, j^\prime_2} \otimes_{23} 
	D^B_{j^\prime_1 j^\prime_2} \\ \nonumber
	=& \sum\limits_{j^\prime_1 j^\prime_2} 
	Z_{j_1,j^\prime_1} \otimes_1 \left(\int_3 Z_{j_2 j_{3v}, j^\prime_2} \right) \otimes_{2} 
	D^B_{j^\prime_1 j^\prime_2} \\
	=& \sum\limits_{j^\prime_1 j^\prime_2} 
	Z_{j_1,j^\prime_1} \otimes_1 (\colorbox{blue!50}{{$\displaystyle \delta_{j_3\bar{j}_2}$}}-\colorbox{purple!50}{{$\displaystyle \delta_{j_3j_2}$}}-\colorbox{green!50}{{$\displaystyle \delta_{j_3\bar{j}'_2}$}}+\colorbox{orange!50}{{$\displaystyle \delta_{j_3j_2'}$}})Z_{j_2j_2^\prime} \otimes_{2} 
	D^B_{j^\prime_1 j^\prime_2}.
\end{align}
The other term in the ``(permutations)'' has the same structure, but with $1 \leftrightarrow 2$.

We express the second term in Eq.~\eqref{eq:NumSum} in terms of bare quantities using Eq.~(70) from Ref.~\cite{Diehl:2018kgr}:
\begin{align} \nonumber
    -& (N_{j_3v}+\delta_{j_3\bar{j}_1}+\delta_{j_3\bar{j}_2}-\delta_{j_3j_1}-\delta_{j_3j_2})D_{j_1j_2} \\ \nonumber
    =&  (-\colorbox{red!50}{{$\displaystyle N_{j_3v}$}}-\colorbox{yellow!50}{{$\displaystyle \delta_{j_3\bar{j}_1}$}}-\colorbox{blue!50}{{$\displaystyle \delta_{j_3\bar{j}_2}$}} +\colorbox{yellow!50}{{$\displaystyle \delta_{j_3j_1}$}}+\colorbox{purple!50}{{$\displaystyle \delta_{j_3j_2}$}}) \sum_{j^\prime_1j^\prime_2} Z_{j_1  j^\prime_1} \otimes_1 
	Z_{j_2  j^\prime_2} \otimes_2  
	D^B_{j^\prime_1 j^\prime_2} \\ 
	+& 
(-N_{j_3v}-\delta_{j_3\bar{j}_1}-\delta_{j_3\bar{j}_2} +\delta_{j_3j_1}+\delta_{j_3j_2}) \sum\limits_{j^\prime} 
	Z_{j_1 j_2, j^\prime} \otimes_{12} 
	f^B_{j^\prime}.
\end{align}

As for the momentum sum rule, a number of cancellations now occur. The terms in the \colorbox{red!50}{{$\phantom{1}$}}, \colorbox{green!50}{{$\phantom{1}$}}, \colorbox{blue!50}{{$\phantom{1}$}}, \colorbox{orange!50}{{$\phantom{1}$}}, \colorbox{purple!50}{{$\phantom{1}$}} boxes cancel against each other directly, whilst the terms in the \colorbox{yellow!50}{{$\phantom{1}$}} boxes cancel against the other ``(permutations)'' term that we have not written out explicitly. Once again, the terms with $D^B$ cancel out, and we obtain (using Eq.~\eqref{eq:123simplify}):
\begin{align} \nonumber
    \Delta^{j_1j_2}_{\text{num}} 
    &= (-N_{j_3v}-\delta_{j_3\bar{j}_1}-\delta_{j_3\bar{j}_2} +\delta_{j_3j_1}+\delta_{j_3j_2}) \sum\limits_{j^\prime} 
	Z_{j_1 j_2, j^\prime} \otimes_{12} 
	f^B_{j^\prime}  \\ \nonumber
	&\phantomrel{=} + \sum\limits_{j_1^\prime} 
	Z_{j_1 j_2, j_1^\prime} \otimes_{12} 
		 (N_{j_3v}+\delta_{j_3\bar{j}'_1}-\delta_{j_3j'_1})f^B_{j_1^\prime} \\ \nonumber
	&\phantomrel{=} +
	\sum\limits_{j^\prime_1} 
	\int_3 Z_{j_1 j_2 j_{3v}, j^\prime_1} \otimes_{123} 
	f^B_{j^\prime_1} \\ \nonumber
	&= \sum_{j'_1,k'} \left( \int_3 Z_{j_1 j_2 j_{3v}, j^\prime_1} + \left[-\delta_{j_3\bar{j}_1} -\delta_{j_3\bar{j}_2} +\delta_{j_3j_1}+\delta_{j_3j_2} +\delta_{j_3\bar{j}'_1}-\delta_{j_3j'_1} \right] Z_{j_1j_2,j_1'} \right) \\
	& \qquad \quad \otimes_{12}  Z^{-1}_{j_1^\prime k^\prime} \otimes f_{k^\prime}.
\end{align}
According to a similar argument as was used for $\Delta_{\text{mtm}}$, we find that $\Delta_{\text{num}}=0$ -- that is, the number sum rule holds for the renormalized tPDFs. We also obtain a number sum rule for the $Z_{ijk,l}$ factors:
\begin{align}
    \int_3 Z_{j_1 j_2 j_{3v}, j^\prime_1} = \left( \delta_{j_3\bar{j}_1} +\delta_{j_3\bar{j}_2} -\delta_{j_3j_1}-\delta_{j_3j_2} -\delta_{j_3\bar{j}'_1}+\delta_{j_3j'_1} \right) Z_{j_1j_2,j_1'}.
\end{align}

\subsection{Check of the GS sum rules for tPDFs constructed with \pythia code}
\label{ss:results_tPDFs_check}

In Sections~\ref{ss:bare_tPDFs_sum_rules} - \ref{ss:renorm_tPDFs_sum_rules} we proved the GS sum rules generalized for the case of tPDFs.
Now let us demonstrate how one can build tPDFs which approximately obey  Eq.~\eqref{eq:triple_sum_rule_momentum} and Eq.~\eqref{eq:tiple_sum_rule_number}.
As in Section~\ref{ss:results_dPDFs_asym} we begin the discussion with the case of the asymmetric tPDFs which are given by
\begin{eqnarray}
	T_{j_1 j_2 j_3}(x_1, x_2, x_3, Q) &=& f^{r}_{j_1}(x_1, Q) ~f^{m\leftarrow j_1, x_1}_{j_2}(x_2, Q) ~f^{m\leftarrow j_1, x_1; j_2, x_2}_{j_3}(x_3, Q).
	\label{eq:pythia_tPDFs}
\end{eqnarray}
As for the dPDFs,  we only check the sum rules when integrating over the final parton (parton 3 here) -- this is the case in which we expect the sum rules to be satisfied the best. 
To construct tPDFs we use the \pythia code as it is explained in Appendix~\ref{appendix:how_to_call_modified_PDFs}. 
In this section we always set all scales in the tPDF to be equal (this is the case for which the tPDF sum rules should hold), and set the value of this common scale to $91$ GeV.

First of all let us check the momentum sum rule given by Eq.~\eqref{eq:triple_sum_rule_momentum}. 
Inserting  Eq.~\eqref{eq:pythia_tPDFs} into Eq.~\eqref{eq:triple_sum_rule_momentum}, we get:
\begin{eqnarray}
	\frac{1}{1 - x_1 - x_2}\,\sum\limits_{j_3}\int\limits^{1 - x_1 - x_2}_0 dx_3 ~x_3 ~f^{m\leftarrow j_1, x_1; j_2, x_2}_{j_3}(x_3, Q) \overset{?}{=} 1.
	\label{eq:triple_sum_rule_momentum_int}
\end{eqnarray}
The results of the numerical check of Eq.~\eqref{eq:triple_sum_rule_momentum_int} are given in 
Table~\ref{tab:triple_sum_rule_momentum_test}. 
The problem of oscillations for the \pythia dPDFs mentioned  in Section~\ref{s:results_dPDFs} becomes stronger for the case of tPDFs. 
Therefore, we increase a number of averaging calls by one order of magnitude comparing to the case of dPDFs. 
For our check we choose $j_1=j_2=u$, and vary  $x_1$ between  $10^{-6}$ and $0.8$ intervals while keeping  $x_2$  equal to $10^{-4}$. 
As a baseline for our computation we use naive tPDFs  defined in analogy with Eq.~\eqref{eq:naive_dPDFs} as
\begin{eqnarray}
	T_{j_1 j_2 j_3}(x_1, x_2, x_3, Q) = f^{r}_{j_1}(x_1, Q) ~f^{r}_{j_2}(x_2, Q) ~f^{r}_{j_3}(x_3, Q) ~
	\theta(1 - x_1 - x_2 - x_3).
	\label{eq:naive_tPDFs}
\end{eqnarray}

\begin{table} [b]
    \centering
    \begin{tabular}{ | c | c | c | c | c | c | c | }
    	\hline
        $x_1$ 		& $x_2$	& $j_1$ & $j_2$ &  \pythia tPDFs & Naive tPDFs	\\ \hline
        
		$10^{-6}$ 	& $10^{-4}$	&  $u$  &  $u$ & 0.996 & 0.996	\\ \hline
        $10^{-3}$ 	& $10^{-4}$	&  $u$  &  $u$ & 0.996 & 0.996	\\ \hline
        $10^{-1}$ 	& $10^{-4}$	&  $u$  &  $u$ & 1.008 & 1.106	\\ \hline
        
        0.2      	& $10^{-4}$	&  $u$  &  $u$ & 1.010 & 1.244	\\ \hline
        0.4      	& $10^{-4}$	&  $u$  &  $u$ & 1.011 & 1.649	\\ \hline
        0.8      	& $10^{-4}$	&  $u$  &  $u$ & 1.011 & 4.119	\\ \hline
        
	\end{tabular}
   \captionof{table}{Test of the momentum sum rule for the  tPDFs. In the ideal situation when the momentum sum rule is perfectly satisfied the number in each cell should be equal to unity.} 
   \label{tab:triple_sum_rule_momentum_test}
\end{table}

\begin{figure}
\begin{minipage}[h]{0.49\linewidth}
\center{\includegraphics[width=1.0\linewidth]{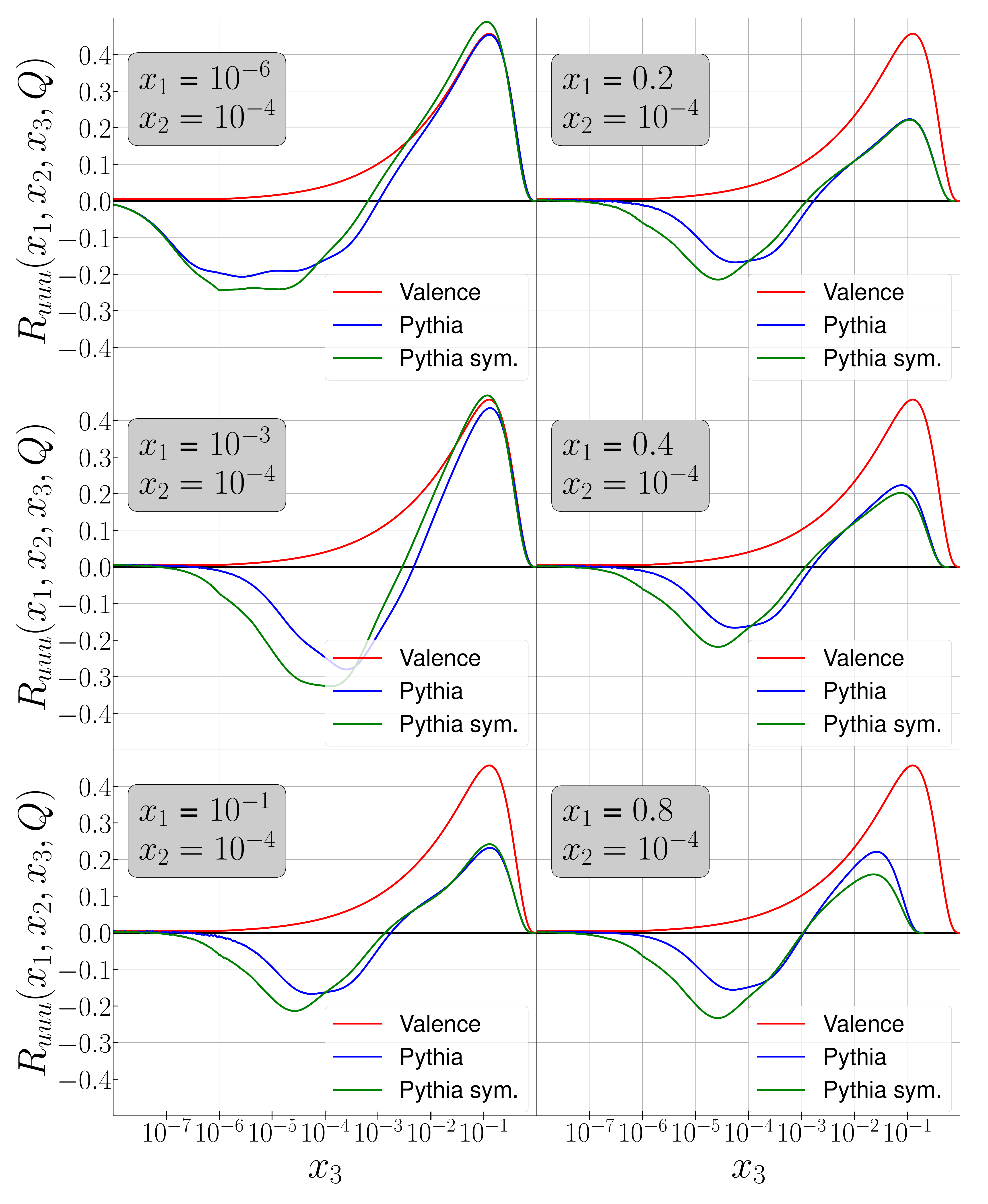}\\a)}
\end{minipage}
\hfill
\begin{minipage}[h]{0.49\linewidth}
\center{\includegraphics[width=1.0\linewidth]{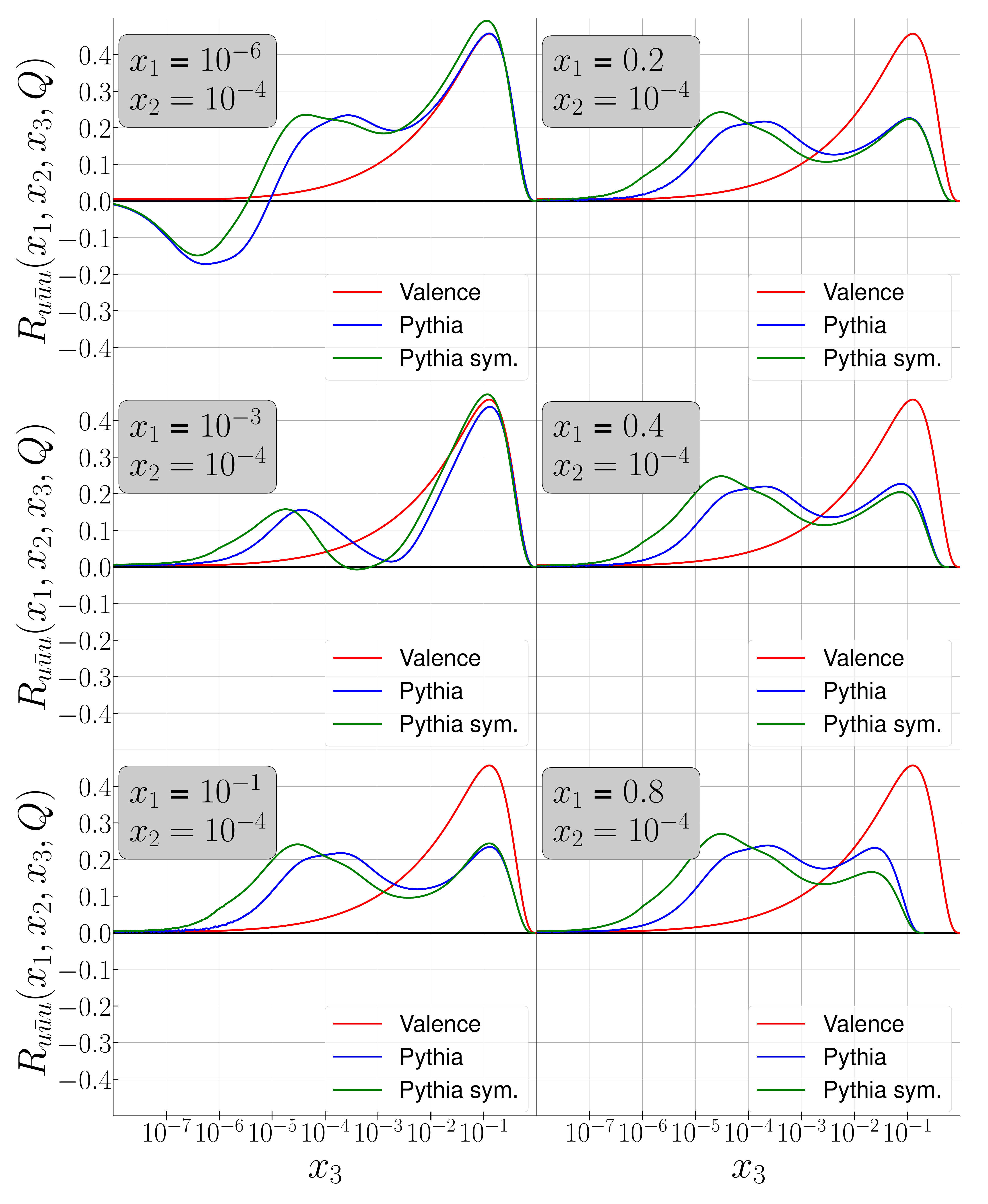}\\b)}
\end{minipage}
\cprotect\caption{The responses of the valence $u$-quark sPDF computed with asymmetric (blue) and symmetric (green) tPDFs:   a) $uuu$ flavour combination and  
b) $u\bar{u}u$  flavour combination.}
\label{fig:Pythia_sum_rules_Ruuu_Rubaru_sym}
\end{figure}

Now let us check the number rule given by Eq.~\eqref{eq:tiple_sum_rule_number}. 
In order to do that, similarly to Eq.~\eqref{eq:responce_function_def}, we define
\begin{eqnarray}
	R_{j_1 j_2 j_3}(x_1, x_2, x_3, Q) \equiv
	x_3 \, 
	\frac{ 
		T_{j_1 j_2 j_3}(x_1, x_2, x_3, Q) - T_{j_1 j_2 \bar{j_3}}(x_1, x_2, x_3, Q) 
	}
	{
		f_{j_1}^{r}(x_1, Q) f_{j_2}^{m\leftarrow j_1, x_1}(x_2, Q)
	},
	\label{eq:responce_function3_def}
\end{eqnarray}
which can be seen as a response of the valence sPDF $f_{j_{3v}}(x_3, Q)$ to the two interactions involving partons $j_1$ and $j_2$ with momentum fractions $x_1$ and $x_2$ scales respectively.
Due to the large number of different flavour combinations we restrict ourselves to a few illustrative cases. Let us first look at the $uuu$ and $u\bar{u}u$ combinations.
The results of our numerical checks are given
in Tables~\ref{tab:test_pythia_Ruuu} - \ref{tab:test_pythia_Ruubaru} and the corresponding response functions are given by blue lines in 
Fig.~\ref{fig:Pythia_sum_rules_Ruuu_Rubaru_sym}. 
We see that the aforementioned $1\rightarrow2$ splitting mechanism now give rise to a non-trivial modification of the valence $u$-quark sPDF after the first two  interactions. 
As a consequence, the  modified valence $u$-quark sPDF can possess several minima and maxima.
\begin{table}
    \begin{minipage}{1.0\linewidth}
    \centering
    \begin{tabular}{ | c | c | c | c | }
        \hline
        $x_1$ 		& $x_2$ 		& $N_{u_{v2}}$ \pythia & $N_{u_{v2}}$ Naive\\ \hline
	  	
		$10^{-6}$ & $10^{-4}$  & 0.025 & 2.006 \\ \hline
		$10^{-3}$ & $10^{-4}$  & 0.011 & 2.006 \\ \hline
		$10^{-1}$ & $10^{-4}$  & 0.007 & 2.005 \\ \hline
		0.2       & $10^{-4}$  & 0.006 & 2.005 \\ \hline
		0.4       & $10^{-4}$  & 0.005 & 1.997 \\ \hline
		0.8 	  & $10^{-4}$  	& 0.002 & 1.708 \\ \hline
	\end{tabular}
  	\caption{Integration over $R_{uuu}$ response function with respect to $x_3$ at fixed $x_1$, $x_2$. In the ideal situation when the GS sum rules are perfectly satisfied $N_{u_{v2}} = 0$.}
	\label{tab:test_pythia_Ruuu}
    \end{minipage}%
\end{table}
\begin{table}
    \begin{minipage}{1.0\linewidth}
    \centering
    \begin{tabular}{ | c | c | c | c | }
        \hline
        $x_1$ 		& $x_2$ 		& $N_{u_{v2}}$ \pythia & $N_{u_{v2}}$ Naive	\\ \hline
		
		$10^{-6}$ & $10^{-4}$  & 2.019 & 2.006 \\ \hline
		$10^{-3}$ & $10^{-4}$  & 2.005 & 2.006 \\ \hline
		$10^{-1}$ & $10^{-4}$  & 2.001 & 2.005 \\ \hline
		0.2       & $10^{-4}$  & 2.000 & 2.005 \\ \hline
		0.4       & $10^{-4}$  & 1.999 & 1.997 \\ \hline
		0.8       & $10^{-4}$  & 1.995 & 1.708 \\ \hline
	\end{tabular}
    \caption{Integration over $R_{u\bar{u}u}$ response function with respect to $x_3$ at fixed $x_1$, $x_2$. In the ideal situation when the GS sum rules are perfectly satisfied $N_{u_{v2}} = 2$.} 
    \label{tab:test_pythia_Ruubaru}
    \end{minipage} 
\end{table}

According to Eq.~\eqref{eq:tiple_sum_rule_number} the integrations over the
$uuu$ and $u\bar{u}u$ response functions should yield
\begin{eqnarray}
	\int\limits^{1 - x_1 - x_2}_0 \frac{dx_3}{x_3} ~R_{u u u}(x_1, x_2, x_3, Q) &=& 
	N_{u_v} - \delta_{u u} - \delta_{u u} + \delta_{\bar{u} u} + \delta_{\bar{u} u} = 0,
	\label{eq:R_u_u_u_num_test}\\
	\int\limits^{1 - x_1 - x_2}_0 \frac{dx_3}{x_3} ~R_{u \bar{u} u}(x_1, x_2, x_3, Q) &=& 
	N_{u_v} - \delta_{u u} - \delta_{u \bar{u}} + \delta_{\bar{u} u} + \delta_{\bar{u} \bar{u}} = 2.
	\label{eq:R_u_ubar_u_num_test}
\end{eqnarray}
We see that  for most of the cases the results of the numerical integration given in Table~\ref{tab:test_pythia_Ruuu} and Table~\ref{tab:test_pythia_Ruubaru} agree  with Eq.~\eqref{eq:R_u_u_u_num_test} and Eq.~\eqref{eq:R_u_ubar_u_num_test} up to a few per mille.

\begin{figure}
\center{\includegraphics[width=0.65\linewidth]{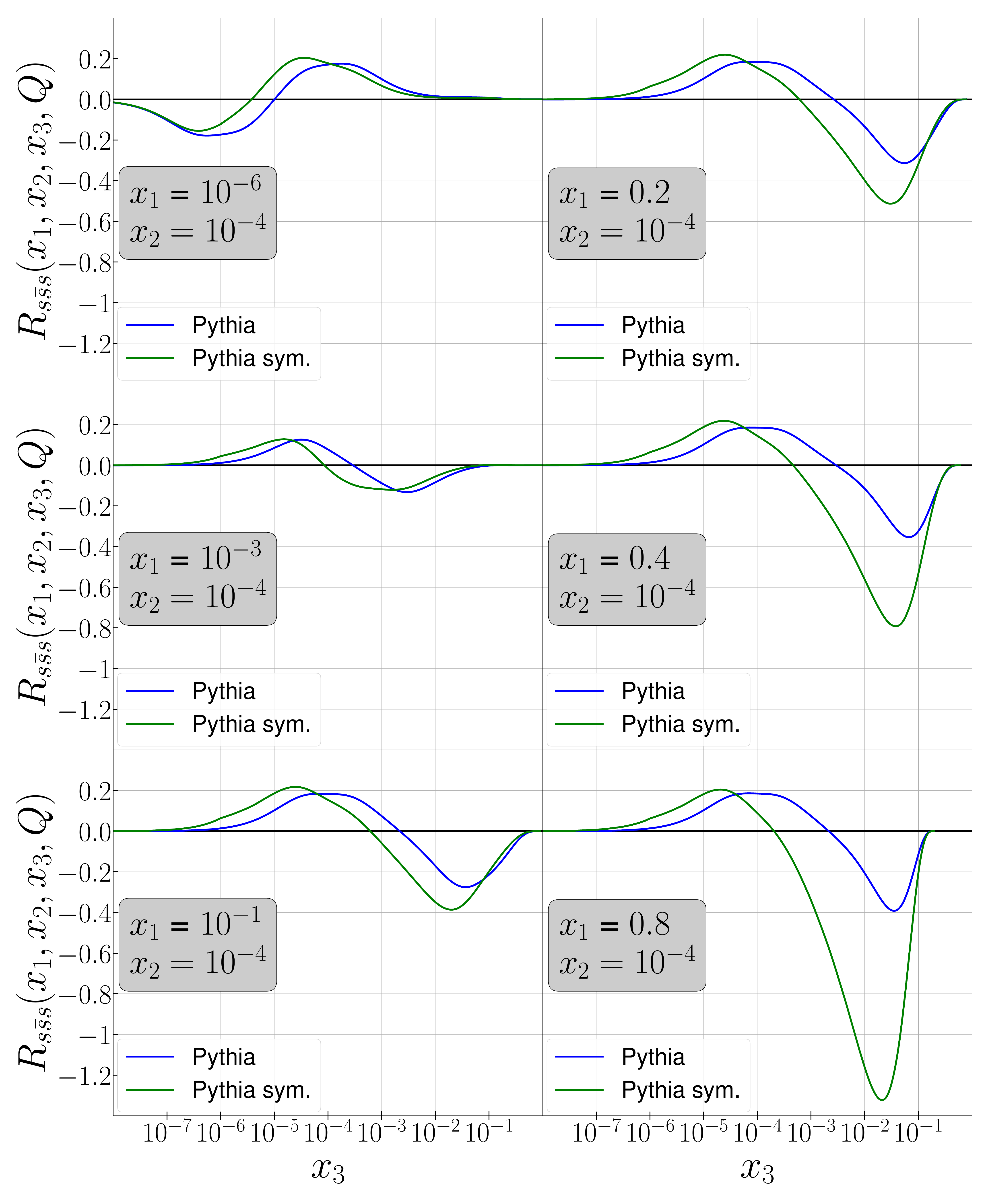}}
\cprotect\caption{Same as in Fig.~\ref{fig:Pythia_sum_rules_Ruuu_Rubaru_sym} but for the $s\bar{s}s$ flavour combination.}
\label{fig:Pythia_sum_rules_Rssbars_sym}
\end{figure}

Very similar results are obtained for the other flavour combinations. For example, for the $s\bar{s}s$ case the number sum rule reads as follows:
\begin{eqnarray}
	\int\limits^{1 - x_1 - x_2}_0 \frac{dx_3}{x_3} ~R_{s \bar{s} s}(x_1, x_2, x_3, Q) &=& 
	N_{s_v} - \delta_{s s} - \delta_{s \bar{s}} + \delta_{\bar{s} s} + \delta_{\bar{s} \bar{s}} = 0.
	\label{eq:R_s_sbar_s_num_test}
\end{eqnarray}
The response functions for this case are plotted in Fig.~\ref{fig:Pythia_sum_rules_Rssbars_sym}, and numerically we find that the integral on the left hand side of Eq.~\eqref{eq:R_s_sbar_s_num_test} does give zero up to deviations of order 0.001, for the same $x$ values as were probed in the $uuu$ and $u\bar{u}u$ cases. In Fig.~\ref{fig:Pythia_sum_rules_Rssbars_sym} we see once again that the companion quarks produced via the $g \rightarrow q \, \bar{q}$ splitting introduce additional strange and anti-strange quarks to the beam remnant, leading to peaks and troughs appearing in the response function.

Now let us move to discuss symmetrized tPDFs -- we recall that the tPDFs should be symmetric when the flavour, $x$ fraction and scale arguments for parton $i$ are exchanged with those of parton $j$ (where $i,j \in \left\{1,3\right\}$), and that the (equal-scale) tPDFs should satisfy the sum rules when integrating over any of the partons, not just the final one.
Similarly to Eq.~\eqref{eq:pythia_dPDFs_sym} we define symmetrized \pythia tPDFs as 
\begin{eqnarray}
	T^{\rm sym}_{j_1 j_2 j_3}(x_1, x_2, x_3, Q) = 
	\frac{1}{3!} \sum\limits_{\{1, 2, 3\}} T_{j_1 j_2 j_3}(x_1, x_2, x_3, Q),
	\label{eq:tPDFsym_def}
\end{eqnarray}
where the sum on the right hand side of Eq.~\eqref{eq:tPDFsym_def} runs over all partonic permutations. 
For the  sake of simplicity we omit the factorization scale label in what follows (recall we take \mbox{$Q = 91$ GeV} in the numerics).
The momentum rule Eq.~\eqref{eq:triple_sum_rule_momentum}, therefore, becomes
\begin{eqnarray}
    \frac{1}{1-x_1-x_2}
	\sum_{j_3} \int\limits^{1 - x_1 - x_2}_0 dx_3 x_3 ~
	\frac{T^{\rm sym}_{j_1 j_2 j_3}(x_1, x_2, x_3)}{D^{\rm sym}_{j_1 j_2}(x_1, x_2)}  &\overset{?}{=} 1.
    \label{eq:tiple_sum_rule_number_sym}
\end{eqnarray}

The numerical check of Eq.~\eqref{eq:tiple_sum_rule_number_sym} for the $j_1 = j_2 = u$ 
case is given in Table~\ref{tab:triple_sum_rule_sym_momentum_test}.  We see that when both $x_1$ and $x_2$ are small the sum rule is pretty well satisfied (with deviations of the order of only $4\%$), but the violations become larger when one $x$ fraction approaches $1$, reaching the $70\%$ level when $x_1 = 0.8, x_2 = 10^{-4}$.

\begin{table}
    \centering
	\begin{tabular}{ | c | c | c | c | c | c | c | c |}
    	\hline
        $x_1$ & $x_2$	& $j_1$ & $j_2$ &  \pythia tPDFs & \pythia tPDFs sym. & Naive tPDFs\\ 
        \hline

		$10^{-6}$ 	& $10^{-4}$ &  $u$  &  $u$  & 0.996 & 0.965 & 0.996	\\ \hline
        $10^{-3}$ 	& $10^{-4}$ &  $u$  &  $u$  & 0.996 & 0.967 & 0.996	\\ \hline
        $10^{-1}$ 	& $10^{-4}$ &  $u$  &  $u$  & 1.008 & 0.998 & 1.106	\\ \hline
        0.2      	& $10^{-4}$ &  $u$  &  $u$  & 1.010 & 1.029 & 1.244	\\ \hline
        0.4      	& $10^{-4}$ &  $u$  &  $u$  & 1.011 & 1.117 & 1.649	\\ \hline
        0.8      	& $10^{-4}$ &  $u$  &  $u$  & 1.011 & 1.719 & 4.119	\\ \hline

	\end{tabular}
   \captionof{table}{Test of the momentum sum rule for the  tPDFs. In the ideal situation when the momentum sum rule is perfectly satisfied the number in each cell should be equal to unity.} 
   \label{tab:triple_sum_rule_sym_momentum_test}
\end{table} 

Similarly to the case of the symmetrized dPDFs we define the symmetrized  response functions as 
\begin{eqnarray}
	R^\ast_{j_1 j_2 j_3}(x_1, x_2, x_3) \equiv
	x_3 \, 
	\frac{ 
		T^{\rm sym}_{j_1 j_2 j_3}(x_1, x_2, x_3) - T^{\rm sym}_{j_1 j_2 \bar{j_3}}(x_1, x_2, x_3) 
	}
	{
		D^{\rm sym}_{j_1 j_2}(x_1, x_2)
	},
	\label{eq:responce_function3_sym_def}
\end{eqnarray}
where $D^{\rm sym}_{j_1 j_2}$ are  symmetrized \pythia dPDFs  defined by Eq.~\eqref{eq:pythia_dPDFs_sym}.
The numerical checks for $R^\ast_{u u u}$ and $R^\ast_{u \bar{u} u}$ are given in Tables~\ref{tab:test_pythia_Ruuu_sym} and \ref{tab:test_pythia_Ruubaru_sym} 
and in Fig.\ref{fig:Pythia_sum_rules_Ruuu_Rubaru_sym}.
We see that for $R^\ast_{u u u}$ and $R^\ast_{u \bar{u} u}$, the number sum rules are preserved reasonably well -- at most of the $x_1,x_2$ points tested, we have deviations of order $0.2-0.3$. At certain points we see larger deviations up to $\sim 0.5-0.6$ -- for the $uuu$ case, this occurs when one of the $x$ values gets close to $1$, but interestingly, for the $u\bar{u}u$ case, this occurs when the $x$ fractions are at the smallest values tested ($x_1 =10^{-6},x_2=10^{-4}$).  
\begin{table}
\begin{center}
	\begin{tabular}{ | c | c | c | c | c |}
    	\hline
        $x_1$ & $x_2$ & $N_{u_{v2}}$ \pythia & $N_{u_{v2}}$ \pythia sym. & $N_{u_{v2}}$ Naive\\ \hline
	  	
		$10^{-6}$ 	& $10^{-4}$  & 0.025 & 0.108 & 2.006 \\ \hline
		$10^{-3}$ 	& $10^{-4}$  & 0.011 & -0.276 & 2.006 \\ \hline
		$10^{-1}$ 	& $10^{-4}$  & 0.007 & -0.232 & 2.005 \\ \hline
		0.2			& $10^{-4}$  & 0.006 & -0.242 & 2.005 \\ \hline
		0.4			& $10^{-4}$  & 0.005 & -0.317 & 1.997 \\ \hline
		0.8			& $10^{-4}$  & 0.002 & -0.589 & 1.708 \\ \hline
								
  	 \end{tabular}
  	 \captionof{table}{Same as in Table~\ref{tab:test_pythia_Ruuu} but for the symmetrized $R^\ast_{uuu}$ response function. In the ideal situation when the GS sum rules are perfectly satisfied $N_{u_{v2}} = 0$.}
	 \label{tab:test_pythia_Ruuu_sym}
\end{center}
\end{table}
\begin{table}
\begin{center}
	\begin{tabular}{ | c | c | c | c | c |}
    	\hline
        $x_1$ & $x_2$ & $N_{u_v}$ \pythia & $N_{u_v}$ \pythia sym. & $N_{u_v}$ Naive\\ \hline
	  	
		$10^{-6}$	& $10^{-4}$  & 2.019 & 2.542 & 2.006 \\ \hline
		$10^{-3}$	& $10^{-4}$  & 2.005 & 2.154 & 2.006 \\ \hline
		$10^{-1}$	& $10^{-4}$  & 2.001 & 2.188 & 2.005 \\ \hline
		0.2 		& $10^{-4}$  & 2.000 & 2.189 & 2.005 \\ \hline
		0.4			& $10^{-4}$  & 1.999 & 2.161 & 1.997 \\ \hline
		0.8 		& $10^{-4}$  & 1.995 & 2.079 & 1.708 \\ \hline
	\end{tabular}
    \captionof{table}{Same as in Table~\ref{tab:test_pythia_Ruubaru} but for symmetrized $R^\ast_{u\bar{u}u}$ response function. In the ideal situation when the GS sum rules are perfectly satisfied $N_{u_{v2}} = 2$.} 
    \label{tab:test_pythia_Ruubaru_sym}						
\end{center}
\end{table}

Whereas the symmetrized tPDFs in Fig.~\ref{fig:Pythia_sum_rules_Ruuu_Rubaru_sym} show relatively small violations of the GS sum rules the problem of the large peaks discussed in  Section~\ref{s:results_dPDFs}  also shows for certain tPDFs, for instance for the $s\bar{s}s$ combination in Fig.~\ref{fig:Pythia_sum_rules_Rssbars_sym}.
Here we see a large violation of the number sum rule when $x_1$ is very large, $x_1 = 0.8$, similar to what we saw for the dPDF case in Fig.~\ref{fig:Pythia_sum_rules_Rssbar_sym}.

\section{Summary and conclusions}
\label{s:summary}

We have demonstrated how one can use the MPI model of the \pythia event generator to construct asymmetric and symmetric dPDFs which approximately obey the Gaunt-Stirling (GS) number and momentum sum rules. 
The asymmetric dPDFs constructed in this way obey the GS sum rules  at a few percent accuracy level in one of the arguments. 
We found that the symmetrized dPDFs obey the GS sum rules at about 20$\%$ accuracy level for $x$ fractions smaller than 0.4. 
This is not perfect, but we remind the reader that the task of constructing dPDFs that even approximately satisfy the sum rules in this way is highly nontrivial - for example, in the GS09 paper \cite{Gaunt:2009re}, the sum rules are only satisfied ``to better than $25\%$ precision (for $x<0.8$)''. 
The concept of a ``response function'' was introduced for both the momentum and number sum rules, a function of $x_1$ and $x_2$ constructed out of dPDFs and sPDFs that, when integrated over $x_2$, should give a constant value according the sum rules. The precise definitions of these response functions are given in Eqs.~\eqref{eq:responce_function_def} and \eqref{eq:momentum_sum}. An interesting observation we made is that even though only the ``areas'' under the response function curves in the $x_2$ direction are fixed by the sum rules, the overall response functions computed with \pythia and GS09 dPDFs show very similar behaviour.

We also compared the GS09 and \pythia dPDFs in terms of their predictions for the double Drell-Yan cross section, focussing in particular on the distribution in the maximal rapidity separation between the leptons ($\Delta \rm Y$). We observed differences in shape between the GS09 and \pythia $\Delta \rm Y$ distributions, with discrepancies up to the $10\%$ level. This indicates that although the response function curves are rather similar between the two models, there are notable differences between the dPDFs themselves.

Additionally, we generalized the GS sum rules to the case of tPDFs. We demonstrated that these hold at the bare level using the light-cone wave function approach, and showed that they also hold at the renormalized level following the approach of Ref.~\cite{Diehl:2018kgr}. We demonstrated how one can use the \pythia code to construct the asymmetric and symmetric tPDFs which approximately obey the corresponding version of the GS sum rules. 
We argue that in the absence of available sets of tPDFs one can use \pythia tPDFs as a good first  approximation for phenomenological studies of  the TPS phenomena.

Finally, we shall note that both GS09 and \pythia approaches do not take into account transverse dependence ($\bm{y}$-dependence) between partons, and that the cross section formulae for DPS or TPS actually involve the quantities that include this transverse dependence, the double parton distributions (DPDs) or triple parton distributions (TPDs). Theoretical studies of DPS phenomena \citep{Blok:2011bu, Blok:2013bpa, Ryskin:2011kk,  Ryskin:2012qx, Manohar:2012pe, Diehl:2017kgu, Diehl:2021gvs} have demonstrated that proper account of the $\bm{y}$-dependence results in contributions with $1\rightarrow2$ partonic splittings  being numerically more important than in approaches neglecting transverse dependence of dPDFs, \textit{e.g.} GS09.
Even though dPDFs and tPDFs do not appear directly in the rigorous treatment of the DPS and TPS cross section formulae, they are nonetheless linked to the DPDs and TPDs which do -- the former can be obtained from the latter via an appropriate integral over transverse separation parameter $\bm{y}$.
One could, presumably, extend the \pythia approach to construct DPDs and TPDs approximately satisfying sum rules, following a  similar approach to Ref.~\citep{Diehl:2020xyg}. 
The DPDs and TPDs constructed in a such way would be a useful low-scale input for new tools (\textit{e.g.} dShower \citep{Cabouat:2019gtm, Cabouat:2020ssr}) that follow the theoretical rigorous approach to multiple scattering.

\section*{Acknowledgments}
The work of OF has received funding from the European Union's Horizon 2020 research and innovation programme as part of the Marie Sk\l{}odowska-Curie Innovative Training Network MCnetITN3 (grant agreement no. 722104), the Deutsche Forschungsgemeinschaft (DFG) through the Research Training Group ``GRK 2149: Strong and Weak Interactions - from Hadrons to Dark Matter'' and the curiosity-driven grant ``Using jets to challenge the Standard Model of particle physics'' from Universit\`{a} di Genova.  The work of
JRG is supported by the Royal Society through Grant URF\textbackslash R1\textbackslash 201500.

We  acknowledge Torbj\"{o}rn Sj\"{o}strand for providing detailed instructions on how to access the values of modified sPDFs used inside of the MPI model of \pythia, as well as proposing the concept of response functions that we used in this article. OF  also  acknowledges Johannes Bellm,  Anna Kulesza and  Leif L\"onnblad for useful and fruitful discussions and  thanks  all members of the theoretical particle physics group of the Lund University for the warm hospitality and friendly atmosphere. 

All diagrams in this paper were created with the JaxoDraw code~\citep{Binosi:2003yf}. 
All plots in this paper were created utilizing Matplotlib~\citep{Hunter:2007ouj} and NumPy~\citep{NumPy} libraries. 
For our computations and simulations involving sPDFs we were using LHAPDF6 library~\citep{Buckley:2014ana} and a central value of the
MSTW2008 LO PDF set~\citep{Martin:2009iq}.
Our scripts and analysis files can be obtained on request.

\appendix
\section{How to access modified sPDFs in \pythia}
\setcounter{figure}{0}    
\label{appendix:how_to_call_modified_PDFs}
By design the values of modified sPDFs being used to simulate DPS and MPI processes are not included in the standard  \pythia output. 
However, with the help of  the methods of the class \verb|BeamParticle| one can calculate the value of the sPDF being used for the second hard interaction $f_{j_2}^{m\leftarrow j_1, x_1}(x_2, Q_2)$. 
In Listing~\ref{fig:accessing_dPDFs} we provide an example of a C++ function which allows to get access to the sPDFs  modified after the first interaction. 
As  arguments it takes the reference \verb|&beam|
to the object of the \verb|BeamParticle| type, the Particle Data Group ID numbers, \mbox{Bjorken-$x$'es} $x_1$, $x_2$ and factorization scales  $Q_1$, $Q_2$. 
The reference to the beam object  can be obtained by setting  \verb|BeamParticle &beam = pythia.beamA;| in the analysis code 
after the initialization of the \pythia object.
The dPDF, therefore, is given by Eq.~\eqref{eq:pythia_dPDFs}.

\begin{figure}[!h]
\begin{tiny}
\lstset{language=C++}          
\begin{lstlisting}[frame=single,  basicstyle=\footnotesize]  % Start your code-block

double second_xPDF (BeamParticle &beam, int id1, int id2,  
				double x1, double x2, 
				double Q1, double Q2) {
		    
    double res = 0.;
    
    // to avoid clash of notation
    int ida = (id1 == 0) ? 21 : id1;
    int idb = (id2 == 0) ? 21 : id2;
    
    // set the first sPDF
    beam.clear();
    beam.append (0, ida, x1);
    beam.xfISR  (0, ida, x1, Q1 * Q1);
    beam.pickValSeaComp();
    
    // get the second sPDF
    res = beam.xfMPI (idb, x2, Q2 * Q2);
	
    return res;
}
\end{lstlisting}
\end{tiny}
\renewcommand{\figurename}{Listing}
\caption{An example of a C++ function to access sPDFs being used for the second hard interaction.}
\label{fig:accessing_dPDFs}
\end{figure}

On can easily extend this approach to get sPDFs after $n$ subsequent MPI events.
For example, if we need to get the sPDF being used to simulate the third interaction we need to append the lines given in Listing~\ref{fig:accessing_tPDFs} to the function defined in Listing~\ref{fig:accessing_dPDFs}. 
The tPDFs are then given by Eq.~\eqref{eq:pythia_tPDFs}.
\begin{figure}
\begin{tiny}
\lstset{language=C++}         
\begin{lstlisting}[frame=single,  basicstyle=\footnotesize]  % Start your code-block
  
int idc = (id3 == 0) ? 21 : id3;
	
// Take into account changes of the beam remnant after second hard interaction
beam.append (1, idb, x2);
beam.xfISR  (1, idb, x2, Q22);
beam.pickValSeaComp();
	
// Get third sPDF
double x_pdf3 = beam.xfMPI (idc, x3, Q23);
		
res = x_pdf3;       

\end{lstlisting}
\end{tiny}
\renewcommand{\figurename}{Listing}
\caption{Modification of the function in Listing~\ref{fig:accessing_dPDFs} to the case of TPS.}
\label{fig:accessing_tPDFs}
\end{figure}

\section{\pythia setup we use.}
\label{appendix:set_up_we_use}
Here we provide the setup we have used for our simulations of the double Drell-Yan production.

\begin{table}[!hbt]
      \begin{minipage}{.5\linewidth}
      \centering
      \begin{tabular}[t]{ | l | c |}
			\hline
			
			PYTHIA FLAG							& VALUE \\ \hline
			\verb|Beams:frameType|			 	& 1		\\ \hline
			\verb|WeakSingleBoson:ffbar2gmZ| 	& on	\\ \hline
			\verb|HardQCD:nQuarkNew|  			& 4		\\ \hline
			\verb|SecondHard:generate| 			& on	\\ \hline
			\verb|SecondHard:SingleGmZ| 		& on	\\ \hline
			\verb|SigmaProcess:renormScale1| 	& 2		\\ \hline
			\verb|SigmaProcess:factorScale1| 	& 2		\\ \hline
			\verb|SigmaProcess:renormScale2| 	& 5		\\ \hline
			\verb|SigmaProcess:factorScale2| 	& 5		\\ \hline
			\verb|SigmaProcess:renormFixScale| & 8281.0\\ \hline
			\verb|SigmaProcess:factorFixScale| & 8281.0\\ 

			\hline
		\end{tabular}
		\caption*{}
	 \label{}
     \end{minipage}%
     \begin{minipage}{.5\linewidth}
     \centering
     \begin{tabular}[t]{ | l | c |}
    	\hline
	    	PYTHIA FLAG							& VALUE \\ \hline

			\verb|23:onMode| 					& off 	\\ \hline
			\verb|23:onIfAny| 					& 11, 13 \\ \hline
			\verb|22:onMode| 					& off	\\ \hline
			\verb|22:onIfAny| 					& 11, 13	\\ \hline
			\verb|PDF:pSet| 					& 5		\\ \hline
			\verb|PartonLevel:ISR| 				& off	\\ \hline
			\verb|PartonLevel:FSR| 				& off	\\ \hline
			\verb|PartonLevel:MPI| 				& off	\\ \hline
			\verb|HadronLevel:all| 				& off	\\ \hline
			\verb|PartonLevel:Remnants|  		& off	\\ \hline
			\verb|Check:event| 					& off	\\ 
	    		    	
		\hline
   		\end{tabular}
   	\caption*{} 
    \label{}
    \end{minipage} 
    \caption{\pythia settings we use}
    \label{tab:Pythia_settings_new_vs_old_DPS}
\end{table}

\clearpage
\bibliography{dps_sum_rules_v1_0_1}

\end{document}